%% file: main.tex
\documentclass[11pt,fleqn]{book} 

\input{structure.tex}

\input{Acrodef.tex}
\begin{document}
\newcounter{boxcount}
\setcounter{boxcount}{1}
\input{Cover_TOC.tex}
\label{part:RaD}
\pagenumbering{roman}

\newpage
\input{Introduction.tex}
\input{Facilities_and_Infrastructures.tex}
\input{Newtonian_Noise}
\input{Suspension_and_Isolation}

\input{Cryogenics}
\input{Core_Optics.tex}

\input{coatings.tex}

\input{Light_Sources}
\input{Quantum_Enhancements}
\input{Aux_Optics.tex}
\input{Sim_and_Control}
\input{calibration.tex}
\input{Outlook.tex}
\appendix
 \titleformat{\chapter}[display]
    {\normalfont\huge\bfseries}{\chaptertitlename\ \thechapter}{20pt}{\Huge}
\clearpage
\pagenumbering{arabic}
\renewcommand*{\thepage}{A\arabic{page}}
\input{Appendix_Core_Optics.tex}
\clearpage
\pagenumbering{arabic}
\renewcommand*{\thepage}{B\arabic{page}}
\input{Appendix_Coating.tex}

\clearpage
\pagenumbering{arabic}
\renewcommand*{\thepage}{AC\arabic{page}}
\input{acronyms}
\bibliographystyle{unsrtshort}
\bibliography{GWrefs}

\end{document}

%% file: structure.tex
\usepackage[top=2.54cm,bottom=2.54cm,left=2.54cm,right=2.54cm,headsep=10pt,letterpaper]{geometry} 

\usepackage{xcolor} 
\definecolor{ocre}{RGB}{52,177,201} 

\usepackage{acro}
\usepackage{authblk}

\usepackage{avant} 
\usepackage{mathptmx} 

\usepackage{microtype} 
\usepackage[utf8]{inputenc} 
\usepackage[T1]{fontenc} 

\usepackage{titlesec} 

\usepackage{graphicx} 
\graphicspath{{Pictures/}{figures/cosmo/}} 

\usepackage{lipsum} 

\usepackage{tikz} 

\usepackage[english]{babel} 

\usepackage{enumitem} 
\setlist{nolistsep} 

\usepackage{booktabs} 

\usepackage{eso-pic} 

\usepackage{titletoc} 

\contentsmargin{0cm} 
\titlecontents{chapter}[1.25cm] 
{\addvspace{15pt}\large\sffamily\bfseries} 
{\color{ocre!60}\contentslabel[\Large\thecontentslabel]{1.25cm}\color{ocre}} 
{}
{\color{ocre!60}\normalsize\sffamily\bfseries\;\titlerule*[.5pc]{.}\;\thecontentspage} 
\titlecontents{section}[1.25cm] 
{\addvspace{5pt}\sffamily\bfseries} 
{\contentslabel[\thecontentslabel]{1.25cm}} 
{}
{\sffamily\hfill\color{black}\thecontentspage} 
[]
\titlecontents{subsection}[1.25cm] 
{\addvspace{1pt}\sffamily\small} 
{\contentslabel[\thecontentslabel]{1.25cm}} 
{}
{\sffamily\;\titlerule*[.5pc]{.}\;\thecontentspage} 
[]

\titlecontents{lsection}[0em] 
{\footnotesize\sffamily} 
{}
{}
{}

\titlecontents{lsubsection}[.5em] 
{\normalfont\footnotesize\sffamily} 
{}
{}
{}

\usepackage{fancyhdr} 

\fancypagestyle{plain}{\fancyhead{}} 

\makeatletter
\renewcommand{\cleardoublepage}{
\clearpage\ifodd\c@page\else
\hbox{}
\vspace*{\fill}
\thispagestyle{empty}
\fi}

\usepackage{amsmath,amsfonts,amssymb,amsthm} 

\newtheoremstyle{ocrenumbox}
{0pt}
{0pt}
{\normalfont}
{}
{\small\bf\sffamily\color{ocre}}
{\;}
{0.25em}
{\small\sffamily\color{ocre}\thmname{#1}\nobreakspace\thmnumber{\@ifnotempty{#1}{}\@upn{#2}}
\thmnote{\nobreakspace\the\thm@notefont\sffamily\bfseries\color{black}---\nobreakspace#3.}} 

\newtheoremstyle{blacknumex}
{5pt}
{5pt}
{\normalfont}
{} 
{\small\bf\sffamily}
{\;}
{0.25em}
{\small\sffamily{\tiny\ensuremath{\blacksquare}}\nobreakspace\thmname{#1}\nobreakspace\thmnumber{\@ifnotempty{#1}{}\@upn{#2}}
\thmnote{\nobreakspace\the\thm@notefont\sffamily\bfseries---\nobreakspace#3.}}

\newtheoremstyle{blacknumbox} 
{0pt}
{0pt}
{\normalfont}
{}
{\small\bf\sffamily}
{\;}
{0.25em}
{\small\sffamily\thmname{#1}\nobreakspace\thmnumber{\@ifnotempty{#1}{}\@upn{#2}}
\thmnote{\nobreakspace\the\thm@notefont\sffamily\bfseries---\nobreakspace#3.}}

\newtheoremstyle{ocrenum}
{5pt}
{5pt}
{\normalfont}
{}
{\small\bf\sffamily\color{ocre}}
{\;}
{0.25em}
{\small\sffamily\color{ocre}\thmname{#1}\nobreakspace\thmnumber{\@ifnotempty{#1}{}\@upn{#2}}
\thmnote{\nobreakspace\the\thm@notefont\sffamily\bfseries\color{black}---\nobreakspace#3.}} 
\makeatother

\newcounter{dummy}
\numberwithin{dummy}{section}
\theoremstyle{ocrenumbox}
\newtheorem{theoremeT}[dummy]{Theorem}

\newtheorem{exerciseT}{Exercise}[chapter]
\theoremstyle{blacknumex}
\newtheorem{exampleT}{Example}[chapter]
\theoremstyle{blacknumbox}

\newtheorem{definitionT}{Definition}[section]
\newtheorem{corollaryT}[dummy]{Corollary}
\theoremstyle{ocrenum}

\RequirePackage[framemethod=default]{mdframed} 

\newmdenv[skipabove=7pt,
skipbelow=7pt,
backgroundcolor=black!5,
linecolor=ocre,
innerleftmargin=5pt,
innerrightmargin=5pt,
innertopmargin=5pt,
leftmargin=0cm,
rightmargin=0cm,
innerbottommargin=5pt]{tBox}

\newmdenv[skipabove=7pt,
skipbelow=7pt,
rightline=false,
leftline=true,
topline=false,
bottomline=false,
backgroundcolor=ocre!10,
linecolor=ocre,
innerleftmargin=5pt,
innerrightmargin=5pt,
innertopmargin=5pt,
innerbottommargin=5pt,
leftmargin=0cm,
rightmargin=0cm,
linewidth=4pt]{eBox}

\newmdenv[skipabove=7pt,
skipbelow=7pt,
rightline=false,
leftline=true,
topline=false,
bottomline=false,
linecolor=ocre,
innerleftmargin=5pt,
innerrightmargin=5pt,
innertopmargin=0pt,
leftmargin=0cm,
rightmargin=0cm,
linewidth=4pt,
innerbottommargin=0pt]{dBox}

\newmdenv[skipabove=7pt,
skipbelow=7pt,
rightline=false,
leftline=true,
topline=false,
bottomline=false,
linecolor=gray,
backgroundcolor=black!5,
innerleftmargin=5pt,
innerrightmargin=5pt,
innertopmargin=5pt,
leftmargin=0cm,
rightmargin=0cm,
linewidth=4pt,
innerbottommargin=5pt]{cBox}

\makeatletter
\renewcommand{\@seccntformat}[1]{\llap{\textcolor{ocre}{\csname the#1\endcsname}\hspace{1em}}}
\renewcommand{\section}{\@startsection{section}{1}{\z@}
{-2ex \@plus -1ex \@minus -.4ex}
{1ex \@plus.2ex }
{\normalfont\large\sffamily\bfseries}}
\renewcommand{\subsection}{\@startsection {subsection}{2}{\z@}
{-1ex \@plus -0.1ex \@minus -.4ex}
{0.5ex \@plus.2ex }
{\normalfont\sffamily\bfseries}}
\renewcommand{\subsubsection}{\@startsection {subsubsection}{3}{\z@}
{-1ex \@plus -0.1ex \@minus -.2ex}
{.2ex \@plus.2ex }
{\normalfont\small\sffamily\bfseries}}
\renewcommand\paragraph{\@startsection{paragraph}{4}{\z@}
{-1ex \@plus-.2ex \@minus .2ex}
{.1ex}
{\normalfont\small\sffamily\bfseries}}

\usepackage[colorlinks=true,citecolor=blue,linkcolor=blue]{hyperref}

\newcommand{\thechapterimage}{}
\newcommand{\chapterimage}[1]{\renewcommand{\thechapterimage}{#1}}

\def\thechapter{\arabic{chapter}}
\def\@makechapterhead#1{
\thispagestyle{empty}
{\centering \normalfont\sffamily
\ifnum \c@secnumdepth >\m@ne
\if@mainmatter
\startcontents
\begin{tikzpicture}[remember picture,overlay]
\node at (current page.north west)
{\begin{tikzpicture}[remember picture,overlay]
\node[anchor=north west,inner sep=0pt] at (0,0) {\includegraphics[width=\paperwidth]{\thechapterimage}};
\draw[anchor=west] (2.46cm,-6cm) node [rounded corners=20pt,fill=ocre!10!white,text opacity=10,draw=ocre,draw opacity=1,line width=1.5pt,fill opacity=.6,inner sep=12pt]{\LARGE\sffamily\bfseries\textcolor{black}{\thechapter. #1\strut\makebox[22cm]{}}};
\end{tikzpicture}};
\end{tikzpicture}}
\par\vspace*{130\p@}
\fi
\fi}

\def\@makeschapterhead#1{
\thispagestyle{empty}
{\centering \normalfont\sffamily
\ifnum \c@secnumdepth >\m@ne
\if@mainmatter
\begin{tikzpicture}[remember picture,overlay]
\node at (current page.north west)
{\begin{tikzpicture}[remember picture,overlay]
\node[anchor=north west,inner sep=0pt] at (0,0) {\includegraphics[width=\paperwidth]{\thechapterimage}};
\draw[anchor=west] (2.46cm,-6cm) node [rounded corners=20pt,fill=ocre!10!white,fill opacity=.6,inner sep=12pt,text opacity=1,draw=ocre,draw opacity=1,line width=1.5pt]{\LARGE\sffamily\bfseries\textcolor{black}{#1\strut\makebox[22cm]{}}};
\end{tikzpicture}};
\end{tikzpicture}}
\par\vspace*{120\p@}
\fi
\fi
}
\makeatother

\usepackage{graphicx}
\graphicspath{{./figures/}}
\usepackage{appendix}
\usepackage{latexsym,amsmath,amsfonts,amssymb,booktabs}
\usepackage[font=small]{caption}
\usepackage{slashed,upgreek,amscd,cancel,tensor,color}
\usepackage{adjustbox}
\usepackage[numbers,compress,square]{natbib}
\usepackage{epsfig,latexsym}
\usepackage{url}
\numberwithin{equation}{section}
\usepackage{doi}
\usepackage{subcaption}
\usepackage{mathtools}
\usepackage{textcomp}
\usepackage{upgreek}
\usepackage{stfloats}
\usepackage{afterpage}
\usepackage{multirow}

\definecolor{antiquewhite}{rgb}{0.98, 0.92, 0.84}
\definecolor{aliceblue}{rgb}{0.94, 0.97, 1.0}
\definecolor{blanchedalmond}{rgb}{1.0, 0.92, 0.8}
\definecolor{beaublue}{rgb}{0.74, 0.83, 0.9}
\definecolor{beaubluedark}{rgb}{0.79, 0.85, 0.95}
\definecolor{amber(sae/ece)}{rgb}{1.0, 0.49, 0.0}
\definecolor{airforceblue}{rgb}{0.36, 0.54, 0.66}
\definecolor{aliceblue}{rgb}{0.94, 0.97, 1.0}
\definecolor{alizarin}{rgb}{0.82, 0.1, 0.26}
\definecolor{almond}{rgb}{0.94, 0.87, 0.8}
\definecolor{amaranth}{rgb}{0.9, 0.17, 0.31}
\definecolor{amber}{rgb}{1.0, 0.75, 0.0}
\definecolor{amber(sae/ece)}{rgb}{1.0, 0.49, 0.0}
\definecolor{americanrose}{rgb}{1.0, 0.01, 0.24}
\definecolor{amethyst}{rgb}{0.6, 0.4, 0.8}
\definecolor{anti-flashwhite}{rgb}{0.95, 0.95, 0.96}
\definecolor{antiquebrass}{rgb}{0.8, 0.58, 0.46}
\definecolor{antiquefuchsia}{rgb}{0.57, 0.36, 0.51}
\definecolor{antiquewhite}{rgb}{0.98, 0.92, 0.84}
\definecolor{ao}{rgb}{0.0, 0.0, 1.0}
\definecolor{ao(english)}{rgb}{0.0, 0.5, 0.0}
\definecolor{applegreen}{rgb}{0.55, 0.71, 0.0}
\definecolor{apricot}{rgb}{0.98, 0.81, 0.69}
\definecolor{aqua}{rgb}{0.0, 1.0, 1.0}
\definecolor{aquamarine}{rgb}{0.5, 1.0, 0.83}
\definecolor{beaver}{rgb}{0.62, 0.51, 0.44}
\definecolor{blizzardblue}{rgb}{0.67, 0.9, 0.93}
\definecolor{azure}{rgb}{0.0, 0.5, 1.0}

\usepackage{pdfpages}
\usepackage[squaren]{SIunits}
\usepackage{etoolbox}
\usepackage{multicol}
 \usepackage{vwcol}
\usepackage{placeins}
\usepackage[absolute]{textpos}
\usepackage{ulem} 
\usepackage{tcolorbox}
\usepackage{wrapfig}

\newtcolorbox{DetBox}[2][]{%
standard jigsaw,colframe=green!40!black,colback=blanchedalmond!10!white,opacityback=0.6,coltext=black,oversize, size=small, title=Infobox~\theboxcount: #2,#1}

\newtcolorbox{Infobox}[2][]{%
standard jigsaw,colframe=ocre,colback=blanchedalmond!10!white,opacityback=0.6,coltext=black, size=small, title=Infobox~\theboxcount: #2,#1}

\newcommand{\ao}{Applied Optics}

%% file: Acrodef.tex
\DeclareAcronym{3G}{short = 3G , long = third generation}
\DeclareAcronym{2G}{short = 2G , long = second generation}
\DeclareAcronym{1G}{short = 1G , long = first generation}
\DeclareAcronym{ET}{short = ET , long = Einstein Telescope}
\DeclareAcronym{CE}{short = CE , long = Cosmic Explorer}
\DeclareAcronym{KAGRA}{short = KAGRA , long = Japanese underground gravitational wave detector}
\DeclareAcronym{LIGO}{short = LIGO , long = Laser Interferometer Gravitational wave Observatory}
\DeclareAcronym{aLIGO}{short = aLIGO , long = advanced LIGO}
\DeclareAcronym{Virgo}{short = Virgo , long = Virgo gravitational wave detector}
\DeclareAcronym{AdVirgo}{short = AdVirgo , long = advanced Virgo detector}
\DeclareAcronym{LIGOIndia}{short = LIGO India , long = a copy of the advanced LIGO detectors placed in India}
\DeclareAcronym{USA}{short = USA , long = United States of America}
\DeclareAcronym{a+LIGO}{short = LIGO A+ , long = upgraded advanced Laser Interferometer Gravitational wave Observatory}
\DeclareAcronym{AdVirgo+}{short = AdVirgo+ , long = upgraded advanced Virgo detector}
\DeclareAcronym{Voyager}{short = Voyager , long = {Voyager, a detector with 3G technology in the 2G infrastructure}}
\DeclareAcronym{NEMO}{short = NEMO , long = {Neutron Star Extreme Matter Observatory, a plan for a high frequency  GW detector in a 2G scale facility}}

\DeclareAcronym{2.5G}{short = 2.5G , long = an improved version of the second detector generation with incrementally new technology}
\DeclareAcronym{RaD}{short = R\&D , long = research and development}
\DeclareAcronym{GW}{short = GW , short-plural = s , long = gravitational waves}
\DeclareAcronym{GWD}{short = GWD , short-plural = s, long = gravitational wave detector}
\DeclareAcronym{GWIC}{short = GWIC , long = gravitational wave international committee}
\DeclareAcronym{GWAC}{short = GWAC , long = gravitational wave agency correspondents}
\DeclareAcronym{ML}{short = ML , long = maturity level}
\DeclareAcronym{ET-LF}{short = ET-LF , long = Einstein Telescope low frequency interferometer}
\DeclareAcronym{ET-HF}{short = ET-HF , long = Einstein Telescope high frequency interferometer}
\DeclareAcronym{FP}{short = FP , long = {Fabry-Perot cavity, a linear optical resonator}}
\DeclareAcronym{NN}{short = NN , long = {Newtonian noise, a gravitational coupling of surrounding mass (especcially of moving mass due to seismic waves or infra-sound) to the mirrors}}
\DeclareAcronym{SQZ}{short = SQZ , long = {squeezing, a technology to modify quantum noise distribution over the two quadratures of the light field }}
\DeclareAcronym{PR}{short = PR , long = Power Recycling}
\DeclareAcronym{SR}{short = SR , long = Signal Recycling}
\DeclareAcronym{QRPN}{short = QRPN , long = quantum radiation pressure noise}
\DeclareAcronym{SN}{short = SN , long = shot noise}
\DeclareAcronym{UHV}{short = UHV , long = Ultra High Vacuum}
\DeclareAcronym{NSF}{short = NSF , long = National Science Foundation}
\DeclareAcronym{NIST}{short = NIST , long = National Institute of Standards and Technology}
\DeclareAcronym{JLab}{short = JLab , long = Jefferson Lab}
\DeclareAcronym{CERN}{short = CERN , long = European Centre for nuclear research}
\DeclareAcronym{CE2}{short = CE2 , long = Cosmic Explorer phase 2}
\DeclareAcronym{CE1}{short = CE1 , long = {Cosmic Explorer phase 1, using 2G technology}}
\DeclareAcronym{DS}{short = DS , long = Design Study}
\DeclareAcronym{MCz}{short = MCz , long = magnetically assisted Czochralski}
\DeclareAcronym{ITM}{short = ITM , long = Inner Test mass }
\DeclareAcronym{ETM}{short = ETM , long = End Test Mass}
\DeclareAcronym{Heraeus}{short = Heraeus , long = a German company producing ultra-pure fused silica glass}
\DeclareAcronym{TLS}{short = TLS , long = two level system}
\DeclareAcronym{AlGaAs/GaAs}{short = AlGaAs/GaAs , long = Aluminium-Gallium-Arsenide / Gallium Arsenide}
\DeclareAcronym{GaP/AlGaP}{short = GaP/AlGaP , long = Gallium Phosphide / Alluminium Gallium Phosphide}
\DeclareAcronym{T}{short = $T$ , long = Temperature}
\DeclareAcronym{L}{short = L , long = interferometer arm length}
\DeclareAcronym{w}{short = $w$ , long = laser beam radius}
\DeclareAcronym{phi}{short = $\varphi$ , long = mechanical loss angle}
\DeclareAcronym{d}{short = $d$ , long = coating thickness}
\DeclareAcronym{f}{short = $f$ , long = frequency}
\DeclareAcronym{A+}{short = A+ , long = the upgraded \it{advanced} interferometer generation}
\DeclareAcronym{CCR}{short = CCR , long = {Center for Coatings Research, Stanford}}
\DeclareAcronym{LSC}{short = LSC , long = LIGO Scientific Collaboration}
\DeclareAcronym{LVC}{short = LVC , long = LIGO Virgo Collaboration}

\DeclareAcronym{GEO}{short = GEO , long = the GEO collaboration}
\DeclareAcronym{U.K.}{short = U.K.  , long = United Kingdom}
\DeclareAcronym{LMA}{short = LMA , long = Laboratoire des Matériaux Avanc\'es }
\DeclareAcronym{ILM}{short = ILM , long = Insitut Lumi\`ere Materi\`e}
\DeclareAcronym{ANR}{short = ANR , long = Agence nationale de la recherche }
\DeclareAcronym{ViSIONs}{short = ViSIONs , long = a coating project funded by \acs*{ANR}}
\DeclareAcronym{IBS}{short = IBS , long = ion-beam sputtering}
\DeclareAcronym{LIGO Lab}{short = LIGO Lab , long = laboratory running the LIGO project}
\DeclareAcronym{Y}{short = Y , long = the mechanical admittance}
\DeclareAcronym{omega}{short = $\omega$ , long = the angular frequency}
\DeclareAcronym{k_b}{short = $k_b$ , long = the Boltzmann constant}
\DeclareAcronym{DNA}{short = $\sqrt{S_x(\omega)}$ , long = Fourier components of the displacement of the system}
\DeclareAcronym{FNA}{short = $F(\omega)$ , long = Fourier components of the force leading to the displacement of the system }

\DeclareAcronym{CLIO}{short = CLIO , long = cryogenic 100\,m prototype interferometer in the Japanese Kamioka mine }
\DeclareAcronym{LIDAR}{short = LIDAR , long = light detection and ranging}
\DeclareAcronym{PSL}{short = PSL , short-plural = s, long = pre-stabilized laser system}
\DeclareAcronym{HPL}{short = HPL ,  short-plural = s , long = high power laser}
\DeclareAcronym{AEI}{short = AEI , long = Albert Einstein Institute}
\DeclareAcronym{LZH}{short = LZH , long = Laser Zentrum Hannover}
\DeclareAcronym{Artemis}{short = Artemis , long = }
\DeclareAcronym{Alphanov}{short = Alphanov , long = {Centre Technologique Optique et Lasers, France}}
\DeclareAcronym{ICRR}{short = ICRR , long = {institute for cosmic research, University Tokyo}}
\DeclareAcronym{Mitsubishi}{short = Mitsubishi , long = Japanese company}
\DeclareAcronym{LG33}{short = $\rm LG_{33}$ , long = Laguerre Gauss Mode 33}
\DeclareAcronym{IFO}{short = IFO , long = Interferometer }
\DeclareAcronym{Nufern}{short = Nufern , long = Fibre laser company }
\DeclareAcronym{MOPA}{short = MOPA , long = master oscillator power amplifier}
\DeclareAcronym{MIT}{short = MIT , long = Massachusetts Institute of Technology}
\DeclareAcronym{neoVAN4S}{short = neoVAN 4S , long = a solid state laser of the German company NEOLASE }

\DeclareAcronym{IIT}{short = IIT , long = Indian institute of technology}
\DeclareAcronym{QND}{short = QND , long = Quantum Non-Demolition}
\DeclareAcronym{SQL}{short = SQL , long = standard quantum limit}
\DeclareAcronym{SUS}{short = SUS , short-plural = s , long = suspension system, long-plural = s ,}
\DeclareAcronym{SAS}{short = SAS , short-plural = s , long = Seismic attenuation system}
\DeclareAcronym{HEPI}{short = HEPI , long = hydraulic external pre-isolator}
\DeclareAcronym{SA}{short = SA , long = Super Attenuator}
\DeclareAcronym{FEA}{short = FEA , long = finite element analysis}
\DeclareAcronym{GEO600}{short = GEO\,600 , long = the German/British 600m gravitational wave detector in Germany}
\DeclareAcronym{GWADW}{short = GWADW , long = Gravitational Wave Advanced Detector Workshop}
\DeclareAcronym{LASTI}{short = LASTI , long = LIGO advanced system test interferometer at MIT}
\DeclareAcronym{SIS}{short = SIS , long = seismic isolation system}
\DeclareAcronym{AWC}{short = AWC , long = Active Wavefront Control}
\DeclareAcronym{TCS}{short = TCS , long = thermal compensation system}
\DeclareAcronym{EOM}{short = EOM , short-plural = s, long = electro optic modulator}
\DeclareAcronym{IFI}{short = IFI , long = input Faraday isolator}
\DeclareAcronym{TGG}{short = TGG , long = Terbium gallium garnet}
\DeclareAcronym{IMC}{short = IMC , long = input Mode cleaner}
\DeclareAcronym{OFI}{short = OFI , long = Output Faraday isolator}
\DeclareAcronym{FC}{short = FC , short-plural = s, long = filter cavity}
\DeclareAcronym{OMC}{short = OMC , short-plural = s, long = output mode cleaner}
\DeclareAcronym{BHD}{short = BHD, short-plural = s , long = balanced homodyne detector}
\DeclareAcronym{RF}{short = RF , long = radio frequency }
\DeclareAcronym{SPI}{short = SPI , long = suspension point interferometer}
\DeclareAcronym{FI}{short = FI , short-plural = s,long = Faraday isolator}
\DeclareAcronym{GWINC}{short = GWINC , long = Gravitational wave interferometer noise calculator}
\DeclareAcronym{FFT}{short = FFT , long = fast Fourier transform}
\DeclareAcronym{SISb}{short = SIS , long = Stationary Interferometer Simulation tool }
\DeclareAcronym{OSCAR}{short = OSCAR , long = optical FFT code to simulate Fabry Perot cavities with arbitrary mirror profiles }
\DeclareAcronym{DarkF}{short = DarkF , long = an optical simulation code in FORTRAN 90}
\DeclareAcronym{MIST}{short = MIST , long = Matlab based fast modal simulation of paraxial optical systems }
\DeclareAcronym{Finesse}{short = Finesse , long = Frequency-domain interferometer simulation with higher-order spatial modes}
\DeclareAcronym{Optickle}{short = Optickle , long = a general model for the electro-opto-mechanical
part of a GW detector}
\DeclareAcronym{SimPlant}{short = SimPlant , long = virtual interferometer for commissioning }
\DeclareAcronym{IfoCAD}{short = IfoCAD , long = 3D ray tracing tool using Gaussian beams}
\DeclareAcronym{CAD}{short = CAD , long = computer aided design}
\DeclareAcronym{OptoCAD}{short = OptoCAD , long = 2D ray tracing tool using Gaussian beams}
\DeclareAcronym{MIMO}{short = MIMO , long = {multiple in, multiple out}}
\DeclareAcronym{3D}{short = 3D , long = three dimensional}
\DeclareAcronym{DARM}{short = DARM , long = differential arm motion}
\DeclareAcronym{RMS}{short = RMS , long = root mean square }
\DeclareAcronym{FPGA}{short = FPGA , short-plural = s, long = field programmable gate array}
\DeclareAcronym{GPU}{short = GPU ,  short-plural = s,long = graphics processing unit}
\DeclareAcronym{SNR}{short = SNR , short-plural = s, long = signal to noise ratio}
\DeclareAcronym{BNS}{short = BNS , long =binary neutron star}
\DeclareAcronym{GR}{short = GR , long = general relativity }
\DeclareAcronym{MCMC}{short = MCMC , long = Markov-Chain-Monte-Carlo method}
\DeclareAcronym{VCRaD}{short = VCR\&D , long = Virgo Coating R\&D collaboration}
\DeclareAcronym{UWS}{short = UWS , long = University of the West of Scotland}
\DeclareAcronym{MBE}{short = MBE, long = Molecular Beam Epitaxy }
\DeclareAcronym{GeNS}{short = GeNS , long = Gentle Nodal Support Sytem}
\DeclareAcronym{CSIRO}{short = CSIRO , long =  Commonwealth Scientific and Industrial Research Organisation (an Australian federal government agency responsible for scientific research)}
\DeclareAcronym{ANU}{short = ANU , long = Australian National University}
\DeclareAcronym{CNRS}{short = CNRS  , long = Centre national de la recherche scientifique (the French national research centre for scientific research }

%% file: Cover_TOC.tex

\begingroup
\thispagestyle{empty}

\AddToShipoutPicture*{\put(0,0)\centering{\includegraphics[width=1.45\textwidth]{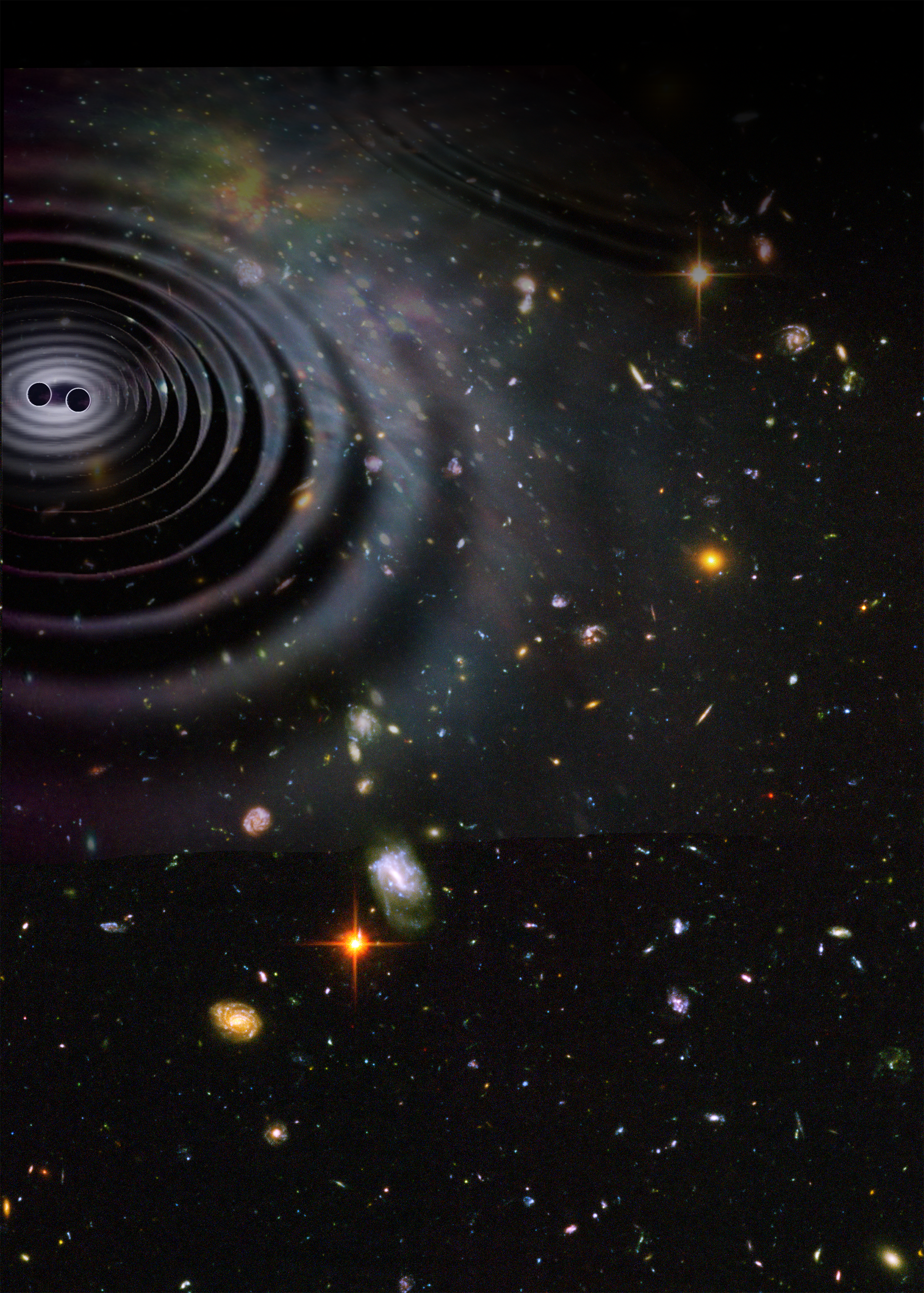}}} 
\centering
\par\normalfont\fontsize{1}{1}\sffamily\selectfont
\textcolor{black}{\textbf{.}}\\ 
\par\normalfont\fontsize{40}{40}\sffamily\selectfont
\textcolor{white}{\textbf{3G R\&D}}\\
\par\normalfont\fontsize{40}{40}\sffamily\selectfont
\textsc{\textcolor{white}{{R\&D for the next generation of ground-based gravitational-wave detectors}}}\par 
\vskip12.5cm
\raggedright
\par\normalfont\fontsize{18}{18}\sffamily\selectfont
\textcolor{white}{GWIC}\par 
\textcolor{white}{April 2021}\par 
\endgroup


\newpage
\thispagestyle{empty}

\noindent \textbf{DETECTOR RESEARCH AND DEVELOPMENT SUBCOMMITTEE}

\noindent David McClelland, Australian National University, Australia (Co-chair)

\noindent Harald Lueck, AEI, Hannover, Germany (Co-chair)

\noindent Rana Adhikari, Caltech, USA

\noindent Masaki Ando, University of Tokyo, Japan

\noindent GariLynn Billingsley, Caltech, USA

\noindent Geppo Cagnoli, ILM, Lyon, France

\noindent Matt Evans, MIT, USA

\noindent Martin Fejer, Stanford University, USA

\noindent Andreas Freise, University of Birmingham, UK

\noindent Paul Fulda, University of Florida, USA

\noindent Eric Genin, Virgo, Italy

\noindent Gabriela Gonz\'{a}lez, Louisiana State University, USA

\noindent Jan Harms, Universit\`a degli Studi di Urbino, Italy

\noindent Stefan Hild, University of Glasgow, UK

\noindent Giovanni Losurdo, INFN Pisa, Italy

\noindent Ian Martin, University of Glasgow, UK

\noindent Anil Prabhakar, IIT Madras, India

\noindent Stuart Reid, University of Strathclyde, UK

\noindent Fulvio Ricci, Universit\`a La Sapienza, and INFN Roma, Italy

\noindent Norna Robertson, Caltech, USA

\noindent Jo van den Brand, Nikhef, Netherlands

\noindent Benno Willke, AEI, Hannover, Germany

\noindent Mike Zucker, MIT, USA\\

\noindent 
\textbf{STEERING COMMITTEE}

\noindent Michele Punturo, INFN Perugia, Italy (Co-chair)

\noindent David Reitze, Caltech, USA (Co-chair)

\noindent Peter Couvares, Caltech, USA

\noindent Stavros Katsanevas, European Gravitational Observatory

\noindent Takaaki Kajita, University of Tokyo, Japan

\noindent Vicky Kalogera, Northwestern University, USA

\noindent Harald Lueck, AEI, Germany

\noindent David McClelland, Australian National University, Australia

\noindent Sheila Rowan, University of Glasgow, UK

\noindent Gary Sanders, Caltech, USA

\noindent B.S. Sathyaprakash, Penn State University, USA and Cardiff University, UK

\noindent David Shoemaker, MIT, USA (Secretary)

\noindent Jo van den Brand, Nikhef, Netherlands\\

\noindent \textsc{Gravitational Wave International Committee}\\

\noindent \textsc{}\\ 

\noindent This document was produced by the GWIC 3G Committee, the GWIC 3G R\&D Team and the International 3G Science Team Consortium\\ 

\noindent \textit{Final release, April 2021} 
\newpage
\noindent \textbf{ADDITIONAL AUTHORS}

\noindent Alessandro Bertolini, Nikhef, Netherlands

\noindent Stefan Danilishin, Maastricht University, and Nikhef, Netherlands

\noindent Francesco Fidecaro, University of Pisa, and INFN Pisa, Italy

\noindent Gianluca Gemme, INFN Genova, Italy

\noindent Giles Hammond, SUPA, University of Glasgow, UK

\noindent James Lough, AEI, and Institut fur Gravitationsphysik der Leibniz Universitat Hannover, Germany

\noindent Ettore Majorana, Università di Roma, Italy

\noindent Ando Massaki, The University of Tokyo, Japan

\noindent Joshua Smith, California State University Fullerton, California, USA

\noindent Helios Vocca, University of Perugia, and INFN Perugia, Italy

\noindent Krishna Venkateswara, University of Washington, USA

\noindent Robert L. Ward, Australian National University, Australia

\noindent Kazuhiro Yamamoto, University of Toyama, Japan\\

\newpage

\chapterimage{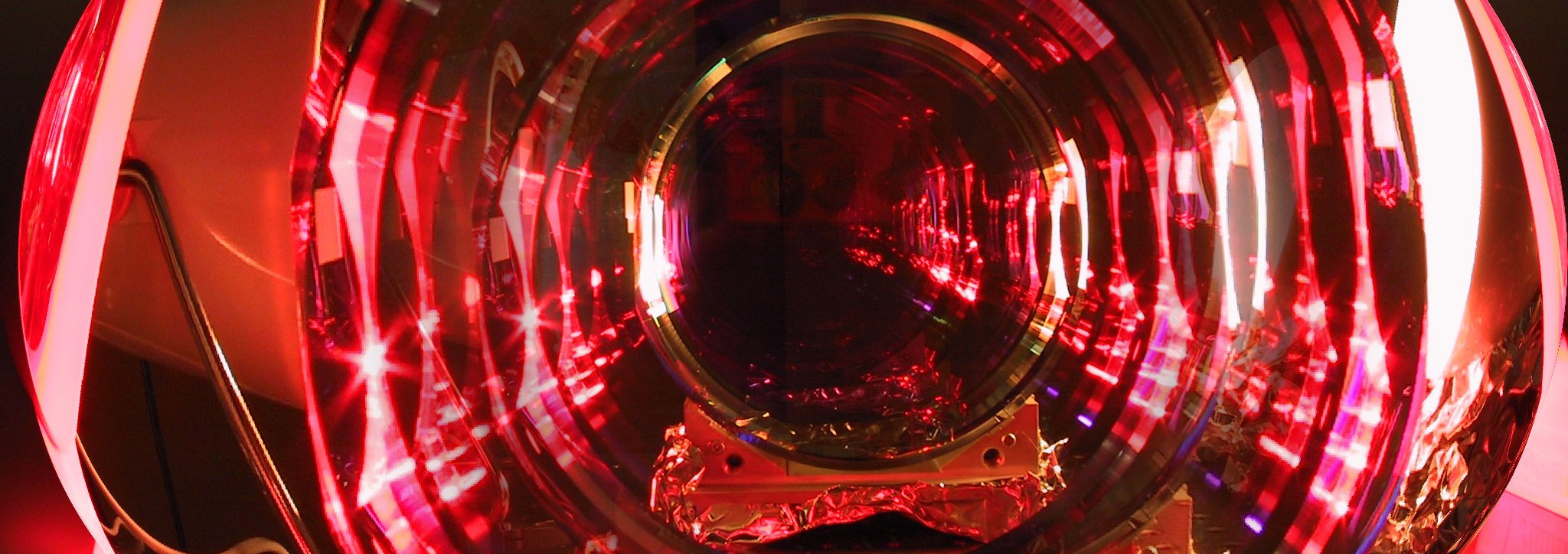} 
\pagestyle{empty} 
\tableofcontents 
\pagestyle{fancy} 

%% file: Introduction.tex
\chapterimage{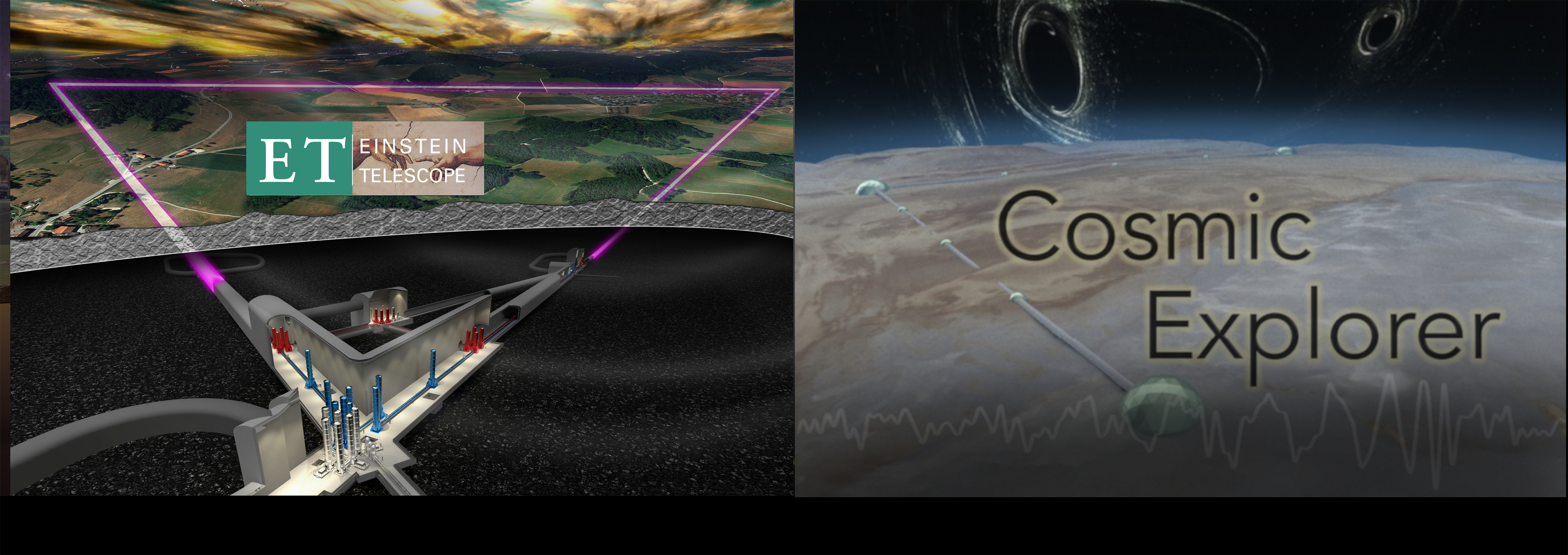} 
\chapter{Introduction}
\label{sec:Intro}
\pagenumbering{arabic}
\setcounter{page}{1}
In this report we review the \ac{RaD} needed to construct massive new facilities, far larger then current facilities,  and operate in them detectors with "\ac{3G}" sensitivity.
The new \ac{3G} facilities will be designed to accommodate successive generations of detectors with increasing sensitivity as the technology evolves and new ideas emerge as has happened with existing facilities. The infrastructures which initially housed the \ac{1G} \textit{\ac{LIGO}}~\cite{Abbott2009} and \textit{\ac{Virgo}}~\cite{VirgoStatus2008} detectors, now house the second generation \textit{\ac{aLIGO}}~\cite{AdvancedLIGO2015} detectors, (\textit{\ac{AdVirgo}}~\cite{AdvancedVirgo2015} along with LIGO India and KAGRA.  There are plans for further upgrades to these \ac{2G} detectors -  \ac{a+LIGO}, \ac{AdVirgo+} \cite{Zucker:LIGOAplus, Cagnoli:VirgoAplus}.  Further more completely new detectors, such as  \ac{Voyager}\cite{VoyagerDCC2018} and NEMO \cite{NEMO2020} may  later be installed in the existing facilities or facilities of a similar scale . These detectors are referred to as 2.5G class detectors.

 Section~\ref{sec:Fac_Inf} will cover the \ac{RaD} needed to select suitable sites and build such large, long-lifetime facilities in a cost efficient way. The remainder of the chapter will focus on the \ac{RaD} required to deliver the first detectors operational in these \ac{3G} facilities.  Currently, there are two main concepts for these detectors, \acf{ET}~\cite{ET2011}, a 10-km triangular underground detector, and \acf{CE}~\cite{CosmicExplorer2017}, a 40-km above-ground L-shaped detector. These concepts, along with Voyager are highlighted in Box~\ref{Box:GWOs}.  
 
 Currently (circa 2021)  Advanced LIGO and Advanced Virgo have completed the third observing run. These detectors are being upgraded toward LIGO A+ and AdVirgo+ operations and possible modest sensitivity  improvements will be implemented beyond these designs. Following these detector upgrades by about 10 years we envision a network of new detectors, CE and ET in longer-baseline \ac{3G} facilities, along with detectors that use \ac{3G} technology within 2G scale facilities.
 
\begin{wrapfigure}{r}{0.65\textwidth}
\centering
\includegraphics*[width= 0.64\textwidth]{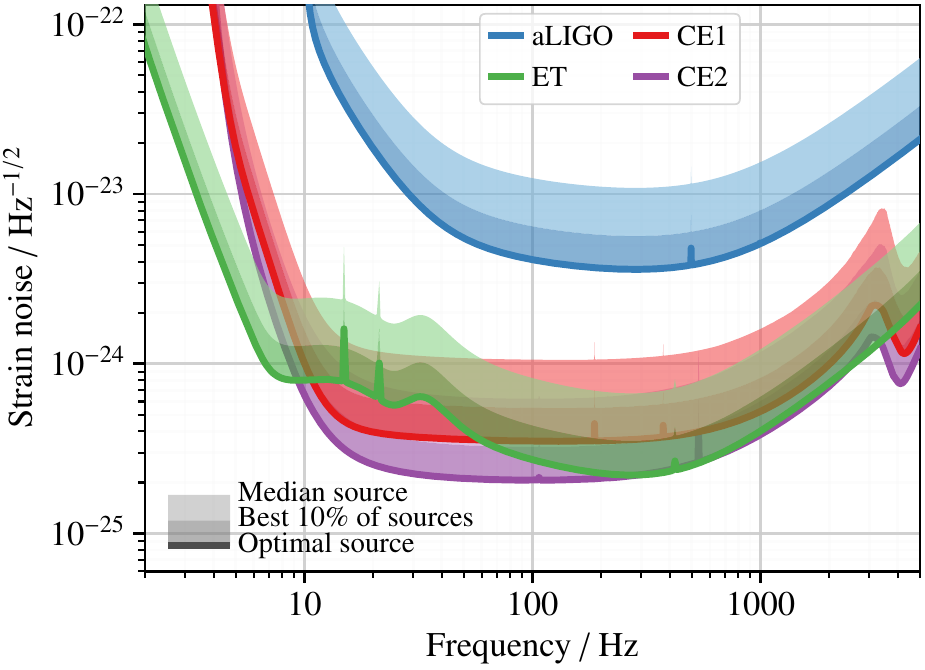}
\caption{Target sensitivity curves for \ac{3G} gravitational-wave detectors \acf{ET}~\cite{ET2011}, shown in green, and \acf{CE}~\cite{CosmicExplorer2017}, shown in pink, compared with the design sensitivity of Advanced LIGO, shown in blue. 
The shades of the curves represent sensitivity to sources with differently distributed locations.}
\label{fig:3GSens}
\end{wrapfigure}

The first instruments to be installed in the new \ac{3G} Observatories will be 10 to 20 times more sensitive than the current \ac{2G} instruments above 100\,Hz (Fig.~\ref{fig:3GSens}).  The improvement factor exceeds 100 around 10\,Hz and thousands below 10\,Hz. The key parameters for \ac{2G} and \ac{3G} instruments are summarized in Table~\ref{Tab:FutIfos}. The \ac{3G} facilities are being designed with lifetimes on the order of  50 years in order to house detectors far more sensitivity that the initially  proposed ET and CE  designs.  Strategies for \ac{3G} will be modified according to observations made with current detectors, evolution in the science case, technology readiness and funds available. This \ac{GWIC} \ac{3G} report represents a milestone community vision for the future of ground based gravitational wave observations. 

Planning for the \ac{3G} detectors began more than 20 years before they were envisioned to become operative. This was based on experience with past and current detectors, for which there was a lead time of 15 years or more from conception to operation. Assuming a 2035 start date for initial \ac{3G} operations, preceded by five years of construction and five years of commissioning, it is likely that only technologies with mature \ac{RaD} in 2025 will feed into final design and engineering for the initial \ac{3G} detectors. 

Despite their differences in design, the \ac{3G} detectors \ac{ET} and \ac{CE} rely on similar `enabling technologies' -- the main pillars on which the predictions of sensitivity are based. These technologies are used to mitigate `fundamental noise sources' affecting the instruments, in particular: \textbf{Quantum noise} associated with the laser light fields is modified by high laser power, quantum squeezing, massive mirrors and interferometer topology; \textbf{Thermal noise} in mirror substrates, coatings and suspensions is modified by temperature and material properties (and their behaviour as a function of temperature); and \textbf{Newtonian noise} caused by the gravitational forces of moving masses (such as air, the ground, and machinery) is modified by the location of the sites and subtraction schemes.  Similarly supporting technologies such as control systems and methods to  mitigate  a forest of technical noise  are issues common to all interferometer designs.\\

These considerations are interdependent and have implications for the detector designs. For example, using low temperatures to reduce Brownian noise requires a departure from the fused silica optics used in Advanced LIGO and Advanced Virgo. Sapphire and Silicon are promising low temperature materials. Silicon would require changing the operating wavelength to 1.5 -- 2\,\micro m, necessitating the development of a new suite of light sources, optical components and detectors. Considering the material properties as a function of temperature and wavelength as well as the ability to handle high optical power while minimizing noise during heat extraction from the core optics leads to four possible operating temperatures: room temperature and the cryogenic temperatures of 123\,K, 20\,K and below 5\,K.  At this time we are not in a position to make the choice. The final technologies for the first interferometers in the \ac{3G} facilities will depend on \ac{RaD} progress across the various subsystems.  It  will require intensive sensitivity trade studies informed by science goals and is beyond the scope of this report.  Indeed different collaborations  may choose different optimizations.   In a conservative scenario, the first instruments could use technologies already proven in  \ac{2G} facilities.  This is the option currently favoured by the Cosmic Explorer collaboration, referred to as \acs{CE1}.





In this chapter, we assess the state of \ac{RaD} for \ac{3G} enabling and supporting technologies  toward readiness before the end of the decade while also looking ahead to the \ac{RaD} that will continue as a necessary preparation for subsequent \ac{3G} upgrades. Each section contains an outlook for the activity and recommendations on how to best  progress the \ac{RaD} globally.  We begin by presenting a brief overview on the basics of gravitational wave detection, and introduce the main subsystems that make up an interferometer. Then in section \ref{sec:Fac_Inf} we address requirements and design aspects of the \ac{3G} infrastructure. 

In subsequent sections we consider the state of the art in the various technologies, the \ac{RaD} needed in each area, the level of resources needed (broadly bracketed as high, medium, low), and how the \ac{RaD} should be focused in order to deliver fully tested subsystems for timely installation in new \ac{3G} facilities.  The analysis suggests prototyping \ac{3G} technology including new test facilities and use of existing long baseline facilities. 

Section~\ref{sec:Newtonian_Noise} describes Newtonian noise and its connection with facility choices and reliance on modelling and subtraction schemes. Section~\ref{sec:Suspensions_Isolation} covers suspensions and seismic isolation systems while Section~\ref{sec:Cryogenics} addresses cryogenics. Sections ~\ref{sec:Core_optics} and \ref{sec:Coatings} describe the closely related subjects of core optics and coatings. A common theme among these chapters is thermal noise, described in Box~\ref{Box:Thermal}, which is the primary consideration for many choices in optics, coatings, suspensions, and operating temperature. Section~\ref{sec:Light_sources} reviews Light sources, both lasers and squeezed state generators.  This is followed in Section~\ref{sec:Quantum} with an examination of quantum enhancement techniques. In addition to the fundamental noises described above, a myriad of technical noise sources and control issues can limit interferometer performance: parametric instabilities, scattered light, and noise originating from auxiliary optics and control systems. The current state of the art in these areas, \ac{RaD} needed, and coordination for 3G are reviewed in Sections~\ref{sec:Aux-optics} and \ref{sec:Sim_Controls}. Finally, Section~\ref{sec:Calibration} describes plans for accurately calibrating the instruments to levels that will enable the dramatic science described in the first part of this report. We end in the summary by proposing four broad recommendations designed to optimise global \ac{RaD} resources in order to make timely progress.

\newpage

\begin{DetBox}{\bf Future Gravitational Wave Observatories}
\stepcounter{boxcount}
\label{Box:GWOs}
\begin{tcolorbox}[standard jigsaw,colback=amber!10!white,colframe=red!70!black,coltext=black,size=small,  title=The Einstein gravitational--wave Telescope (ET)] 
\begin{wrapfigure}{r}{0.4\linewidth}
\vspace{-10pt}
\includegraphics*[width=0.4\textwidth]{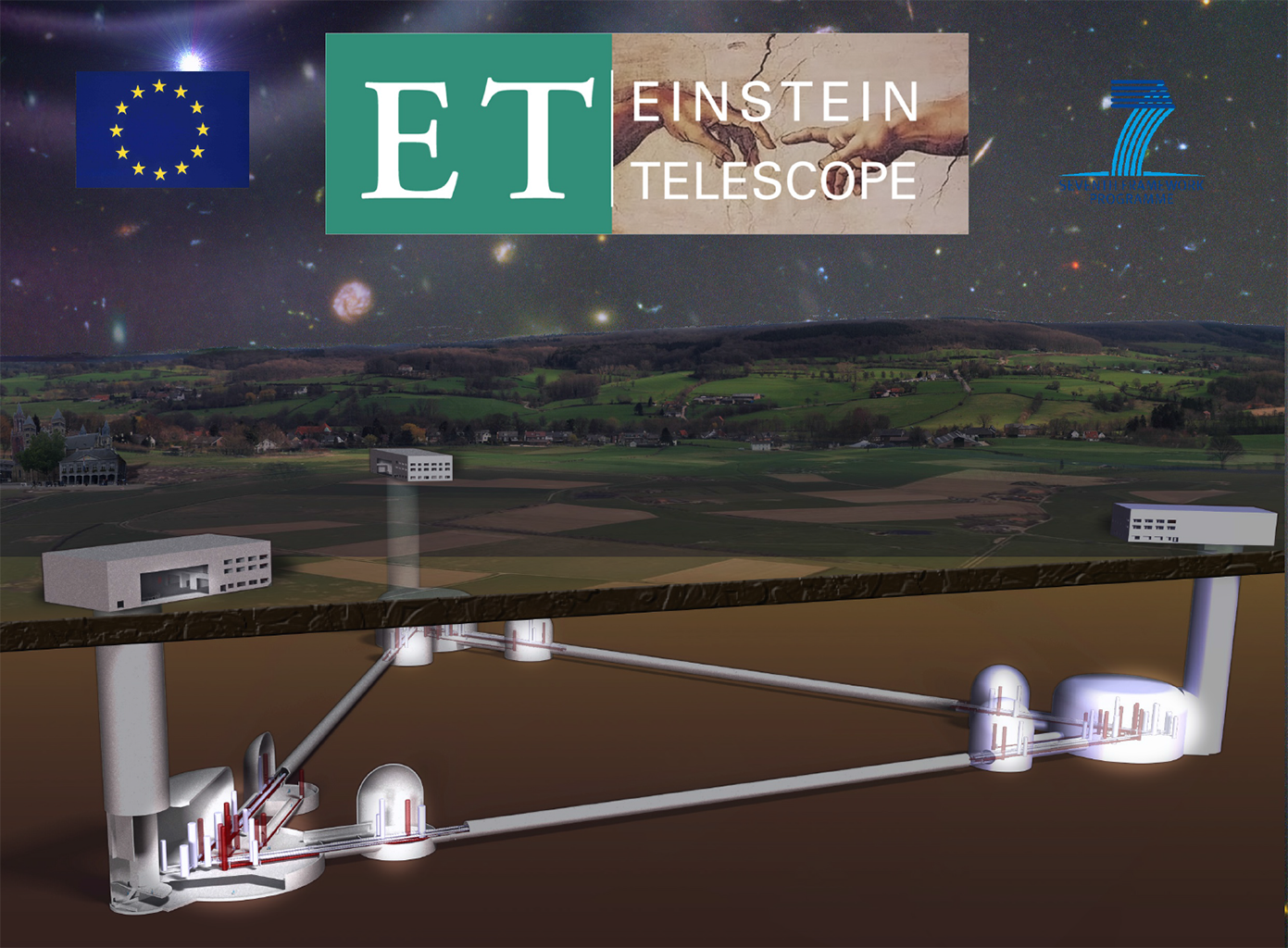}
\label{fig:ET_Thumb}
\vspace{-25pt}
\end{wrapfigure}
ET~\cite{ET2011} is 
 a concept for a third generation gravitational-wave \emph{observatory}
 likely with a location in Europe. 
 To reduce the effects of seismic motion, the \acs{ET} concept calls for the site to be located at a depth of about 100\,m to 200\,m below ground. In its final configuration it shall be arranged as an equilateral triangle of three interlaced detectors, each consisting of two interferometers. The configuration of each detector dedicates one interferometer \acs{ET-LF} to detecting the \textbf{L}ow \textbf{F}requency components of the gravitational-wave signal (2--40\,Hz), while the other one \acs{ET-HF}   is dedicated to the \textbf{H}igh \textbf{F}requency components, called a "xylophone" design. Each interferometer will have a dual-recycled Michelson layout with about 10\,km long \acs{FP}. In \acs{ET-LF} , which operates at cryogenic temperature, thermal, seismic, gravity gradient and radiation pressure noise sources are particularly suppressed; in \acs{ET-HF} , sensitivity at high frequencies is improved by high laser light power circulating in the \acs{FP}  cavities and the use of frequency-dependent squeezed light technologies.
\end{tcolorbox}

\begin{tcolorbox}[standard jigsaw,colframe=antiquefuchsia!80!black,colback=antiquefuchsia!20!white,opacityback=0.6,coltext=black,size=small, title=Cosmic Explorer (CE)] 
\begin{wrapfigure}{r}{0.4\linewidth}
\vspace{-10pt}
\includegraphics*[width=0.4\textwidth]{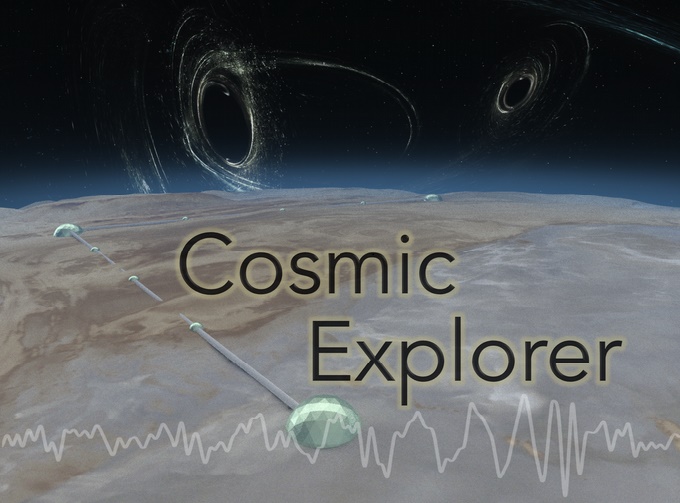}
\label{fig:CE_Thumb}
\vspace{-20pt}
\end{wrapfigure}
CE~\cite{CosmicExplorer2017} is a concept for an L-shaped above-ground observatory with 40\,km arm-length, operating a dual recycled Michelson interferometer with \acs{FP}  arm cavities, with possible site location in the US or/and in Australia. 
Its initial phase, called \acs{CE1}, will employ scaled-up \emph{Advanced LIGO technology} including 440\,kg fused silica test masses, 1.5\,MW of optical power, and frequency-dependent squeezing. 
A major upgrade, \acs{CE2}, will properly exploit the new facility by either using \emph{Voyager technology} such as silicon test masses and amorphous silicon coatings operating at 123\,K, with $1.5$ or $2\,\mu m$ laser light and 3\,MW of optical power in its arm cavities or continue to use room temperature A+ technology pushed to its limits.

\end{tcolorbox}

\begin{tcolorbox}[standard jigsaw,colframe=azure!70!black,colback=azure!20!white,opacityback=0.6,coltext=black, size=small, title=Voyager]

\begin{wrapfigure}{r}{0.4\linewidth}
\vspace{-10pt}
\includegraphics*[width=0.4\textwidth]{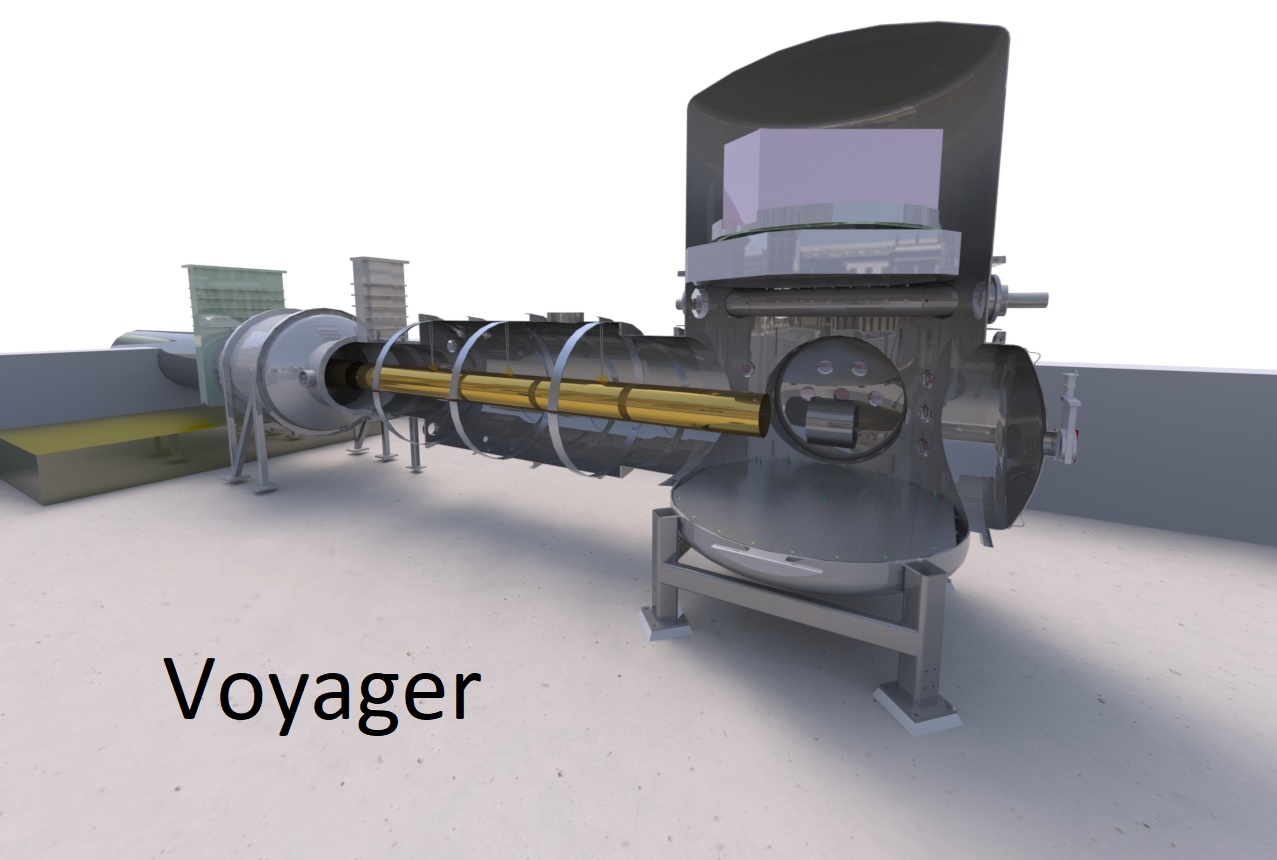}
\label{fig:Voyager_Thumb}
\vspace{-20pt}
\end{wrapfigure}

Voyager~\cite{Voyager:Inst,VoyagerDCC2018, VoyagerDCC2019} is the tentative concept for a new detector in the current facilities, designed to maximize the observational reach of the infrastructure and demonstrate the key technologies to be used for \acs{3G} observatories in new infrastructures.
Voyager proposes to use heavy (ca.\,200\,kg) cryogenic mirrors with improved coatings and upgraded suspensions made of ultra-pure silicon at a temperature of 123\,K in the existing vacuum envelope and a laser wavelength of $\sim1.5\,-\,2\,\mu m$. 
A further factor of 3 increase in BNS range (to 1100\,Mpc) is envisioned along with a reduction of the low frequency cutoff. 
 In the context of this report we use the term \emph{Voyager Technology} for this type of technology irrespective of plans to implement it in existing or future infrastructure.  Recently the \acs{NEMO} \cite{NEMO2020}, has been proposed. It  may use aspect of  Voyager technology but is optimised for the kHz band in order to measure the neutron star state equation of state. 
\end{tcolorbox}

\end{DetBox}

\newpage

\begin{table}[ht]
\centering
\begin{tabular}{|l|l|l|p{1.6cm}|l|l|l|l|}
\hline
 &aLIGO / AdV &A+/V+ &KAGRA &CE 1 &CE 2 &ET-LF &ET-HF\\
\hline
Arm Length [km] & 4 / 3 &4 / 3& 3& 40& 40& 3x10$^*$ & 3x10$^*$\\
\hline
Mirror Mass [kg]& 40 / 42& 40& 23& 440& 470& 211& 200\\
\hline
Mirror Material& silica& silica& sapphire& silica& silicon& silicon& silica\\
\hline
Mirror Temp [K]& 295& 295& 20& 295& 123& 10& 290\\
\hline
Suspension Fiber& 0.6m/0.7m& 0.6m& 0.35m& 2m& 2m& 2m& 0.6m\\
& SiO2& SiO2&Al2O3&SiO2&Si&Si&SiO2\\
\hline
Fiber Type& Fiber& Fiber& Fiber& Fiber& Ribbon& Fiber& Fiber\\
\hline
Input Power [W]& 125& 125& 70& 140& 280& 3& 500\\
\hline
Arm Power [kW]& 710 / 700& 750& 350& 1500& 3000& 18& 3000\\
\hline
Wavelength [nm]& 1064& 1064& 1064& 1064& 2000& 1550& 1064\\
\hline
\acs{NN} Suppression& 1& 1& 1& 2& 10& 1& 1\\
\hline
Beam Size [cm]& (5.5/6.2) / 6& 5.5/6.2& 3.5/3.5& 10/13& 14/18& 9/9& 12/12\\
\hline
\acs{SQZ} Factor [dB]& 0& 6& foreseen& 6& 10& 10& 10\\
\hline
Filter Cavity & none& 300& unknown& 4000& 4000& 10000& 500\\
Length [m] &&&&&&&\\
\hline
\end{tabular}
\caption{Key parameters of the current Advanced detectors, their enhancements, LIGO A+ and AdVirgo +, the two phases of Cosmic Explorer, \ac{CE1} and \ac{CE2}, and the two interferometer types of Einstein Telescope. $^*$\ac{ET} will have an angle of 60$\deg$ between the detector arms}
\label{Tab:FutIfos}
\end{table}

\begin{Infobox}{\bf Thermal Noise in Gravitational Wave Detectors}
\stepcounter{boxcount}
Thermal noise is one of the fundamental noise sources limiting \ac{2G} detectors over a considerable frequency range. The main contributions come from Brownian noise of the mirror suspensions, substrates and coatings and thermo-optic (thermo-elastic plus thermo-refractive) noise of substrates and coatings. The relation between the dissipation and the power spectrum (single sided) of the noise is described by Callen's Fluctuation-Dissipation Theorem~\cite{CaWe1951, Kubo:FDT, Callen:1959} and is given by:
\begin{equation}
S_x(\omega) = \frac{4\,k_B T}{\omega^2} \left| \mathrm{Re} \big[ Y(\omega) \big]\right| .
\end{equation}
\label{eq:FDT}
with \acf{k_b}, \acf{omega} , \acf{T} and \acf{Y}, defined as
\begin{equation}
Y(\omega) = i \omega\frac{X(\omega)}{F(\omega)} \, ,
\end{equation}
where $X(\omega)$ and \acs{FNA} are the Fourier components of the displacement of the system and force applied leading to the displacement, respectively. The real part of the admittance is proportional to mechanical losses, hence low noise requires low mechanical losses.
Operating the mirrors and suspensions at reduced temperature reduces thermal noise since the 
\ac{DNA} scales with $\sqrt{T}$. 
More significantly, many material properties of mirror substrates and coatings depend on temperature, and hence influence the temperature dependence of thermal noise.
Fused silica, like most other glasses, has increased mechanical losses at cryogenic temperatures, making it unsuitable as substrates at cryogenic temperatures.  Crystalline materials (Sapphire, Silicon) are therefore the prime candidates for low temperature operation.
Most amorphous oxide coatings show higher mechanical losses at cryogenic temperatures while the losses of crystalline or semiconductor coatings improve. However, the latter may suffer from other drawbacks, e.g. increased optical absorption.
Further dependence of coating thermal noise on coating parameters is shown in Appendix~\ref{sec:Appendix_Coatings}, Figure~\ref{fig:Thermal_Noise}.
\label{Box:Thermal}
\end{Infobox}


\section{Gravitational Wave Detection Basics}
\begin{figure}[ht]
\includegraphics*[width=\textwidth]{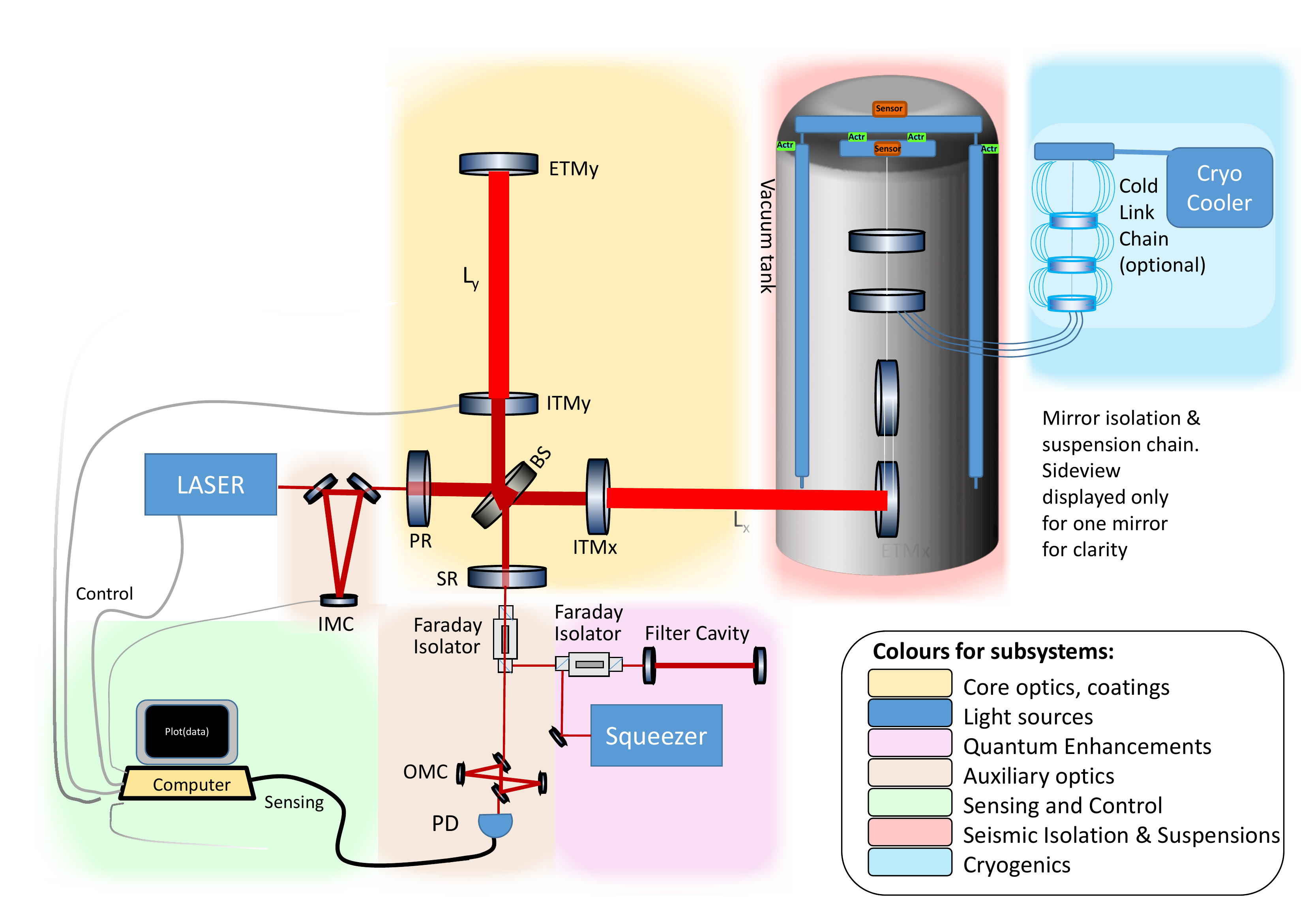}
\caption{Schematic representation of an interferometric gravitational wave detector. Subsystems as used in this report are indicated by the colored backgrounds. The borders between subsystems are not sharply defined and somewhat arbitrary.}
\label{fig:ifo_layout}
\end{figure}

Gravitational waves induce tiny changes in separation between widely spaced `test masses'. The instrumental challenge is to measure these tiny changes. Audio band detectors use laser interferometry, where the interferometer mirrors are the test masses at the ends of long baselines whose length changes are measured. The basis of all present and next generation gravitational wave detectors is a dual recycled Fabry-Perot arm cavity Michelson interferometer as sketched in figure\,\ref{fig:ifo_layout}. 

Being quadrupolar radiation, a passing \ac{GW}  alternately expands and then contracts one arm (e.g. $L_x$) of the interferometer whilst it contracts and then expands the arm perpendicular to it ($L_y$).  The effect is extremely small: expressed as a relative length change, $\delta L/L$, it is less than $10^{-22}$! The arm cavities increase the phase change imposed on the light. Interfering the single frequency light beams from the two arms at the beamsplitter (BS) cancels common noise whilst the signal adds. Extra elements (\ac{PR} and \ac{SR} mirrors, squeezing, filter cavities etc)  further increase sensitivity and optimise the response \cite{InterferometerTechniquesBond2017}.  
The core optics (test mass mirrors) are hung from sophisticated suspensions systems as indicated in figure\,\ref{fig:ifo_layout} (in the vacuum tank on the right) so that, above resonance frequencies, they are effectively free to move \cite{SuspensionsvVeggel2018}. The main optics can be cooled to cryogenic temperatures to minimise thermal noise. In addition to the  core optics there are a host of auxiliary optics to condition  and match (in angle and size) the laser beam into the interferometer  and the signal field out of the interferometer and into the photodetection system. 

\begin{figure}
\centering
\includegraphics[width=0.7\textwidth]{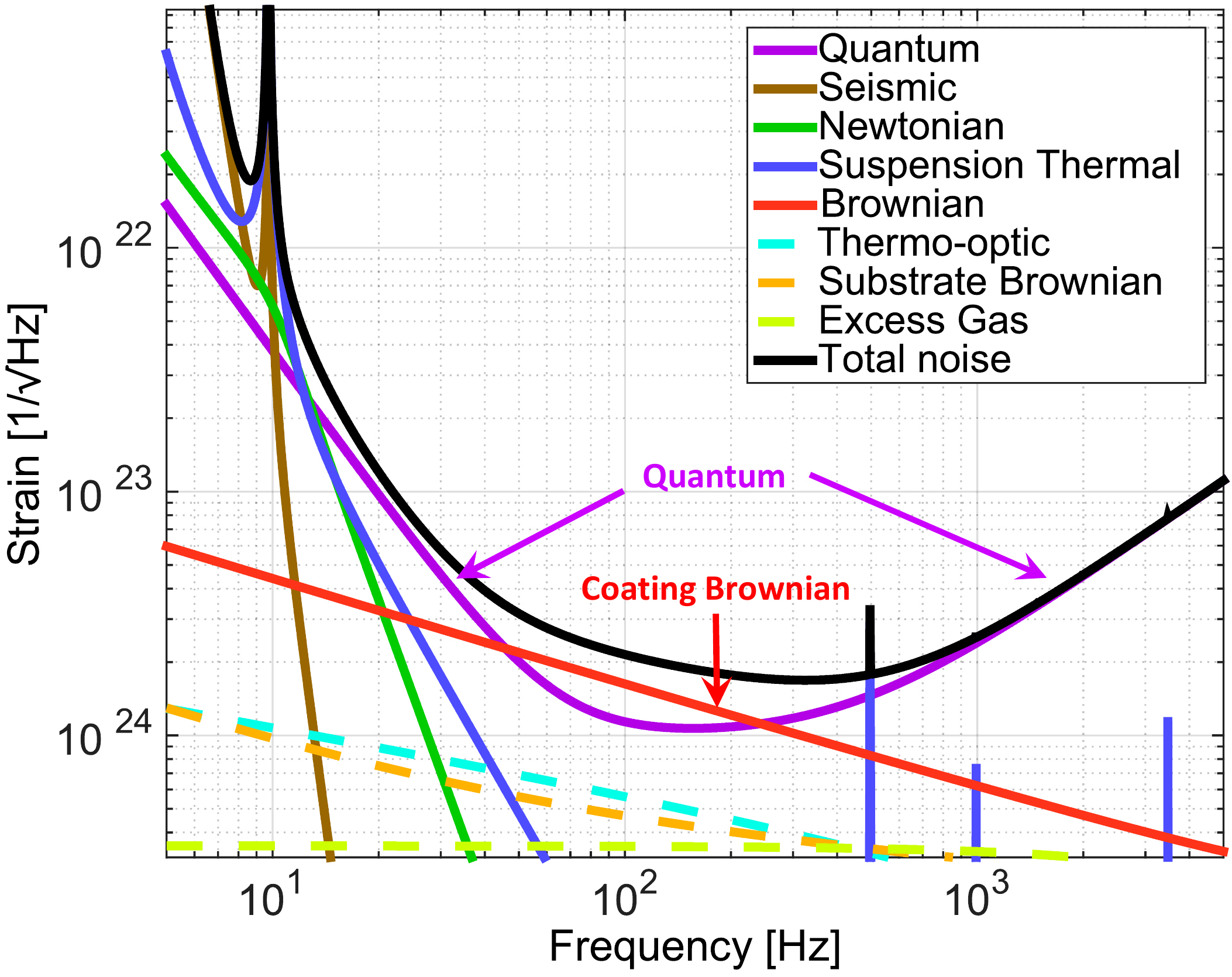}
\caption{Design noise budget for the \acf{a+LIGO} (LIGO Lab/Caltech/MIT)}
\label{A+noisebudget}
\end{figure}

Once technical noises, such as laser frequency and intensity noise, acoustic noise and seismic noise, have been reduced there are three basic processes limiting the interferometer sensitivity: thermal, Newtonian (gravity gradient), and quantum noise. 
In Figure \ref{A+noisebudget} we demonstrate the typical frequency distributions of these processes using the A+ design curve \cite{LIGOA+sensitivity}

Thermal noise \ref{Box:Thermal},\cite{ThermalNoiseHammond2014} is produced by random displacements of the mirror surfaces in response to thermally fluctuating stresses in the mirror coatings, substrates, and suspensions and it typically dominates at midband frequencies  (ca.~50-200\,Hz). Quantum noise can be roughly divided into \ac{QRPN} and quantum phase noise or \ac{SN} \cite{QuantumNoiseBarsotti2018}. \ac{QRPN}, or quantum back-action noise, arises from the random buffeting of the suspended interferometer mirrors by the quantum mechanical amplitude fluctuations of the light field being used to sense the arm-length. Shot noise results from quantum phase uncertainty: there is a limit to how well the phase difference between two light fields returning from the interferometers arms can be determined. Radiation pressure noise dominates at low frequencies (ca. 10Hz to 50\,Hz) while shot noise dominates at higher frequencies (above ca. 200\,Hz). The standard quantum limit is the noise floor for which the \ac{QRPN} and shot noise are equal at a given power level (typically around 100\,Hz). 
\Ac*{NN} arises from the direct gravitational forces exerted on the interferometer mirrors by nearby changing mass distributions primarily caused by density fluctuations of the surrounding earth due to seismic waves as well as low frequency atmospheric density changes \cite{InfrasoundNewtonianNoise2018}. 
The test masses cannot be shielded against these fluctuating forces.  In addition to the noise sources discussed above, another important issue is the suppression of instabilities that arise from photon pressure at high optical power. 

\begin{figure}[ht]
\includegraphics[width=\textwidth]{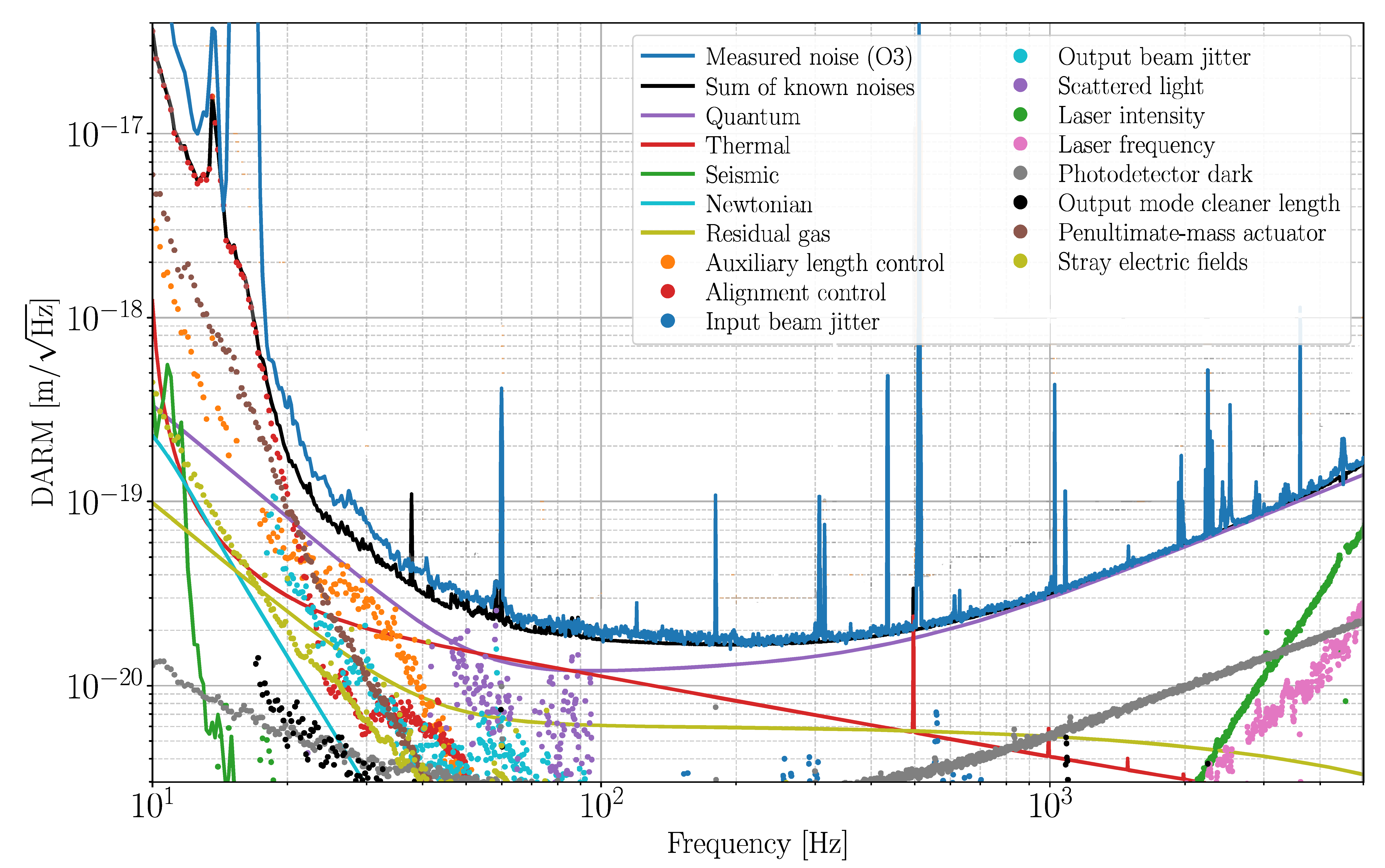}
\caption{Example noise budget for the \acf{aLIGO} interferometer in Livingston, Louisiana, circa March 2020. For details see~\cite{ALIGO_O3noise} }
\label{fig:aLIGO_Noises}
\end{figure}

In reality there are a myriad of technical noise sources that need to suppressed in order to reveal the `fundamental limitations' (scattered light, electronic noise, various other control noises)  This is dramatically demonstrated in  figure \ref{fig:aLIGO_Noises} which shows the noise anatomy for the \acf{aLIGO} detector at Livingston, \ac{USA},  circa March 2020.  Much  of the effort and focus when commissioning  a detector is devoted to suppressing such "technical"  noise.

From Figure \ref{A+noisebudget} it is clear that the impact on the sensitivity of reducing one noise source depends on other noise contributions at relevant frequencies.  
For example, lowering quantum noise around 100\,Hz will have little impact unless coating Brownian noise is similarly reduced.  The low frequency band below 20\,Hz is far more complex.  
In this report we will review each noise source and the R\&D that needs to be done, independently of other constraints.   
It is beyond the scope of this report to reflect on what may happen if progress on a particular subsystem is slower than expected.  
Such impact on the science that can be done will be considered in trade studies that will be carried out in the various detector collaborations.

%% file: Facilities_and_Infrastructures.tex
\chapterimage{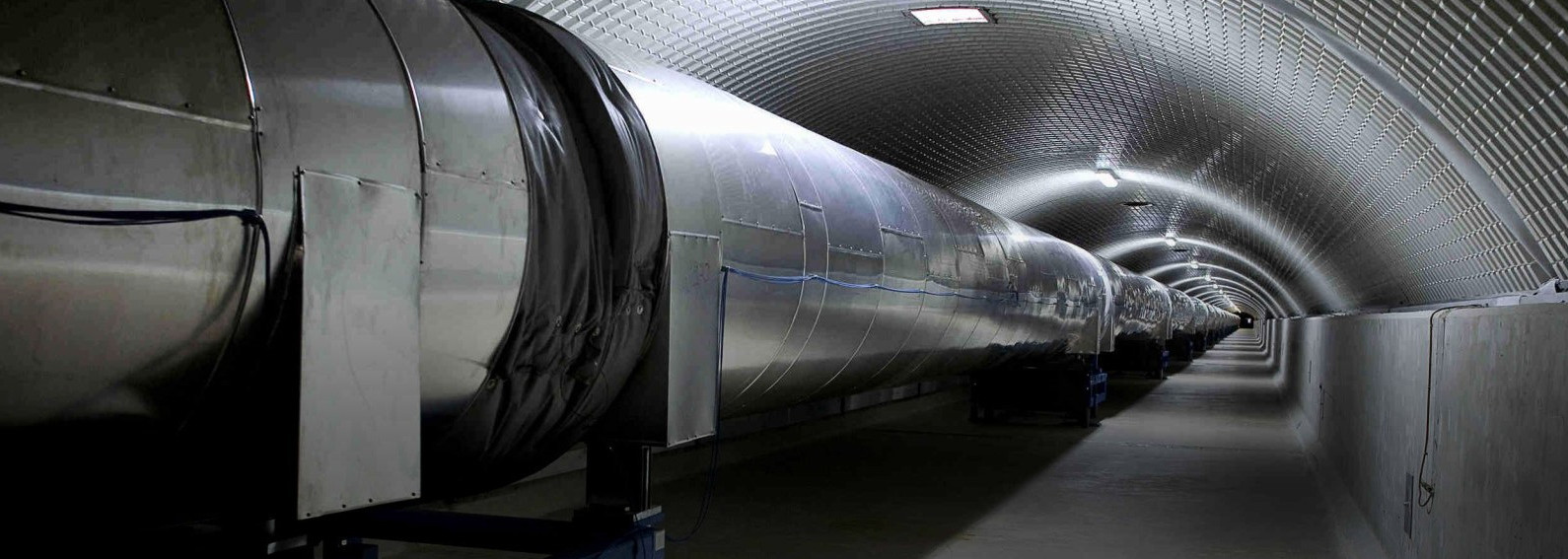} 

\chapter{Facilities and Infrastructures}
\label{sec:Fac_Inf}
\vspace{1cm}
\ac{3G} facility conceptual designs must be mature well before construction, as the infrastructure will dominate both schedule and cost of new observatories, and so carries great importance. 
The sensitivity and observation time achievable by the \ac{3G} detectors will be directly impacted by characteristics of the observatory sites and their infrastructure. 
Local noise (of natural and human origin) is a key parameter for evaluating candidate sites. Ground vibrations will limit the sensitivity directly, through seismic and Newtonian noise couplings, and indirectly by driving scattered light noise and by complicating interferometer control. 
Aspects such as the nature of the rock, abundance of water, and geological stability are particularly relevant for the stability and costs of underground detectors, such as ET. 
Anthropogenic activity can influence both underground and surface installations. The construction of the sites, chiefly tunneling and leveling, together with the civil facilities and vacuum systems will dominate the total project costs. 
Every effort to minimize costs, cost contingencies, and other collateral impacts on society in these domains will reap rewards, potentially pivotal, in approval and support.

%

\section{High Level Design Considerations}
\label{Req:Fac_Inf}
High-level site-related design considerations include: characteristics of seismic noise; surface meteorological conditions, anthropogenic noise and geological stability; site topography, specifically levelness for above ground detectors; rock type and the abundance of water; regional and national permitting and environmental clearances; and livability such as proximity to urban centers.  


For underground detectors, the design of the experimental halls requires special attention. They have to be large enough to facilitate assembly and maintenance of detector components and to accommodate future upgrades. On the other hand, the larger the volume the larger the cost and the engineering challenge. Cavern volume and shape also affect atmospheric Newtonian noise (Section \ref{sec:Newtonian_Noise}), which might limit the sensitivity.

For surface detectors, 
the Earth’s sagitta is of order 30\,m for a 40\,km laser-straight arm, somewhat blurring the concept of “surface” construction. This approach does, however, promote surface topography and surface geology in their priorities as site criteria. The direct effect of wind on above-ground structures also joins its indirect influences on seismicity and gravitational gradients. Effects on the artificial and natural environment, flora and fauna are also concerns, and must be factored differently into site selection, project approval, and design.
Advances in tunnel boring machines and in surface road and pipeline excavation have been seen to drastically reduce the cost per kilometer of public works structures over recent decades, e.g.,~\cite{BoringCompany}. Directed research into applying these modern methods to reduce the cost and collateral impact of \ac{3G} projects should be explored as a high priority. Attention must be also paid to the legal aspects of large-scale civil construction specific to each candidate country, which could impact the timing of the infrastructure realization. 

\section{Impact/Relation to 2G and Upgrades}

The sensitivity of second generation detectors, and their upgrades such as A+ and Virgo+, will continue to improve. 
However, in the coming decade these detectors will reach their infrastructure limit. In other words, the level of environmental noise as well as the detector lengths and even the space available in the buildings will make it too challenging and expensive to improve further. The infrastructure for ET, with 10\,km-long underground triangular caverns, and CE, with a 40\,km-long L-shaped surface footprint, is designed to go well beyond these limits and house instruments that will continue to progress in sensitivity over their planned 50-year lifetime.   
In creating these designs it is crucial to understand the 2G limits and to be sure that all related lessons have been learned. For instance, review of the machine-induced noise in the current infrastructures will allow the \ac{3G} site designs to ensure that otherwise quiet environments (and the benefits of going underground) are not spoiled by such disturbances. A joint effort based on the \ac{LIGO}, Virgo and particularly \ac{KAGRA}  experience on this topic will be beneficial. 

In current practice, incremental upgrades of existing detectors like \ac{LIGO}, GEO600  and \ac{Virgo}  have been an effective way to prove new technologies in context, at full sensitivity, while directly expanding the observing horizon. Enhanced \ac{LIGO}, Advanced \ac{LIGO}, Advanced Virgo, \ac{AdVirgo+}  and \ac{a+LIGO}  are examples of this effective combination of \ac{RaD}  and practical deployment ``on the fly''  for improved observations.  
The incremental upgrade approach has limits, however.  One is the constraint to maintain compatibility with legacy systems and infrastructure; new topologies, footprints, and even wavelengths may be effectively off the table. Less obvious, but potentially more serious, is the growing imperative to minimize interruptions to observing. Upgrades require downtime.  The explosive discoveries of the last three years, particularly the multimessenger astronomy revolution triggered by GW170817,  have raised global desire to keep observing with existing 2G and 2G+ instruments. 
As a result, an increasing portion of \ac{3G} technology demonstration must rely on offline engineering development in subscale demonstrations, reprising the development environment of initial \ac{LIGO}  and Virgo, before any large-scale testbed existed. 

\section{3G Vacuum Systems}
Vacuum systems for planned \ac{3G} detectors are likely to be the largest ultra-high vacuum (UHV) systems built, and will account for a significant part of the cost of building the observatories. Substantial innovation and research went into designing and building the \ac{LIGO}, Virgo, GEO600 and \ac{KAGRA}  vacuum systems within economic constraints; the more stringent technical requirements and much greater size required for \ac{3G} vacuum envelopes threaten to render them infeasible without still further innovation. Many avenues for technical research have been proposed,  centered on themes of either improving an interferometer’s immunity to residual gas, or reducing the cost of suitable installations. Some questions that merit closer investigations include:

\begin{itemize}
\item  Are there economies in using materials other than stainless steel, such as mild steel or aluminum, as the envelope material? 
\item Are nested (e.g., differentially pumped) vacuum systems practical and economical? 

\item Can civil construction mass production techniques, such as extrusion and spiral tube milling, be adapted to future UHV construction?

\item Are there newer mass production surface treatment and cleaning techniques that can be applied to reduce outgassing? 

\item Are there ways to simultaneously meet surface outgassing requirements and possibly distributed pumping together with other physical requirements of the system, such as 
stray light attenuation, vibration damping and particulate mitigation?

\item Can optical pressure gauging and leak detection offer practical advantages for system construction, commissioning and maintenance?

\item Are there new getter materials for pumping and surface treatments for maintaining \ac{UHV} conditions for very large systems with modest gas loads?

\item Are there new ideas for reliable, affordable large gate valves to isolate from atmospheric pressure during construction and service, and to isolate volumes with different requirements?

\item What are the effective methods, or surface treatments, to minimize moisture adsorption during vented system access and to accelerate desorption during pumpdown and recovery?

\item What vacuum tube diameters are optimal taking into account light scattering issues, vacuum conductance etc. This will in turn influence tunnel size and costs.

\end{itemize}

In January 2019, an NSF-sponsored Workshop on Large Ultrahigh-Vacuum Systems was held to identify cost effective technologies for the design, construction and operation of the large vacuum systems required for \ac{3G} observatories~\cite{LLOVacWorkshop2019}. This workshop considered many of the issues above and particularly focused on two concepts: one based on an extrapolation of the single-walled vacuum pipes used by \ac{LIGO}, Virgo, and \ac{KAGRA}  and another using double-walled nested vacuum pipes, with the outer wall handling the atmospheric load and the inner wall supporting ultra-high vacuum. Both of these concepts were found to be viable and recommended to be taken to the next level of detail in follow-on studies. Vacuum pumping and pipe surface treatment solutions were also considered. The participants reported confidence that positive impacts could be achieved for the cost of the construction and operation of vacuum systems for \ac{3G} detectors. They provided recommendations for further study including: investigation of beamtube materials such as mild steel and aluminum; further characterization (outgassing and optical properties) of potential surface coatings; the use of ultra-dry gas purging during venting (even in current detectors); and leveraging partnerships with \ac{NIST}, \ac{JLab}, \ac{CERN}, and industrial contractors. 

\section{Outlook and Recommendations}
Facilities and infrastructure design and costing studies must evolve very soon to studies undertaken by construction engineers. 
\ac{3G} construction must start roughly 5 years before detector installation and ten years before observations. 

We recommend that
\begin{itemize}
\item To allow 2035 observations, the responsible international collaborations and their working groups should be focused on bringing facility-related \ac{RaD} and detailed technical studies to maturity imminently.
\item Careful study of candidate sites considering the above points, as well as studies of drivers of facility decisions that can be made informed by 2G instruments and scaled prototypes, need to be completed now. 
\item  every effort  is made to minimize costs by careful choice of site and methods for facility and infrastructure construction.
\item For vacuum systems, facilities for testing outgassing rates and optical properties of vacuum pipe materials and coatings are needed. Additionally, as design studies progress, facilities to test scaled vacuum concepts may be required.
The NSF vacuum workshop is a good model for leveraging expertise and could be replicated for other areas of facilities and infrastructure work.
\item Collaboration with high energy particle physics community and with industry on facilities and infrastructure research, design, and testing  to leverage experience in building large accelerator facilities and vacuum systems.
\end{itemize}

%% file: Newtonian_Noise.tex
\chapterimage{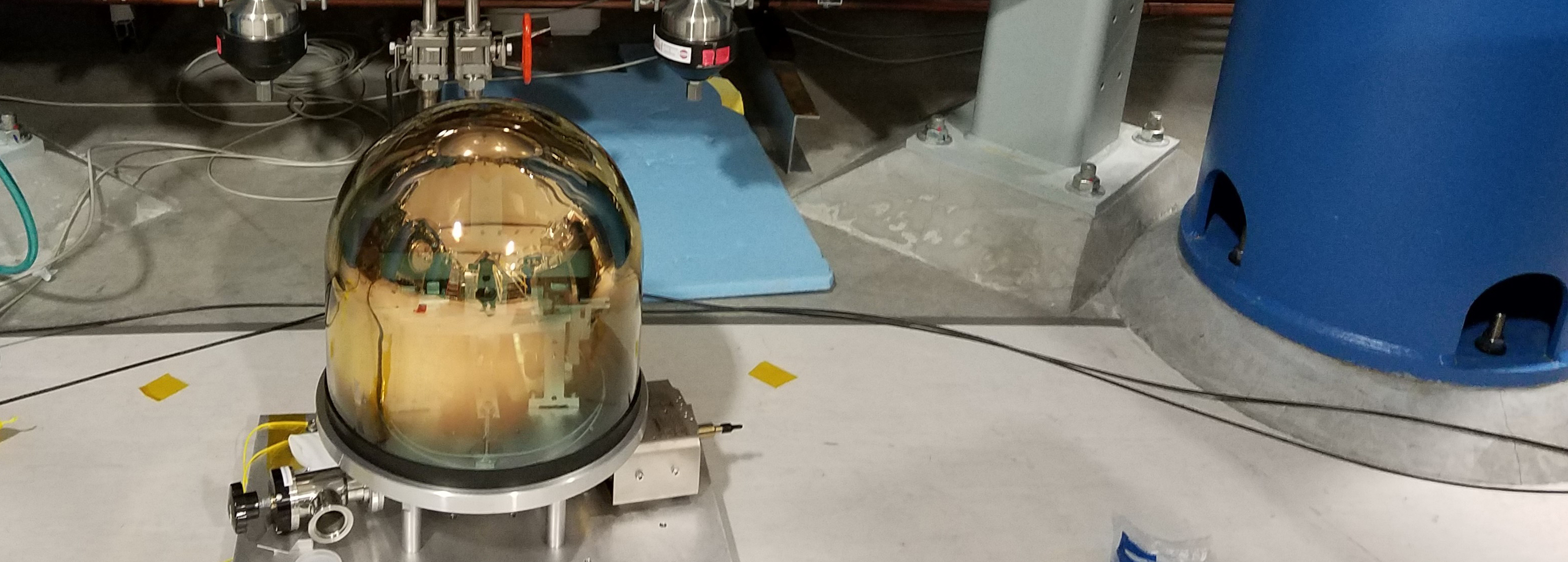} 

\chapter{Newtonian Noise}
\label{sec:Newtonian_Noise}
\vspace{0.5cm}
\Acf{NN} is predicted to be one of the limiting noise sources in third-generation detectors at frequencies below 30\,Hz~\cite{Saulson:NN,Har2015}. Sources of \ac{NN} include seismic fields, atmospheric sound and temperature fields, and vibrating infrastructure~\cite{HuTh1998,BeEA1998,Cre2008,FiEA2018,Har2015}. Mitigation of \ac{NN} can be achieved by suppressing density perturbations in the environment near the test masses~\cite{HaHi2014}, and by subtraction of \ac{NN} estimates from gravitational-wave data using data from environmental sensors~\cite{Cel2000,CoEA2016a}.

\begin{samepage} 

In \ac{2G} detectors, the dominant contribution to seismic and acoustic fields, and therefore to the associated \ac{NN}, is mostly produced by detector infrastructure such as pumps and ventilation fans. The natural environment will become more important in \ac{3G} detectors, since, (a), we have learned and will continue to learn how to build detector infrastructure that does not significantly disturb the environment in the \ac{NN} frequency band, and (b), our targets for \ac{NN} cancellation will be so ambitious that we will care both about the dominant anthropogenic perturbations and the weaker disturbances caused by nature. Therefore, \ac{NN} research has a potentially big impact on site selection~\cite{BeEA2010}.

\section{State of the Art}
Much effort has gone into modeling \ac{NN}. Analytical models of seismic \ac{NN} have been calculated for underground detectors in homogeneous half spaces including the scattering of waves with arbitrary polarizations from spherical caverns~\cite{Har2015} and for specific seismic sources such as point forces and point moments in homogeneous half spaces~\cite{HaEA2015,Har2016}. Numerical simulations have been performed to calculate seismic fields from point forces in laterally homogeneous half spaces~\cite{BeEA2010c}. Analytical models of atmospheric \ac{NN} have been calculated for temperature fields in laminar flows~\cite{Cre2008}, turbulence induced pressure fluctuations~\cite{Har2015}, and homogeneous sound fields~\cite{FiEA2018}. 

\end{samepage} 

Surface seismometer arrays optimized for \ac{NN} cancellation have been calculated using analytic and numerical methods~\cite{Har2015,CoEA2016a}. Analytic expressions were derived for Rayleigh waves to express \ac{NN} and correlations between \ac{NN} and seismometers in terms of observed seismic correlations~\cite{Har2015,CoEA2016a}. The potential impact of surface topography on \ac{NN} cancellation has been investigated~\cite{CoHa2012}. A tiltmeter signal was suppressed by more than an order of magnitude using Wiener filtering with a seismometer array, which serves as proxy for \ac{NN} cancellation~\cite{HaVe2016,CoEA2018}. A factor 1000 suppression of seismic signals in seismometers using Wiener-filtered data was achieved with underground arrays at Homestake ~\cite{CoEA2014} and 100x at \ac{LIGO} Hanford with surface arrays ~\cite{CoEA2018}. Extensive seismic array measurements were performed at Sanford Underground Research Facility, \ac{LIGO} Hanford, and Virgo. Practical work on the mitigation of atmospheric \ac{NN} has only started. Extensive sound correlation measurements have been done at the Virgo site to characterize the sound field.

Some aspects of how the site and infrastructure can impact \ac{NN} are now well known. 
Ambient seismic noise is mostly understood from world-wide long-term observations and studies of how it depends on, for example, geology, and distance to major cities and coast~\cite{CoHa2012b}. Detailed studies of the connection between geology and ambient seismic fields were made at detector sites~\cite{HaOR2011} and at the Sanford Underground Research Facility. Seismic scattering from topography was studied in linear order to estimate the effect of scattering on \ac{NN}~\cite{CoHa2012}. Sound and seismic noise between about 5\,Hz and 50\,Hz are dominated by sources that are part of \ac{LIGO}/Virgo infrastructure (e.g., ventilation). Concrete plans are being developed for infrastructure changes at Virgo to mitigate \ac{NN} from sound fields.

\section{Requirements}
There remain, however, aspects of site and infrastructure influence on \ac{NN} that are not well understood. 
While noise-cancellation systems can possibly enhance sensitivities of \ac{3G} detectors, methods to cancel \ac{NN} from underground seismic fields and the atmosphere have not been developed yet even in theory. Consequently, atmospheric and underground seismic \ac{NN} should currently be considered a fundamental noise limitation of \ac{3G} detectors. Generally, the aim of any new infrastructure and site selection should be to have natural sound and seismic noise levels as close as possible to the global low-noise models~\cite{Pet1993}, especially with the goal to extend the observation band to frequencies below 10\,Hz, and to perturb the natural fields as little as possible with the infrastructure. 

We have seen that the infrastructure of current \acp{GWD} is the main source of seismic and sound disturbances in the frequency band between 10\,Hz and 30\,Hz. It is therefore important to (a) learn from this experience, and develop low-noise infrastructure designs for \ac{3G} detectors and (b) develop a set of tools to characterize the environment, and to understand how it affects \ac{NN}. This part needs to take into account what information about a site can realistically be obtained in a 1 to 2 year site-characterization study. It is recommended to establish globally accepted guidelines of how site-characterization measurements are to be carried out to be complete and of sufficient quality. We need to establish how much \ac{NN} is reduced as a function of detector depth. Finally, (c) improved models especially of atmospheric \ac{NN} are required to be done either with analytical calculations or numerical simulations.
Noise-cancellation systems need to be developed to go beyond the infrastructural noise limitations. (a) Concerning surface detectors, new technologies are required to monitor fluctuations of the atmospheric mass-density field, which is connected to sound and temperature fields. Coherent \ac{LIDAR} has been proposed as a possible sensor. Collaborations with LIDAR groups have started, but it is still unclear whether the required sensitivity can be achieved in the future. Torsion type sensors may offer a solution ~\cite{McManus2017.CQG}.  a This problem plays a minor role in underground detectors, where atmospheric \ac{NN} is strongly suppressed. However, (b) a cancellation system of \ac{NN} from acoustic fields inside buildings and underground caverns might be required also for underground detectors. In this case, a simple microphone array can in principle be used, but how to design it based on sound-correlation studies is still an unsolved problem. These studies rely on (c) new numerical simulations and advanced analytic models of atmospheric fields including effects such as turbulence and sound scattering, especially for underground detectors where the goal is to observe \acp{GW} down to 2-3\,Hertz. (d) cancellation of \ac{NN} from underground seismic fields needs to be developed. Here, the main questions are where seismic sensors should be ideally placed, and what type of sensors are to be used (single-axis or three-axis seismometers, seismic tiltmeters and strainmeters). Special attention should be given to (e) how local geology and above all surface topography increase the complexity of the seismic field through scattering of seismic waves. This will have an important impact on the design of seismometer arrays for seismic \ac{NN} cancellation.

\section{Impact/Relation to 2G and Upgrades}
Newtonian-noise cancellation techniques will evolve continuously from 2G to \ac{3G} surface detectors, and insight gained from 2G \ac{RaD} will be applicable to \ac{3G} detectors. This is true for seismic as well as atmospheric \ac{NN} in surface detectors. This development extends from the current 2G detectors, to their minor and major upgrades, into the \ac{3G} era, whenever the goal is to also achieve low-frequency sensitivity improvements. 

There are aspects of \ac{NN} modeling and cancellation unique to \ac{3G} underground detectors, as for example the question how quickly surface disturbances are suppressed with increasing detector depth, and how to cancel seismic \ac{NN} in underground detectors. The \ac{KAGRA} underground detector in Japan might 
provide some continuity of underground \ac{NN} \ac{RaD} towards the \ac{3G} era, but \ac{NN} is currently not expected to be a limiting noise source for \ac{KAGRA}.

\section{Pathways and Required Facilities}
The main facilities of \ac{NN} \ac{RaD} are the natural environments of \ac{3G} detector candidate sites. In addition, since we have very little experience in observing and modelling underground seismic and sound fields, underground studies done anywhere can provide important information. There are significant differences between sites in terms of sources of environmental disturbance, local geology and topography, so transferring results from one site to another should be done with caution.

Finite-element simulations of environmental fields and algorithms for the optimization of array configurations for noise cancellation are computationally expensive \cite{Andric:2020}. Computing time on high-performance clusters is required to carry out the most demanding simulations and optimizations.

Collaboration between groups is strongly recommended with respect to site characterization. It requires substantial expertise to set up robust, high-quality, environmental measurements to understand which properties of environmental fields are important, and how to analyze environmental data. Collaboration is also encouraged between groups working on numerical simulations to share generic knowledge, such as how to assess the accuracy of numerical simulations, e.g., by comparing simulations using different software, or by running simple simulations that allow comparison with analytic models.

\section{Outlook and Recommendations}

As the level and character of \ac{NN} and the choice of the \ac{3G} sites are strongly interdependent, a solid understanding of the \ac{NN}-relevant characteristics of candidate sites is required imminently. After this, significant \ac{RaD} into the above open questions should continue through the \ac{3G} era such that the \ac{NN} of the chosen site and infrastructure can be reduced as much as possible. 

We recommend that:
\begin{itemize}
\item \ac{NN} criteria be incorporated into any site evaluation,
\item Infrastructure and facilities be designed to mitigate and minimise \ac{NN} 
\item site studies and the knowledge  gained be shared extensively.
\end{itemize}



%% file: Suspension_and_Isolation.tex
\chapterimage{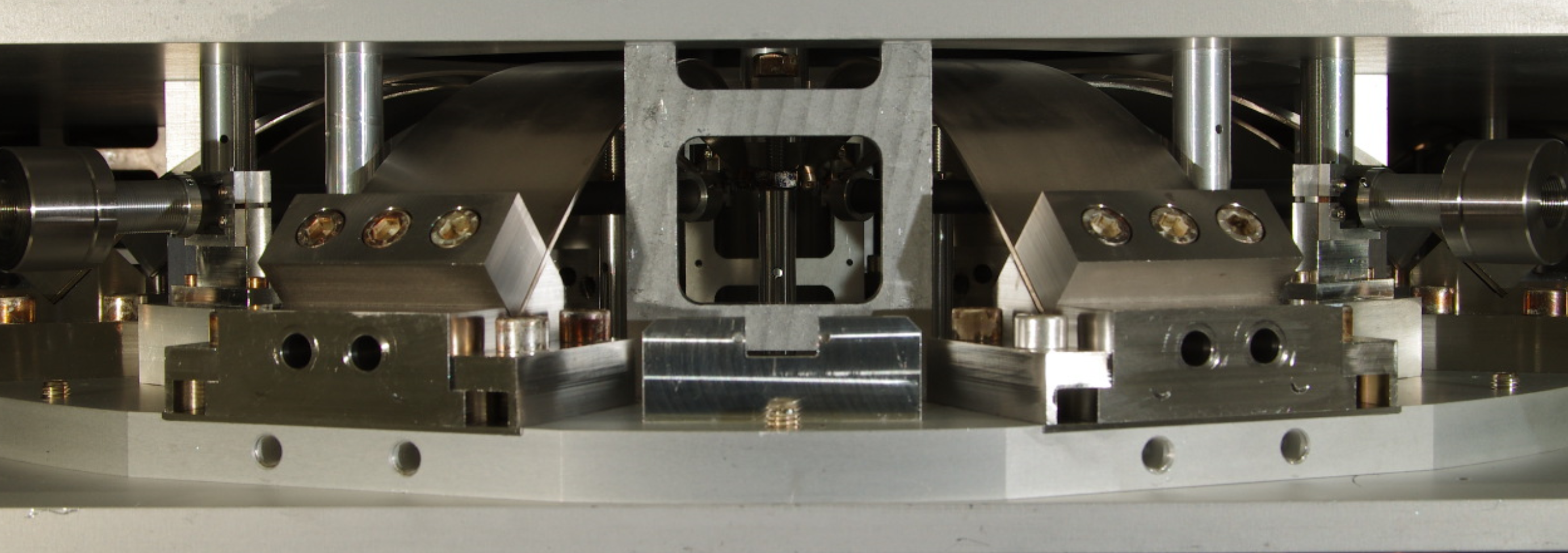} 
\chapter{Suspensions and Seismic Isolation Systems}
\label{sec:Suspensions_Isolation}

\begin{samepage} 

In this section we discuss suspensions and isolation systems for 3rd generation detectors. We address four areas: suspensions (especially the final stage), isolation, damping and control, and interface with cryogenics. The last two areas overlap Sections~\ref{sec:Sim_Controls} and~\ref{sec:Cryogenics} and thus are only covered briefly here.

\section{State of the Art}
The use of fused silica fibers is a well-established technique for the final stage of the suspension of fused silica test masses, leading to a monolithic suspension which minimises suspension thermal noise. Currently there are four detectors operating with fused silica suspensions at room temperature: two \ac{aLIGO} detectors, \ac{AdVirgo} and \ac{GEO600}. Research is ongoing on silica within these collaborations.
The \ac{KAGRA} project is pursuing the use of cryogenic sapphire suspensions (i.e. sapphire fibres supporting a sapphire test mass) working at about 20K. 
Current detectors use different combinations of active and passive stages to achieve the necessary level of isolation and control. In \ac{aLIGO}, overall isolation is achieved using three sub-systems: the \ac{HEPI} for low frequency alignment and control, a two-stage hybrid active and passive isolation platform and a quadruple pendulum suspension system that provides passive isolation above a few Hz and supports the test mass~\cite{Matichard_2015}. \ac{AdVirgo} employs a combination of a tall inverted pendulum which is actively controlled, a passive seismic attenuation chain (the \ac{SA}), and a double pendulum supporting the test mass~\cite{AdvancedVirgo2015}. \ac{KAGRA} also uses a passive attenuation chain supporting the cryogenic payload suspension system~\cite{KAGRA2013}. All detectors incorporate active damping for various resonances within their systems and their suspensions incorporate means to apply signals (magnetic, electrostatic) for global alignment and arm length control.

\section{Requirements, Challenges and current/Planned R\&D}
Suspension thermal noise and residual seismic noise are two of the dominant noise sources which limit the low frequency performance and define the low-frequency cut-off for ground based gravitational wave detectors. Thus the requirements of the suspension and isolation systems are to a great extent set by what the target low end of the operating frequency band is chosen to be, as well as by the intrinsic seismic levels of the chosen sites.
For reference, the design low frequency cut-off for \ac{aLIGO} and \ac{AdVirgo} is at $\sim$10\,Hz (see figure \ref{fig:3GSens}).

\end{samepage} 

\noindent{\bf Suspensions, including cryogenic aspects} For future detectors at room temperature the aim will be to further reduce the suspension thermal noise. This is likely to involve suspension of heavy mirrors, up to several hundred kg, making the fibers as long as practicable, and making them relatively thinner (and thus requiring them to support higher stress) to push the bounce modes down in frequency and push the violin modes up~\cite{Heptonstall:2014, Bell:2014,aisa:2016advanced, Tokmakov:2012, Amico:2002_monolithic}.  Testing of individual elements and fully assembled prototypes will be required, as will upgrading current methods for pulling and welding fibers and of assembly procedures~\cite{Hammond:2014,Travasso:2018}. Ensuring that robust techniques are developed to handle the delicate fibers and heavier masses through assembly and installation will be an engineering challenge. For a general overview regarding reducing suspension thermal noise see~\cite{Hammond:2014, Hammond:2012}.

The work of \ac{KAGRA} is ground-breaking for the understanding and application of cryogenic techniques, and will lead to the first full stage sapphire suspensions operating at low temperature \cite{Kumar:2016_KAGRA}. The community will have a much better feel of where effort needs to be applied based on \ac{KAGRA}'s experience.
In general for detectors operating at cryogenic temperature, silicon or sapphire are the materials of choice for suspensions and test masses for low thermal noise, and these will have different challenges compared to the use of fused silica (see e.g., \cite{Cumming:2014Silicon, nawrodt:2013,Haughian:2016, Alshourbagy:2006_thermoelastic, Alshourbagy:2006,amico:2004, Alshourbagy:2005,Khalaidovski:2014}). A significant challenge with 20\,K operation is the need to extract any deposited power via the fibers, which in turn drives their cross sectional area and vertical stiffness, and requires knowledge of their thermal properties. Suspending heavy mirrors with thick fibers/ribbons will need a smart design to soften the vertical and horizontal modes. Operation at 123\,K is less challenging for heat extraction since radiative cooling can be used.

All aspects of the monolithic assembly process will need development for \ac{3G} detectors. This includes hydroxy-catalysis bonding, which is already successfully used in \ac{GEO}, \ac{LIGO} and \ac{Virgo} (silica-silica) and \ac{KAGRA}, (sapphire-sapphire)~\cite{Haughian:2016,dari:2010, Amico:2002, vanVeggel:2014}. Properties of Si-Si bonds are being investigated, and indium or gallium bonding may also have applications in certain areas~\cite{Hofmann:2015, Murray:2015Low_Temp}.  Fiber fabrication of sapphire or silicon material with circular or ribbon geometries will be demanding and require significant \ac{RaD}: laser heated pedestal growth, micro-pull down, machining or etching are all possible techniques to be investigated~\cite{Cumming:2014Silicon, Alshourbagy:2005}. Assembly processes for sapphire and silicon including welding will need to be developed.

Other aspects which will need consideration include excess losses like clamping or bonding losses. The challenge is to have suspension dissipation dominated by the material thermal noise and not by thermoelastic or other losses.
Simulations and modelling of suspensions will be important for understanding overall behavior, including dynamics of fiber suspensions, violin mode splitting and long term stability. \Ac{FEA} is also an important tool as a cross check to design the best strategy to produce and realize the lower stage suspension \cite{Lorenzini:2010, Sorazu2017Sus}.
Consideration should also be put into upper stages of the suspensions to ensure they do not limit thermal noise performance of the final stage, for example due to noise from the maraging steel blade springs. Lower loss materials such as sapphire and silica could be used. Indeed \ac{KAGRA} already incorporates sapphire springs at the final stage. Achieving a robust design with high breaking stress, low mechanical loss and good thermal conductivity, with the possible use of protective coatings are areas to be studied. 

As regards cryogenic operation, we have already noted that extracting power via heat conduction through the suspension elements is a major consideration for the design of a 20\,K detector. Finite element analysis with the various geometries, losses and thermal parameters will be valuable. The design of upper stages needs to be compatible with cryogenic operation and allow efficient heat extraction. A cryogenic suspension is by definition `out of thermal equilibrium' and this needs to be evaluated. Cooling time is also a potential issue, as the timeline to commission such a detector may be driven by the several weeks to cool/warm up between vents. Currently there are efforts to work on mechanical heat links that can be removed. A cooling exchange gas is also a possibility, although may not be used due to concerns about residual vacuum level. Additionally, the development of vibration-free cryogenic suspensions is a pressing challenge. While results from \ac{KAGRA} may help inform this work, it is deserving of more attention.


\noindent{\bf Isolation } For the \ac{3G} detectors, combinations of active and passive stages will be used to achieve the required isolation set by the site locations and the detectors' target sensitivities.
These considerations will also influence the overall height of the isolation systems and the number and type of stages it uses.
The design goal for better sensitivities at lower frequencies requires the use of less noisy inertial sensors to be used in loop for the control of the active platforms.
Additionally, increased vertical isolation will be needed for \ac{3G} since vertical to horizontal coupling increases with arm length. 

More specifically, the \ac{ET} design~\cite{ET2011} aims to reduce the low frequency sensitivity cutoff to 1.8\,Hz, in part using a longer (17\,m) superattenuator. However, it would be valuable either to relax further this cutoff and/or to reduce the overall height of the vibration isolation system in order to save money for the realization of the underground caverns. Toward this end, a modified design of the superattenuator, using two cascaded inverted pendulums is being considered. On the other hand, it would be interesting, and open additional collaboration channels, to study the possibility of a merging of the technologies used so far by \ac{AdVirgo} and \ac{aLIGO}. 
The \ac{CE}  isolation system follows a scaled up approach from \ac{aLIGO} and \ac{Voyager}, with a more relaxed cut-off of 8\,Hz. For a 40\,km arm length detector such as \ac{CE}, requirements for vertical seismic isolation will be particularly challenging to achieve, especially at low frequencies, given the larger cross-coupling from the vertical to horizontal direction due to the curvature of the Earth on the longer baseline. Addressing this will require dedicated \ac{RaD} efforts.

\noindent{\bf Controls} The design of control systems for suspension and isolation systems has evolved significantly, and is expected to use modern controls that are now being tested, such as noise subtraction, automatic filter design and supervised machine learning and neural networks for feedback optimization. 
Noise subtraction is best done by directly actuating on the suspensions and seismic isolation systems; this needs careful design and modeling of actuation mechanisms for guaranteed dynamic range and low noise.  To achieve this, lower noise sensors may be required and the robustness of operation will have to be tested. Controls are discussed further in Section~\ref{sec:Sim_Controls}.

\section{Outlook and Recommendations}


Suspensions and isolation systems are a central part of the \ac{3G} interferometer designs and their \ac{RaD} must be mature roughly a decade before systems installation such that facilities and related subsystem choices can be made. It has typically taken 10-15 years to take suspension and seismic isolation hardware from prototype designs to interferometer installation. Thus prototypes for \ac{3G} suspension and isolation systems are an essential step to take in the near future. These can start with small scale (bench top) prototypes but purpose built facilities will be needed.  Several small-scale collaborations across detector groups in different countries already exist and we expect these to continue. Between them ideas and results can be shared and discussed at existing meetings such as \ac{GWADW} or other meetings such as \ac{ET} workshops. No additional larger collaborations are currently envisioned. 

We recommend that:
\begin{itemize}
\item full-scale \ac{3G} \acs{SIS}  test facilities be developed by upgrading \ac{2G} facilities such as \ac{LASTI}, the 40m detector (Caltech), the 1500 W hall (EGO) and Gingin (Western Australia) 
\item new prototyping facilities, such as the new \ac{ET} Pathfinder in Maastricht, and the refurbished Glasgow facility  be developed  for integrating \ac{SUS} with Cryogenics.
\item the performance  of \ac{KAGRA} be analysed broadly and lessons learned shared widely broadly with the community.
\end{itemize}

%% file: Cryogenics.tex
\chapterimage{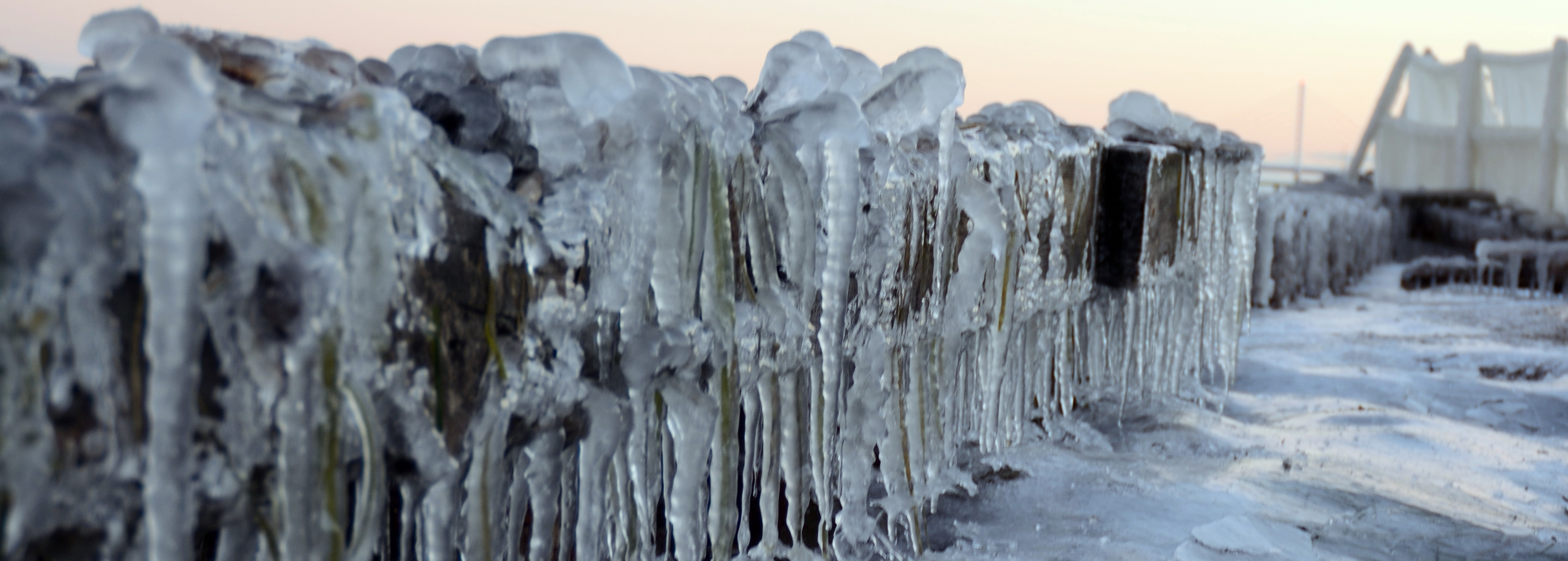} 
\chapter{Cryogenics}
\label{sec:Cryogenics}
\vspace{1cm} 


To reduce thermal noise (see Box~\ref{Box:Thermal}), the Einstein Telescope~\cite{ET2011}, Voyager~\cite{Voyager:Inst} and Cosmic Explorer~\cite{CosmicExplorer2017} (\ac{CE2}) are designed to operate with their mirrors and suspensions at cryogenic\footnote{Here we use the term cryogenic to refer to temperatures significantly below room temperature, including 123\,K.} temperatures (see Table~\ref{tab:CryoTemps}).

Gravitational-wave physics has a history of using cryogenics to improve the sensitivity of detectors, starting with resonant mass (bar) detectors~\cite{ColdBars}. Currently, the first cryogenic laser interferometers \ac{CLIO}~\cite{CLIO:2008, Agatsuma:2010, Uchiyama:2012} and the 3\,km gravitational-wave detector \ac{KAGRA}~\cite{KAGRA2013}, both in Japan, are in operation and testing the cryogenic performance of mirrors and suspensions. 
In CLIO, the mirrors were cooled to cryogenic temperatures without observing significant additional noise, suggesting active cryo-coolers as potential candidates for use in future interferometers. \ac{KAGRA}, which will be the first cryogenic interferometer large and sensitive enough to detect gravitational waves, will provide a valuable test bed for uncovering potential, unknown problems with sensitive cryogenic interferometry, a key step toward \ac{3G} cryogenic detectors. Several other laboratories worldwide operate cryogenic optical cavities for precision measurements~\cite{Mueller:03}, atomic clocks~\cite{JunYeGroup:2019} and tests of quantum mechanics~\cite{CaltechIQIM} and are valuable partners for the development of technologies in cryo-cooling and low noise cryostats. 

\begin{table}[h]
\centering
\begin{tabular}{ |l||l|l|l|  }
 \hline
 Interferometer & Mirror Temperature [K] & Mirror Material & Suspension \\
 \hline
 CLIO           &   20\,K           & sapphire     &  steel wires \\
 KAGRA          &   20\,K           & sapphire  &  sapphire fibers  \\
 ET-LF             &   20 or 123\,K    & silicon   &  silicon  \\
 CE2             &   123\,K          & silicon   &  silicon ribbons \\
 Voyager        &   123\,K          & silicon   &  silicon ribbons \\
 \hline
\end{tabular}
\caption[Cryo IFOs]{Parameters of the cryogenic interferometers}
\label{tab:CryoTemps}
\end{table}

In addition to thermal noise improvement, there are a number of possible operational advantages due to a low a low temperature environment:
\begin{enumerate}
\item Increased thermal conductivity in crystalline substrates and at cryogenic temperatures; this dramatically reduces the thermal gradients in the mirror, and thereby, the induced wavefront distortions due to thermo-elastic deformations of the mirror surface and thermo-refractive lensing in the substrate bulk.
\item Zero thermal expansion in silicon at 18 and 123\,K~\cite{Touloukian_Brett6,Wiens:14}. Not only does this suppress thermo-elastic noise for stress-free structures, but it also reduces greatly the thermally induced thermo-elastic deformation of the mirror surface from any residual temperature gradients.
\item Cryo-pumping of the residual gas by the cold (colder than the mirrors) shields surrounding the mirror reduces the gas pressure and thereby the presence of squeeze-film damping~\cite{Cavalleri:09,Bao:07}.
\item As temperature approaches 0\,K, all thermally induced parameter fluctuations (e.g. thermo-elastic, thermo-refractive, etc.) tend to zero, as a consequence of the Nernst theorem.
\item Electronic noise of sensors and actuators used to control the mirrors can be reduced by low temperature operation.
\end{enumerate}


\section{Requirements}
A sketch of possible requirements for operating future detectors with their mirrors and the final stage of their suspensions at cryogenic temperatures are as follows. 
\begin{enumerate}
\item Maintain the target cryogenic temperature with sufficient accuracy:
      \begin{itemize}
        \item for silicon at 123\,K, and 18\,K, the fluctuations must be kept below 0.1\,K
        \item for sapphire at $\sim$20\,K, the requirement is less stringent since there is no null in the material properties to reach
        \item the cryocooling mechanism must be able to compensate for the power absorbed by the mirror as well as the laser power scattered by the mirror into wide angles. In a single interferometer design with typical \ac{3G} arm powers of around 3\,MW, and typical values of absorption (1\,ppm) and scattering (30\,ppm) this implies 3\,W of cooling for the test mass and 100\,W of cooling for the baffles. In ET's xylophone design, the injected power in the cold interferometer is very low and the cooling power is less critical. 
      \end{itemize}

\item The time required to cool from room temperature to cryogenic operation must be fast enough to avoid delays in operations. Currently, the cooling time in \ac{KAGRA} is about one month. For \ac{3G} detectors, the goal is to have the cooldown time be less than the vacuum pumpdown time, which is expected to be about one week.  


\item Steady-state cooling should be accomplished with minimal disturbance:
      \begin{itemize}
        \item the surrounding cold shields must be designed to have low backscatter (into the main cavity mode) and low vibration (relative to the mirror), such that the combined amplitude/phase noise does not degrade the sensitivity of \ac{3G} detectors
        \item in designs where the suspension fibers/ribbons are used for the steady-state cryocooling, the suspension thermal noise, see Chapter~\ref{sec:Suspensions_Isolation}, must not degrade the sensitivity of \ac{3G} detectors,
        \item auxiliary cold links must be designed so as to not produce seismic shortcuts to ground of the suspension system nor increase the overall suspension thermal noise;
        \item the required cryogenic machinery must be quiet or isolated enough to not increase the acoustic and gravity gradient noise.
      \end{itemize}
\end{enumerate}

%

\section{Required Facilities and Collaborations}

The successful operation of cryogenics in \ac{3G} instruments relies on continued \ac{RaD}
to meet the above requirements, while also demonstrating the practicality of integration in gravitational-wave detectors. Specifically, methods for reliable and fast heat extraction and ways to achieve the required temperature stability must be demonstrated. Cryogenic systems must be integrated with seismic isolation, suspensions, and vacuum systems.  
Other subsystems, such as auxiliary optics for wavefront control, must be modified to take into account the changed material and optical properties at cryogenic temperatures. Silent and vibration-free refrigeration machinery (e.g., using Pulse tube cryocoolers with symmetric cold heads) need to be designed and constructed (collaboratively with industry and the high energy physics community). Furthermore, the use of cryogenics in the electronics and magnetic actuators~\cite{cryo:OSEM} of gravitational-wave detectors has potential benefit and should be explored. In all of these endeavors, cooperation with industry and high-energy particle physics, where comparable requirements are addressed, could create valuable synergies. 


This \ac{RaD} will be carried out in tabletop experiments, operating detectors, upgrades to detectors, and prototype systems. \ac{KAGRA} will provide an outstanding demonstration of many of these considerations, for sapphire core optics at 20\,K. However, it will not have as stringent requirements as \ac{3G} detectors.
A number of facilities are planned or operational to test silicon core optics at 18\,K and 123\,K. 
Existing facilities at Stanford, Caltech and Gingin are currently being reconfigured for such tests. A new facility, ET Pathfinder, in Maastricht, the Netherlands, has been designed with an L-shaped vacuum system that will investigate both 123\,K and  18\,K, separately in each arm. Much of the technology required for cryogenic operation will also be tested in the near future in experiments exploring the material properties of mirrors, coatings, suspensions and in-vacuum materials. Tight coordination between all these groups is recommended to learn as much as possible about cryogenic operations to ensure rapid progress and to focus efforts on the most important open questions.

Ultimately, full system integration tests must be done in present multi-km facilities or large, sensitive prototypes to demonstrate sustained high performance in situations that are reliably scalable to \ac{3G} systems. New prototypes, such as ET Pathfinder, are very valuable, but are also large endeavors that require global coordination for the necessary resources and execution. Similarly, large investments in low temperature operation in 2G facilities, such as the \ac{Voyager} concept, would not only demonstrate the technical readiness, but might also serve to expand the astronomical scope of the global 2G \ac{GWD} network.


Cryogenic interferometers can be operated with a variety of temperatures, materials, and wavelengths, each with interdependence and implications for the facilities design. Current plans call for initial Cosmic Explorer, \ac{CE1}, to operate at room temperature with fused silica optics. Its major upgrade, \ac{CE2}, would operate at 123\,K with silicon, technologies that the Voyager concept would test and exploit. Einstein Telescope is designed to operate at 18\,K with silicon. However, there is flexibility in these plans to account for technological readiness and feasibility. Thus, planning must become more solid in the coming years to allow timely \ac{3G} operations. 

\section{Outlook and Recommendations}
We recommend that 
\begin{itemize}
\item \ac{3G} community develop a realistic roadmap for \ac{3G} cryogenic operation over the next 2-5 years to achieve the requirements mentioned above to test and improve the best candidate cooling technologies in terms of cooling power, vibration level, safety (with special regard to underground operation) and ready these technologies for implementation. 
\item  cooperation with industry and international laboratories where cryogenics are routinely used, such as \ac{CERN}, is strongly pursued. 
\item \ac{KAGRA} and new research facilities be used to learn about major issues with cryogenic suspensions  on a full-scale facility.
\end{itemize}

%% file: Core_Optics.tex
\chapterimage{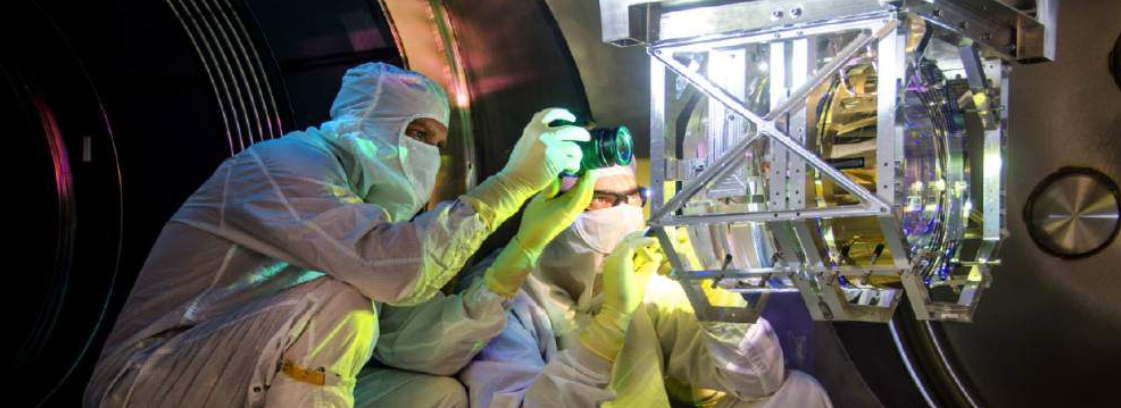} 
\chapter{Core Optics}
\label{sec:Core_optics}
\vspace{0.5cm}

The substrate materials that will be used for core optics in the \ac{3G}  detectors are interdependent on the operating temperatures of those detectors (Section~\ref{sec:Cryogenics}) and the laser wavelengths to be used (Section~\ref{sec:Light_sources}). 
Thermal noise plays a strongly limiting role in all current gravitational-wave detectors. At room temperature, fused silica, which is used in \ac{LIGO} and \ac{Virgo}, is an excellent substrate material due to its very low thermoelastic effect and ultra low optical~\cite{GEO_Absorption} and mechanical losses~\cite{Ageev_2004}. Many planned detectors
including \ac{KAGRA}~\cite{KAGRA2013}, \ac{Voyager}~\cite{VoyagerDCC2018}, the Einstein Telescope low-frequency detector (\ac{ET-LF})~\cite{ET2011} and the upgrade to Cosmic Explorer (CE2)~\cite{CosmicExplorer2017} will operate at cryogenic temperatures to reduce thermal noise. 
Fused silica is not a suitable mirror substrate material at cryogenic temperatures due to strongly increased low-temperature mechanical loss~\cite{Travasso_2007}.
In general, crystalline materials have much less Brownian noise at cryogenic temperatures than fused silica, due to their ordered lattice structure.
Sapphire, which is used in \ac{KAGRA} at 20\,K~\cite{Hirose_2014a, Hirose:2020}, and silicon, which is planned for use in \ac{Voyager}, \ac{ET}, and \ac{CE2}, have especially promising performance including low mechanical loss at cryogenic temperatures. 

\ac{3G}  detector core optics must also meet a number of other challenging requirements. 
Uniform and pure masses of several hundred kg (see Table~\ref{Tab:FutIfos}), significantly more than currently in use, are required to reduce radiation pressure noise and accommodate larger diameter laser beams in order to better reduce (through averaging, see figure~\ref{fig:Thermal_Noise}) thermal noise. Optical absorption and scatter must both be low. The \ac{ET}  Design Study~\cite{ET2011} specifies a scatter loss of 37.5\,ppm per mirror surface. We note that there is excess scatter loss in the arm cavities of the advanced detectors, and further understanding of this will be important to reach the \ac{3G}  scatter requirements. Finally, requirements for test mass cooling for Voyager and \ac{ET-LF}   lead to requirements for the optical absorption of the input test mass substrates to be less than $\sim$10\,ppm/cm.

The major open questions for core optics materials that must be addressed by \ac{3G}  \ac{RaD}   are described below with a focus on fused silica, silicon and sapphire. More details are given in Appendix~\ref{sec:Appendix_Core_optics}.

\section{Fused Silica} 
For fused silica, homogeneity of the refractive index is an important consideration. For the substrates of the cavity mirrors, excellent homogeneity is only important in the two dimensions perpendicular to the beam axis. This can be achieved even for large volumes (corresponding to a total mass of several hundred kilograms). However, the beam splitter requires a very high homogeneity in all three dimensions and this can currently only be guaranteed by the manufacturers for masses up to 40 kg (diameter 55 cm, thickness 7 cm). The company Heraeus has planned some tests to push this limit to about 100 kg. Whether such large beam splitters are required depends on the optical layout of the detector and is under investigation. Other issues with fused silica include charging and relatively low Young's modulus. Significant progress has been made on aluminium doped zinc oxide and gallium doped zinc oxide coatings, which are close to combining the required electrical conductivity and optical absorption to mitigate charge-related noise associated with silica test masses. Further optimisation, including studies on aging and scatter, are ongoing.

\section{Silicon}
Silicon has a low mechanical loss at cryogenic temperatures (similar below 10 K to fused silica's at room temperature).
In addition, the thermal expansion coefficient of silicon is zero at $\sim$123\,K and 18\,K which allows, with temperature control, the suppression of substrate thermoelastic noise and thermal expansion effects due to absorbed laser power.

Silicon is opaque at the currently used wavelength of 1064\,nm. Initially, the telecommunications wavelength of 1550\,nm was proposed for use with silicon mirrors, due to wide availability of high-powered lasers and optical components. More recently, there has been growing interest in using a wavelength close to 2000\,nm. A major driver towards 2000\,nm is the development of amorphous silicon as a possible low thermal noise cryogenic coating material. Amorphous silicon exhibits significantly lower absorption (a factor of $\sim$7) at 2000\,nm than at 1550\,nm~\cite{Steinlechner_2018_aSi}. It seems likely, therefore, that the choice of mirror coatings will be a major factor in the choice of wavelength for future detectors. 

Sufficiently low optical absorption can be obtained from silicon refined with the Float Zone technique, however, the maximum diameter for this method is $\sim$200\,mm. This is too small for the requirements of future detectors (e.g. \ac{ET-LF} requires 450\,mm diameter, 550\,mm thick optics and \ac{CE2} up to 700\,mm diameter optics). While larger diameter silicon pieces can be produced using the Czochralski method, the optical absorption of this type of silicon is too high, due to impurities related to the production method. A \ac{MCz} exists, in which a magnetic field is used to reduce the impurity concentration in the centre of the ingot. This process can produce diameters of up to 450 mm, and a production line for manufacturing silicon of this diameter does exist at the company Shin Etsu, but is currently not operational. Initial studies of the optical absorption have shown low values at room temperature of $\sim$3\,ppm/cm at 1550\,nm and $\sim$5\,ppm/cm at 2000\,nm. The measurements showed an increase towards lower temperatures, reaching approximately 10\,ppm/cm at 50\,K, meeting the requirement for cryogenic silicon mirrors of the \ac{ET}  design study. However, initial studies indicate that the absorption of this material can vary significantly, both along the radius and along the length of an ingot, and more studies of the homogeneity of the absorption and its dependence on the thermal history of the sample are required. Composite test masses built from silicon segments have been considered, however the optical losses and possible excess mechanical dissipation associated with the joints (\textit{e.g.} silicate or direct bonding) are expected to be problematic.

Polishing silicon surfaces can increase their optical absorption~\cite{SiliconSurfaceAbsorpBell2017}, and it has been shown that a proprietary polishing process can be used which does not produce this effect. 
While this has been consistently demonstrated, further work is required to test whether a silicon surface can be polished to the specifications required for a \ac{GW}   detector without resulting in surface absorption.

Non-linear absorption in silicon is not expected to set a major limit to performance, contributing <\,0.5\,ppm/cm absorption for a wavelength of 2\,$\mu$m at the light intensity assumed for inside the \ac{Voyager} \ac{ITM}. At the significantly lower intensity within an \ac{ET-LF}   \ac{ITM}, these effects are even less significant. 
Two-photon absorption generates free carriers in silicon: the absorption due to these free carriers depends crucially on the carrier life time. 
It will be important to test the optical scattering from \ac{MCz}   silicon, particularly as the \ac{MCz}   growth process is known to produce a high void content in the material. 
Initial scattering estimates
suggest that the scattering is higher than in fused silica, but is likely to be within the required limits~\cite{SiliconScatter2017}.

\section{Sapphire}
Sapphire is transparent at 1064\,nm and hence does not require changing the currently used laser wavelength. It has low mechanical loss at room temperature~\cite{Rowan_2000a} and even lower loss at cryogenic temperatures~\cite{uchiyama1999mechanical}. 
Sapphire's elastic constants are about 3 times higher than silicon's, helping to reduce thermal noise (\ref{fig:Thermal_Noise}). The high Young's modulus has two additional advantages: fewer parametric instabilities~\cite{Yamamoto:2008} and higher internal resonance frequencies. The thermal conductivity of sapphire increases with decreasing temperature and reaches a peak of several $10^3$\,W/(m \,K) around 20-40K. Thermoelastic noise, which is high at room temperature, is quite low at cryogenic temperatures due to this increase in thermal conductivity. The high conductivity (and the low temperature coefficient) make the thermal lens effect negligible~\cite{Tomaru:2002}. 
The optical absorption of sapphire has been found to vary strongly from crystal to crystal and for crystals from different suppliers. 
For \ac{KAGRA}, it was necessary to develop sapphire crystals with low absorption ($<$50~ppm/cm) by working closely with the crystal manufacturers. 
Theoretical work on scattering in sapphire sets a lower limit of 0.21\,ppm/cm, with higher measured values of around 13 ppm/cm being attributed to impurities and vacancies.
Sapphire's high Mohs hardness of 9 and its crystalline structure, resulting in orientation dependent machinability, makes it harder to process sapphire substrates. 
Finally, sapphire is birefringent, though \ac{KAGRA} has taken steps to address this including alignment of the c-axis with the beam axis, making the crystal isotropic, and locally adjusting the substrate thickness with ion-beam figuring.  

\section{Outlook and Recommendations} 
The choice of core optic materials is strongly dependent on wavelength, temperature, and coatings for \ac{3G}  detectors and has a strong impact on most other subsystems. In addition, much of the \ac{RaD}   described here has long lead times and high cost. Steering the direction needs decision points well ahead of time. Decisions on core optics, based on the community's ongoing \ac{RaD}, must be made a decade before \ac{3G}  construction, i.e., in the coming six years. 

We recommend that
\begin{itemize}
\item Laboratories dedicated to the development and optimization of large-size crystal growth be identified or established.
This effort has already started: For crystalline sapphire a facility will be built at the Universit\'e de Lyon aiming to develop 450 mm diameter, 200 mm thick substrates of GW quality.
\item International consortia should be formed to work on core optics challenges together with industry
partners, in a globally coordinated way.  This is an area where there could be large investments needed with industry, and strong international collaboration could ease the burden and be much more cost effective.
\item For fused silica, continued work with \acf*{Heraeus} to ensure that the homogeneity of the refractive index will be sufficient for future detectors, and in parallel to explore the use of telescopes to use small beam-splitters but still have large beams in the arms. 
\item For silicon, the community should continue to study scatter, absorption, and absorption uniformity, at the temperatures and wavelengths of interest for \ac{3G} . 
\item Sources of phase noise in silicon optics, thermo-refractive noise and carrier density noise, should be studied experimentally  to ensure that they do not set unexpected limits of silicon mirror performance. 
\item For sapphire, as much as possible should be learned from the \ac{KAGRA} experience, and efforts to reduce absorption in large sapphire crystals should continue. 
\end{itemize}

%% file: coatings.tex
\chapterimage{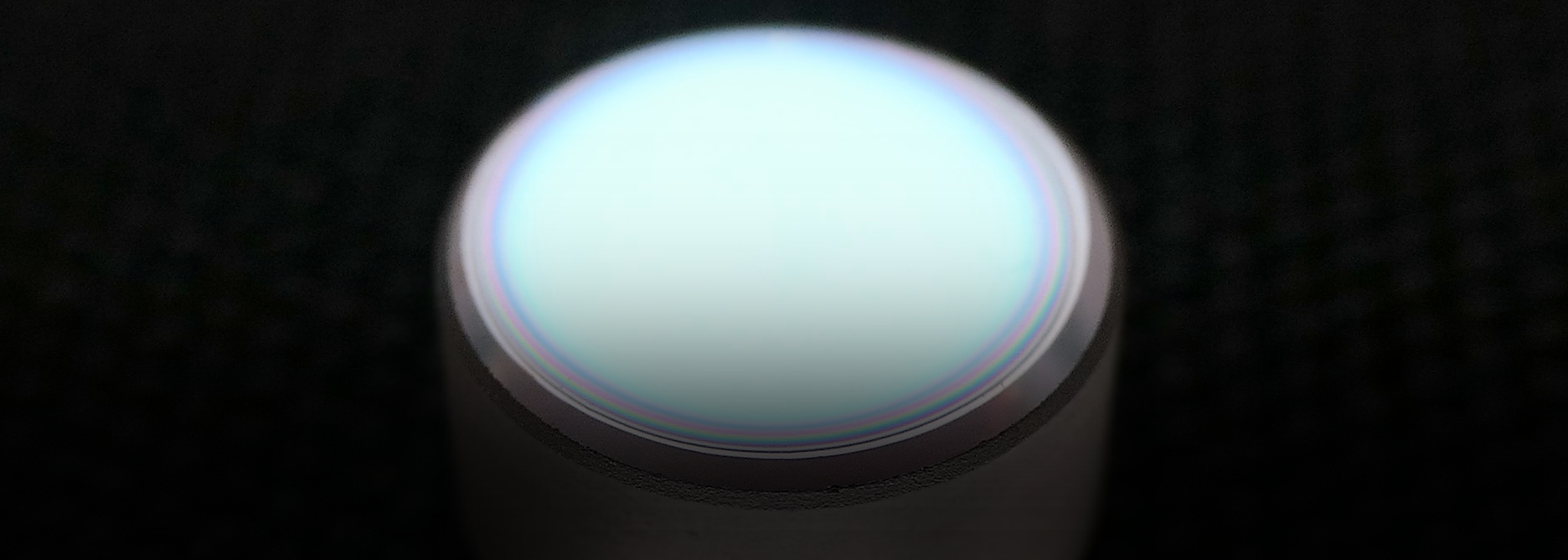} 
\chapter{Coatings}
\label{sec:Coatings}

\vspace{1cm}



Coating thermal noise limits the mid-band design sensitivity of current (\ac{aLIGO}, \ac{AdVirgo})~\cite{ AdvancedLIGO2015,AdvancedVirgo2015}, future enhanced versions of those detectors (\ac{a+LIGO}, \ac{AdVirgo+}) \cite{Zucker:LIGOAplus, Cagnoli:VirgoAplus}, \ac{2.5G} detectors (\ac{Voyager})~\cite{VoyagerDCC2018}, and \ac{3G} detectors (\ac{ET}, \ac{CE})~\cite{ET2011,CosmicExplorer2017}.  While \ac{a+LIGO} and \ac{AdVirgo+} will operate at room temperature with 1.06\,$\mu$m lasers, the designs of the \ac{2.5G} and \ac{3G} detectors are still evolving, so therefore are the requirements on the mirror coatings. In the following discussion we assume the parameters for future detectors as set out in Table.\,\ref{Tab:FutIfos}. 



Thermal noise is fundamentally connected to the mechanical and optical properties of the mirror coatings through the fluctuation-dissipation theorem (see Box~\ref{Box:Thermal}). For coatings, the thermal noise is proportional to the square root of the operating temperature \ac{T} of the mirrors, the \ac{phi}  of the coating at that temperature, the \ac{w}, and the \ac{d} ~\cite{levin1998internal}. In terms of strain as measured by the interferometer the coating thermal noise is also inversely proportional to the  \ac{L}, such that,
\begin{equation}
\text{CTN} \propto \frac{1}{L} \, \sqrt{\frac{T}{f} \frac{1}{w^2} \, \varphi \, d   }.
\end{equation}\label{eq:CTN}
where $f$ is the frequency. Each future detector has a specific set of parameters \ac{T}, \ac{w}, and \ac{L} different from the others, hence the requirements on coating material parameters, \ac{phi} and \ac{d}, vary significantly from one detector to the other. For the advanced-plus detectors a 4-fold reduction in mechanical loss is targeted.


Thermal noise is not the only requirement that must be met. The total optical absorption for room temperature interferometers (\ac{A+} and \ac{3G}) must be less than 0.5\,ppm, although some relaxation of this target is possible if required by low thermal noise coatings at the expense of making thermal management more challenging. Recently small defects have been recognized as affecting the performance of current detectors; scattering centres with sizes from tens of nanometers to several microns, and points with high absorption are clearly visible on the mirrors. Coating \ac{RaD}   has to take into account these optical properties as well as mechanical losses in order to meet the stringent requirements comparable to those for current detectors~\cite{AdvancedLIGO2015,AdvancedVirgo2015}. Advancements on the coating deposition technology are also necessary for new detectors, especially when considering the need for larger mirror diameters.

The situation in \ac{ET-LF}    is somewhat different, in that the thermal noise is dominated by the mirror suspension, which must be able to extract the thermal load imposed by the optical power absorbed in the mirror; in this case the optical absorption plays a key role in the thermal noise performance, and must be held in the range of 1\,ppm \cite{HiEA2011}.


Mechanical loss in amorphous materials results from the coupling of elastic energy into low energy excitations of the materials, generally thought of as \ac{TLS}~\cite{braginsky1985systems,bommel1956dislocations}. These \ac{TLS} typically involve motions of several dozen atoms, and different \ac{TLS} are responsible for losses at different temperatures (low barrier heights at low temperatures, higher barriers at higher temperatures)~\cite{hamdan2014molecular,trinastic2016molecular}. Reducing the mechanical loss thus requires reducing the density of two-level systems with the barrier heights pertinent to the operating temperature. 


\section{Current Approaches to Low Mechanical Loss Mirror Coatings}

The sensitivity limit due to coating thermal noise is explained through equation\,\ref{eq:CTN} and in Appendix~\ref{sec:Appendix_Coatings} where the relevant parameters are presented. Among those is mechanical loss, which will be the subject of this section. The density and distribution of the \ac{TLS} responsible for mechanical loss depend on the coating materials, and they can be altered through deposition conditions and post-deposition treatments.

\noindent Promising approaches for reducing mechanical losses of optical coatings are listed below and described in more detail in Appendix~\ref{sec:Appendix_Coatings} and in\,\cite{DawnIV2018}.
\begin{itemize}
    \item\textbf{Improved conventional amorphous oxides:} Increased annealing temperature by suppressing crystallization through mixing.
    \item\textbf{Alternative amorphous materials:} Semiconductors such as amorphous silicon or silicon nitride are being actively considered by several groups.
    \item\textbf{Multi-materials:} Amorphous semiconductor materials currently have adequate mechanical properties but excess optical absorption; Consider coatings with amorphous oxide layers where the optical intensity is highest and amorphous semiconductor layers below.
    \item\textbf{Crystalline coatings:} Alternating layers of \ac{AlGaAs/GaAs} have shown favorable mechanical loss properties; \ac{GaP/AlGaP} should also be explored.
\end{itemize}
Deposition parameters that are worth exploring are: 
\begin{itemize}
    \item\textbf{High temperature deposition:} producing ultrastable glasses where \ac{TLS} are significantly reduced.
    \item\textbf{Nanolayering:} nm-thick layers are able to frustrate crystallization and to modify the \ac{TLS} distribution.
    \item\textbf{Ion energy and deposition rates:} the energy of ions and the deposition rates are the parameters that seem to impact the mechanical losses the most. A model of the physics of deposition is important in order to help clarify the relation between the optical and mechanical parameters of coatings with that of the physical condition of deposition. This is a challenging problem and significant progress is being made by the collaboration.
\end{itemize}
Finally, \textbf{post deposition annealing} is able to change the \ac{TLS} distribution and therefore mechanical loss. The dynamics of the annealing process is poorly understood and should be explored further.
These investigations will benefit from synergy between 1) synthesis of samples, 2) microscopic and macroscopic characterization, and 3) modelling of deposition of the amorphous materials, and loss calculations.

\section{Current Research Programs}

There are several 
programs devoted to developing low-thermal-noise mirror coatings suitable for enhanced \ac{2G}, \ac{2.5G}, and \ac{3G} detectors. Participants are involved in all aspects of coating research, including various deposition methods, characterization of macroscopic properties at room and cryogenic temperatures, and atomic structure modeling and characterization. The recent incorporation 
of several groups involved in coating deposition is particularly important, as the costs and time delays associated with commercial deposition of research coatings have been significant impediments to rapid progress.

There are about ten U.S. university research groups participating in coating research. In 2017, a more coordinated effort and additional funding for these groups were initiated under the \ac{CCR}, jointly funded by the Gordon and Betty Moore Foundation and the NSF, which was extended in 2020 for another 3 years. Work in the U.S. also importantly includes that in the \ac{LIGO} Laboratory. These efforts are complemented by groups not formally affiliated with the \ac{CCR}, notably U. Montreal and U. Laval. The GEO collaboration (GEO) has a major coatings effort within the U.K. and Germany, within five institutions, and we note that the U. Glasgow and U. Strathclyde are establishing a new coatings center aligned to 3G technologies. The efforts of all these groups are coordinated through the \ac{LSC} Optics Working Group.

About ten universities are involved in the Virgo Coatings \ac{RaD}   (\ac{RaD}  ) Project that has its research plan focused on new materials, deposition conditions, post-deposition treatments and metrology. The \acs*{ViSIONs} project, supporting six French laboratories, including \ac{LMA} and \ac{ILM}, is focused on studying the relationship between the physical properties of sputtered or evaporated materials and the structural and macroscopic properties of the deposited films. 
\ac{LMA} is improving the uniformity of their coating deposition to meet the challenges of the Advanced+ detectors. They have also developed plans to upgrade their coaters and tools to accommodate the increased size and weight of the 55\,cm end mirrors being considered for \ac{AdVirgo+}.

\section{Timelines}

The most pressing timeline is for the enhanced \ac{2G} detectors.The necessary research to identify a coating material and process should be completed by May 2021 for \ac{a+LIGO} and by the end of 2021 for \ac{AdVirgo+}. This timeline implicitly assumes that the coating will be a sputtered amorphous oxide or silicon nitride, possibly deposited at elevated temperature, possibly at a lower rate and/or with a higher annealing temperature than conventionally used. It is unlikely that there will be time during this period to identify coatings and develop the necessary equipment for a process with significantly different deposition processes than the current ones.


The efforts of the \ac{LVC} are currently focused on meeting the coating thermal noise requirements of the enhanced \ac{2G} detectors, which requires conducting basic research on a development timeline. It is recognized and recommended by the community that the development of coatings for future detectors should not be subject to such a constraining timeline. Therefore, a portion of the current research effort is devoted to establishing approaches to mirrors for \ac{2.5G} and \ac{3G} detectors. At the moment, it is difficult to set research timelines, since the funding and construction schedules for these detectors are not yet established.

The deposition process required for \ac{2.5G} and \ac{3G} mirrors remains an open question. The further that process deviates from the currently used \ac{IBS} technology, the longer it will take to develop. In any plausible scenario, at least 5 years are available for research into finding a viable \ac{2.5G} coating. The results for these \ac{2.5G} mirrors will then inform available choices for \ac{3G} mirrors. It is therefore too soon to argue for a large investment in scaling deposition tools alternative to \ac{IBS} for \ac{2.5G} and \ac{3G} mirrors.

As plans for future interferometers mature and funding is secured, it will be important to regularly re-evaluate potential mirror technologies that meet specifications, in order to focus research and development efforts. This is especially important for enabling critical decisions on down-selecting coating materials or technology in a timely manner, especially with respect to developing deposition tools that may require long development times and major financial investment.

\section{Outlook and Recommendations}
\label{coatings_Recomm}
We recommend
\begin{itemize}
\item Continued and deeper coordination among all gravitational-wave groups to maximize the possibility of developing new coatings with the requisite \ac{3G} performance in the limited time available;
\item  parallel research lines be developed, making a division of tasks among the research groups necessary;
\item support for several producers of large high quality coatings and to work with them on ongoing research efforts as an essential risk mitigation for the \ac{3G} effort.
\end{itemize}


In 2017, the U.S. efforts in coatings research received a significant funding increase enabling the creation of the \ac{CCR}.  It is important that this level of funding at least be maintained in order to carry out the required research and development for mirror coatings for future interferometers. It is also important that the \ac{LIGO Lab} continue with at least its current efforts, as their contribution to high-throughput mechanical loss characterization, optical scatter and homogeneity measurements, and their overall coordination of sample fabrication, distribution, and characterization is critical to the success of coatings research. Currently, adequate capacity exists for characterization of properties (other than cryogenic mechanical loss); an increase in modeling and synthesis capabilities would enhance our ability to inform research directions and develop new coating materials.

In the \ac{GEO}, there is growing capacity for depositing coatings at Strathclyde, \ac{UWS} and Hamburg. These coatings can be produced at a rate faster than it is possible to characterize their properties at cryogenic temperatures. U. Strathclyde, U. Glasgow and \ac{UWS} have established a variety of IBS systems and are in the process of establishing a new coatings center aligned to 3G detector requirements. Studies of \ac{GaP/AlGaP} crystalline coatings are underway using hardware now installed in an \ac{MBE} chamber at Gas Sensing Solutions Ltd. This is an important parallel research direction for the development of crystalline coatings.

\ac{VCRaD} is responsible for a significant research activity that is complementary to that carried out in the LSC and its planning has to be supported. Production of samples is done in four labs, modelling in two and characterizations at different scale is widely distributed in all the collaboration. The mechanical loss measurement setups known as \ac{GeNS} is one example of fertilization of \ac{VCRaD} to the \ac{LSC} community.

With the imbalance of coating production and coating characterisation capacities in the various collaborations it is prudent to join forces in a globally coordinated way. 

\ac{LMA} is presently the only institution, capable of depositing coatings with the size and quality required by gravitational-wave detectors. \ac{LMA} will continue to be supported as a research and coating facility for future gravitational-wave detectors by the French \ac{CNRS} through EGO. Activities at \ac{CSIRO} in Australia, which was able to provide similar coating quality, had been discontinued but are now being transferred to and revived at \ac{ANU}. Supporting several producers of large high quality coatings and to work with them on ongoing research efforts is an essential risk mitigation for the \ac{3G} effort.

\subsection{Roadmap}
Since the development of low-thermal-noise coatings is in the stage of research rather than development, even for \ac{2.5G} detectors, a conventional roadmap is not the best model for describing the path towards identifying suitable mirror designs and fabricating corresponding full-scale mirrors for \ac{3G} detectors. Considering that the architectures and operating parameters for \ac{3G} detectors remain in flux, that the results for mirrors developed for \ac{2.5G} detectors will have a strong, perhaps decisive, influence on the designs for \ac{3G} mirrors, and that the funding and therefore construction schedules for \ac{3G} detectors are not yet clear, it is best to estimate schedule implications in terms of time requirements before installation. 

The currently plausible approaches to \ac{3G} mirrors fall into three broad categories which have different cost and schedule drivers: amorphous coatings deposited by methods similar to conventional IBS, amorphous coatings deposited by alternative means, e.g. chemical-vapor deposition (CVD), and crystalline coatings. 

\textbf{Amorphous coatings deposited by IBS methods:} 
Currently, IBS is the only mature technology for deposition of mirror coatings suitable for GW interferometers (GWI); the challenge is to find the appropriate combination of material, deposition conditions (rate, ion energy, substrate temperature, etc.) and post-deposition treatments to meet mechanical loss requirements and optical specifications. The time required for this step is unknown; a solution may be found next month or may not exist at all. Once the materials and conditions have been determined, the time to production readiness will depend on the differences in current IBS practice. One year would be sufficient for room temperature deposition at conventional rates. For more extreme conditions with increased substrate temperature, low rate, microwave annealing, etc., perhaps 3-5 years and USD 3-5M would be needed to develop appropriate equipment and processes. Multi-material coatings would fall between these extremes of time and equipment costs. 

\textbf{Amorphous coatings deposited by methods other than IBS:} 
if research shows that the optimal deposition method is other than IBS, e.g. CVD of a-Si/SiNx mirrors, in addition to open-ended research time (similar to the IBS case) sufficient time would be required to develop deposition tools and a scaled-up process suited to GWI requirements. While CVD tools are widely used in the semiconductor industry, adaptation to GWI mirror requirements would be a significant effort. Instrumentation development and the more complex pathfinder process to production readiness might take USD 25-30M and perhaps 10 years. These figures are order of magnitude estimates that can be tightened up by discussions with equipment vendors. 

\textbf{Crystalline coatings:} for AlGaAs crystalline mirrors, the materials research and modeling phase for small scale (ca. 15\,cm) coatings could be completed in perhaps three years. If these results showed high performance, the time and financial costs for substrate and tool development and the more complex pathfinder process to scale to 45\,cm optics might take about USD 25M and 5 years. 

 

%% file: Light_Sources.tex
\chapterimage{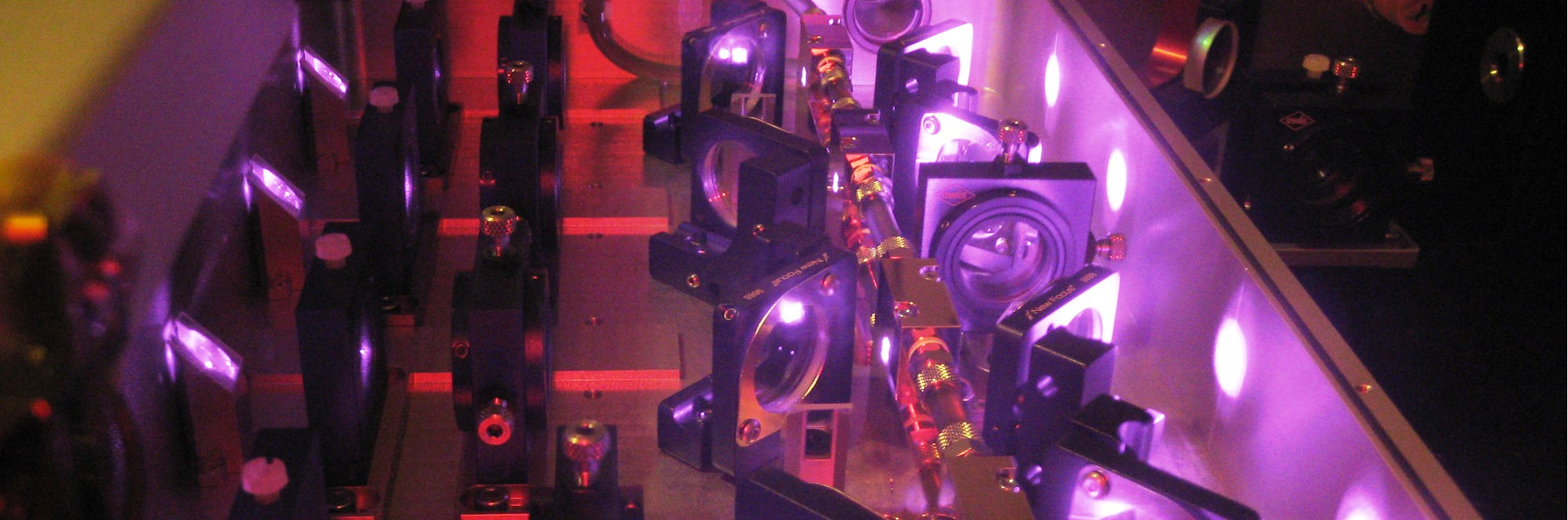} 

\chapter{Light Sources}
\label{sec:Light_sources}

This section describes \ac{3G}  light sources, including the pre-stabilized high power lasers (\acsp{PSL}) and the squeezed light sources. 

\section{Current State of the Art}
All currently operating advanced gravitational-wave detectors were designed to operate with a 200\,W class \ac{PSL}. All \acp{PSL} were independently developed following different approaches for the required \acp{HPL}. A $ 200\,{\rm W} $ injection locked high power oscillator was installed at \ac{LIGO}~\cite{Kwee:12}, a fiber based  \ac{MOPA} design was chosen for \ac{Virgo} and \ac{KAGRA} built on a \ac{MOPA}   design with fiber and solid-state amplifiers. Due to an unexpectedly high pointing jitter noise coupling, \ac{aLIGO} is currently being operated with 70\,W \ac{MOPA} systems using 
commercial solid-state amplifier \acs{neoVAN4S}. A similar system with 100\,W power (neoVAN 4S-HP) was installed in 
\ac{AdVirgo}
as the tested 200\,W fiber \ac{MOPA}   solutions did not work reliably. 
KAGRA is currently operated with a 40\,W \acs{Nufern} fiber amplifier. Commercial seed lasers (Mephisto series, Coherent) are being used in all \ac{PSL} versions. The high power stages are either built by industry or research labs. 

\ac{RaD} is globally underway towards a stable 200\,W light source with sufficiently low power noise, frequency noise and beam pointing jitter. 
The ongoing \ac{RaD} in Nice, at \ac{MIT} and in Hannover is devoted to reliability tests of 200\,W fiber amplifiers. A coherent combination of two 100\,W class solid-state amplifiers is being investigated in Hannover and a combination of two 40\,W class fiber amplifiers followed by a high power solid-state amplifier is being tested in Japan. All of these high power lasers operate at a wavelength of 1064\,nm and are of relevance for the \ac{3G}  laser development at this wavelength. At $ 1.5\, {\rm \mu m}$ and in the $ 2\, {\rm \mu m}$ region, high power laser \ac{RaD} toward developing concepts for \ac{3G}   detectors is being performed in Adelaide, \acs{IIT} Madras, Hamburg and Hannover. \acp{HPL} with approximately 100\,W output power have or will soon be demonstrated by these groups. 

Other \ac{HPL} developments that were not specifically tailored for gravitational-wave applications can be found in the literature. These designs often fall short of one or more of the stringent requirements for \ac{3G}   high power light sources. To the best of our knowledge, there is no commercially available \ac{HPL} with the specifications required for \ac{3G}. A market survey should be carried out to substantiate this statement.

Non-classical light sources at 1064\,nm generating up to 15\,dB of squeezed vacuum have been developed for the currently operating advanced \ac{GWD}. These sources have reached maturity, as shown by GEO600's several years of operation with squeezed light and the use of squeezing in Advanced LIGO and Advanced Virgo. Squeezed light sources at $ 1.5\, {\rm \mu m}$ have reached similar squeezing levels but no system design efforts were undertaken yet to transfer the laboratory systems to prototypes. At $ 2\, {\rm \mu m}$ a few dB of squeezing were demonstrated recently in a laboratory experiment at ANU.

%
%

\section{Requirements and Current/Planned R\&D}



A survey within the gravitational-wave community showed that more than 10 groups currently perform \ac{RaD} on light sources for \ac{3G}   detectors. All relevant research topics are being worked on by at least two groups. A snapshot of which group is working on which topic can be found at~\cite{LightSource_RD_table}. We also analyzed available documentation and presentations on \ac{3G}   detectors to extract requirements for the high power lasers and squeezers.

The requirements for the \ac{PSL} and the squeezed light sources for 3rd generation detectors are not yet well defined. 
The Einstein Telescope
is planned around a 700\,W laser at 1064\,nm 
and a 5\,W laser at a wavelength of 1550\,nm both in the fundamental Gaussian mode.
The current Voyager design plans for a \ac{HPL} with a power of 200\,W and a wavelength of 1550\,nm or longer to minimise absorption in the Silicon test masses. The second phase of Cosmic Explorer~\cite{CosmicExplorer2017}, foreseen to operate at a similar wavelength, will require a \ac{HPL} with much higher power. This power level is currently not well defined but may be as high as 1\,kW~\cite{GWADW2018,ISWP:2018}. The wavelength choice depends on several factors, such as the availability of high power lasers, absorption of the substrates and the high reflective coatings of the test masses, scattering and the availability of photo detectors with high quantum efficiency. As information on several of these factors is missing, a final wavelength choice can not yet be made. Currently wavelengths of 1550\,nm and around $\rm 2\, \mu m $ and $\rm 2.1\, \mu m $ are targets due to promising high power laser concepts for these wavelengths. Hence \ac{RaD} on \ac{HPL} for three different wavelengths (1064\,nm, 1550\,nm and around $\rm 2\, \mu m $) has to be performed until the final wavelengths are selected. 
No information on \acp{PSL} stability requirements for \ac{3G}   detectors exists at present. As a 10 times better sensitivity is aimed for we assume that the power, frequency and beam pointing jitter stability has to be a factor of 10 higher than in advanced detector \acp{PSL}. Concerning the spatial and polarization purity we expect similar requirements as for the advanced detectors.
All \ac{3G}   detector designs currently incorporate 10\,dB of detected squeezing, meaning that squeezed vacuum sources with squeezing levels of $> 15$\,dB are required.

Thus, in this report we assume that \emph{initially} 250\,W at 1064\,nm and 500\,W at  $ 1.5\, {\rm \mu m}$ or in the $ 2\, {\rm \mu m}$ region will be required. We expect that a factor of 10 less noise compared to \ac{2G}  \acp{PSL} and similar spatial and polarization purity as in \ac{2G}  \acp{PSL} will be needed. Furthermore, we assume that a squeezing level of 15\,dB will be sufficient for \emph{initial} \ac{3G}   detector operation. For the \emph{final} \ac{3G}   detectors, we expect that  700\,W at 1064\,nm and 1\,kW at  $ 1.5\, {\rm \mu m}$ or in the $ 2\, {\rm \mu m}$ region again with similar stability and beam purities as in the initial \ac{3G}   phase will be required. A \emph{final} squeezing level goal of 20\,dB is assumed.

%

\section{Pathways and Required Facilities} \label{sec:pathway}
The pathway towards adequate \acp{PSL} for \ac{3G}   detectors has several steps:
\begin{enumerate}
	\item Demonstration of reliable high power generation with required power level, low enough free-running noise (defined by stabilization constraints) and acceptable spatial and polarization purity (functional prototype)
	\item Design, fabrication and test of a \ac{HPL} according to reproducible fabrication steps with the required diagnostic and stabilization actuators. Demonstration of long-term stable operation and conceptual demonstration of the stabilization concept (engineering prototype).
	\item Final design steps as part of a \ac{3G}   project and reliability test of the stabilization concept.
\end{enumerate}

\noindent Item 1 is part of generic laser research and is typically performed by university or laser research laboratories. We expect, that up to a power level of $ \approx 200 \, {\rm W} $ this research will be done in the laboratories listed in \cite{LightSource_RD_table}, but new groups may also join the effort. Together with a coherent combination step this work should be sufficient for the 1064\,nm \ac{3G}   \ac{HPL} development.
A new coordinated \ac{RaD} effort is required to develop a 1\,kW class laser with a wavelength longer than or equal to $ 1.5\, {\rm \mu m}$.

The reliability and reproducibility part of the engineering step needs a dedicated program of a large laser research lab or industrial involvement. The scope of this step is normally not included in the programs of research funding agencies, so a dedicated \ac{RaD} funding program will be required to cover the costs.
Furthermore specific infrastructure and trained staff is required for the fabrication part of the engineering prototype phase. The assembly and stabilization part can be done in one of the laboratories of the \ac{GWD} community (see \cite{LightSource_RD_table}) or in newly joining laser labs. The same holds true for one or several coherent combination steps. Depending on the progress of the wavelength decision, the engineering prototype step needs to be conducted for two or three wavelengths. During this phase several identical \acp{HPL} should be built and characterized in a long term test.

The final design step is part of each specific \ac{3G}   project and should be performed at an early stage of the respective project with project funding.

The pathway towards an adequate squeezed light source will most likely involve university and research lab based \ac{RaD}. The required funding is on a scale that can be covered by regular research grants and the development and improvement of non-classical light sources falls in standard calls of funding agencies. Squeezing levels of $ \approx 15 \, {\rm dB} $ have already been achieved for  $ 1\, {\rm \mu m}$ and  $ 1.5\, {\rm \mu m}$ such that the main technical challenges for these wavelengths are the reduction of loss in optical components and improving the stability and controllability of the squeezing phase. The squeezing research at $ 2\, {\rm \mu m}$ is far less advanced and substantial effort has to be put into the generation of high squeezing levels and low loss components. Several parallel efforts should continue for each wavelength as this approach allows to compare different technical solutions and chose the most appropriate for each project when project funding arrives.


\section{Outlook and Recommendations}
As in the past a strong collaboration between a group within the gravitational-wave community and a laser research lab or industry is required (such as \ac{AEI}/\ac{LZH}, \ac{Artemis}/\ac{Alphanov}, \ac{ICRR}/\ac{Mitsubishi}) to design and build suitable \acp{HPL} at the different wavelengths. As the different wavelengths need different solutions, a loose collaboration between the respective wavelength groups would be sufficient.  No particular collaboration arrangements are required for the squeezing research. The normal exchange of concepts and results at collaboration meetings and conferences seems sufficient. 

We recommend that
\begin{itemize}
\item for high power laser development at least two collaborations work on laboratory prototype solutions for each wavelength to explore different concepts and alternative technical solutions.  These groups should have a strong connection with regular meetings to exchange results and ideas. This approach would possibly avoid a single supplier problem.
\item for squeezing light sources, the existing development effort should be maintained, supported by the development of high efficiency photodetectors and required auxiliary optics for two micron wavelength.  
\end{itemize}

Given the long experience in laser development in the \ac{GW} community, we can lay down a fairly precise roadmap: 

\subsection*{\ac{HPL} 1064\,nm}
\begin{itemize}
	\item 2020 - 2021 : continue development and reliability studies of \ac{2G}  \ac{PSL} systems at the 250\, W level and perform coherent combination demonstration experiments at high powers
	\item 2021 - 2024 : engineering prototype (see section \ref{sec:pathway}) 500\,W \ac{HPL} and conceptual test of stabilization and spatial filter solutions
	\item 2024 - ... : final design and fabrication of \ac{HPL} with the initial \ac{3G}   requirements within specific \ac{3G}   projects, in parallel \ac{RaD} on path toward the final \ac{3G}   requirements 
\end{itemize}

\subsection*{\ac{HPL} ${\bf 1.5 - 2.1 \, {\bf \mu m}}$}
\begin{itemize}
	\item 2020 - 2021 : identify concepts for a 1\,kW \ac{HPL} that fulfills the stringent \ac{3G}   detector \ac{HPL} requirements (most likely several coherently combined stages)
	\item 2022 - 2024 : 1\,kW \ac{HPL} functional prototype phase (see section \ref{sec:pathway})
	\item 2024 - 2028 : 1\,kW \ac{HPL} engineering prototype phase (see section \ref{sec:pathway})
	\item 2028 - ... : final design and fabrication of \ac{HPL} with the initial \ac{3G}   requirements within specific \ac{3G}   projects, in parallel \ac{RaD} on path toward the final \ac{3G}   requirements
\end{itemize}

\subsection*{squeezed light sources for \ac{3G}   \ac{GWD}  }
\begin{itemize}
	\item 2020 - 2026 : continue laboratory based \ac{RaD} on squeezed light sources at all wavelength, potentially involve industrial partners to design and fabricate low loss optical components
	\item 2026 - ... : final design and fabrication within specific \ac{3G}   \ac{GWD} project
\end{itemize}

%% file: Quantum_Enhancements.tex
\chapterimage{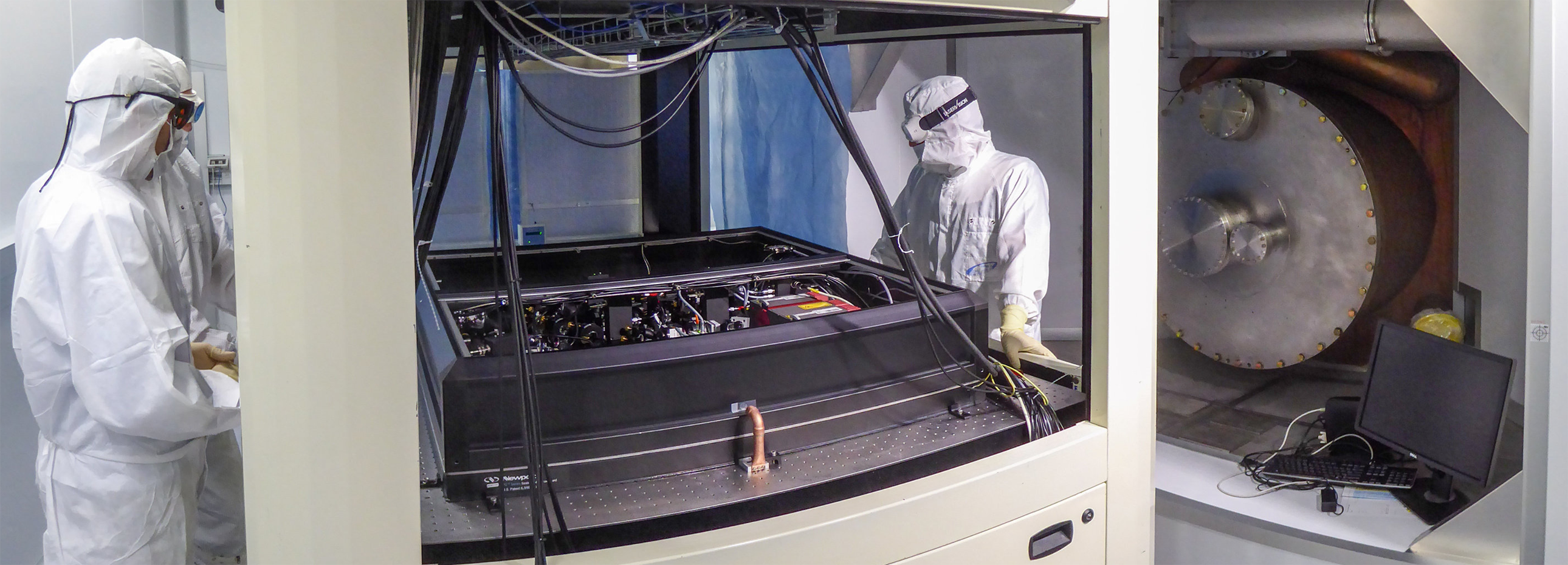} 
\chapter{Quantum Enhancements}
\label{sec:Quantum}

\vspace{1 cm} 
Quantum noise, originating from the quantized nature of light, limits the performance of laser-interferometric gravitational wave detectors over a large part of the observational spectrum. In contrast to many other fundamental noise sources, quantum noise can be strongly influenced and shaped by the interferometer configuration, e.g., the type of interferometer, number of optical components and their masses, mirror reflectivities and optical losses, and the use of additional optical cavities. Advanced interferometer topologies for quantum noise reduction beyond the current state of the art have been proposed~\cite{Danilishin:2019dxq}~cite{NEMO2020}. Such configurations have the potential to significantly improve the quantum noise in a specific frequency region of interest (e.g. at the low-frequency end for more accurate source parameter extraction or the potential of early warning for signals with an expected counterpart; at the high frequency end for improved neutron star physics) or even to provide a broadband improvement across the full observational spectrum. 
With quantum noise reduction schemes developing rapidly, building additional flexibility and space into the 3G infrastructure is advised to allow for enhancements and upgrades to the initial 3G detectors as technology becomes available. 
Broadly speaking, all advanced quantum noise reduction schemes rely on either the addition of long (up to km-scale) optical cavities (with potentially strongly reduced length noise requirements compared to the main interferometer), and/or major modifications to the optical configuration of the output port, i.e., the optical path between the main interferometer and main photodetectors used for reconstruction of the gravitational-wave signal. 
The implementation and use of quantum noise reduction techniques can depend on and affect many subsystems of the interferometer. The associated research and development therefore requires considerable effort, initially with detailed modelling, followed by extensive experimental testing of complete interferometer configurations in suitable environments, covering the entire range from small-scale tabletop experiments to low-noise prototype interferometers. A worldwide coordination of the required R\&D activities will allow an efficient execution of the required research programs and will also support the procurement of the required R\&D funds. 

\section{State of the Art}
The impact of the quantum noise of the light field can be shaped and mitigated by smart optical schemes and interferometer designs. We distinguish between two quantum noise components: shot noise and radiation-pressure noise, which originate from the quantum fluctuations of light and its interaction with the test masses~\cite{Cav1980}. The shot noise contribution to the overall interferometer sensitivity, originating from photon counting noise on the main photo detector, is inversely proportional to the square root of the optical power inside the arm cavity, and is dominant at higher frequencies (above 100 Hz in the case of Advanced LIGO); radiation-pressure noise contribution to the overall interferometer sensitivity, originating from momentum transfer of photons onto the main mirrors, is proportional to the square root of optical power, and is dominant at low frequencies. Radiation-pressure noise is sometimes called quantum `back-action' noise. Due to the Heisenberg Uncertainty Principle, the trade-off between these two, when varying the optical power, leads to the so-called Standard Quantum Limit (SQL). Developing quantum techniques for reducing quantum noise and surpassing the SQL is a very active field of research in the GW community. 
The influence of shot noise can be reduced by any combination of increasing the circulating laser power, by using squeezed light, by improved readout schemes, or by methods which increase the signal response of the instrument in the frequency range of interest~\cite{StMe1991,Mizuno:RSE1993,Osamu:2006}. High laser power poses a number of technical challenges, including thermal distortions of the test masses and optics which can lead to additional optical loss and parametric instabilities where the resonances of the optics are excited by the light~\cite{BSV2001,Evans:2015raa}. Squeezing is anticipated to be applied for all future observatories, either as the main scheme to reduce quantum noise or in combination with more advanced interferometer schemes. High frequency quantum shot noise can also be shaped by enhancing the detector response in a certain frequency range, but usually this involves a trade-off leading to worse sensitivity outside this targeted band.  

Radiation-pressure noise can be addressed by increasing the interferometer test masses and by various schemes adopting aspects of quantum-non-demolition techniques~\cite{KLMTV2001,PuCh2002,Che2003,Braginsky:2004fp}. Theoretical research into radiation pressure noise reduction is very active and has produced a wide range of schemes that show a potential for significant sensitivity improvement. At the same time, table-top experiments on quantum limited systems have become accessible to a wider community. However, the experimental investigations of possible schemes lag behind the theoretical work. We expect the range of interesting options for quantum noise reduction to condense over the coming years based on result from experimental research at small experiments and prototype interferometers.

Quantum noise beyond the SQL already plays an important role in the Advanced detectors~\cite{BuCh2001}. GEO600~\cite{GEO:Squeezing}, Advanced LIGO~\cite{H1:Squeezing,AdvancedLIGO2015} and Advanced Virgo~\cite{AdvancedVirgo2015} are currently demonstrating and improving squeezed light injection, obtaining shot noise reduction of up to 6\,dB. A+ and Advanced Virgo+ will test frequency dependent squeezing with short (300\,m) filter cavities~\cite{Eva2013,TAMA_FDS2016}, targeting 6\,dB effective squeezing over the whole detection band. Further, balanced homodyne readout~\cite{BHD,Stefszky:Balanced2012} will be part of A+. The lessons learned through these implementations will greatly benefit 3G quantum noise designs.

\section{Requirements for 3G}
Dual-recycled Fabry-Perot Michelson interferometers combined with squeezed light injection are the state-of-the-art for current gravitational-wave detectors and serve as a robust, low-risk baseline design for 3G observatories. More innovative interferometer schemes e.g. EPR-based frequency-dependent squeezing, speedmeters or variational readout schemes, less researched so far, 
may offer potentially higher quantum noise suppression factors (in particular at the very low frequency end of the detection band, i.e. below 10\,Hz),  but they require further R\&D 
 in order to first evaluate their principal suitability for application in 3G observatories and then to fully qualify and develop them. 
The following list details the top level  key requirements in terms of quantum noise reduction schemes   for 3G detectors:

\begin{itemize}
\item High circulating light power is an essential ingredient in many quantum noise reduction schemes, for example to reduce shot noise contribution to the overall interferometer sensitivity, or to make use of correlation effects in opto-mechanical systems. Low-loss and stable operation of interferometers at the several megawatt level   is a key requirement for 3G detectors,  which drives requirements for related subsystems, e.g.  for low absorption coatings, suitable cooling systems for cryogenic interferometers, sufficient thermal compensation schemes and mitigation of parametric instabilities.  
\item  Long-term  stable,  low-maintenance  and well controllable squeezed light sources (with the laser wavelengths used by 3G detectors, e.g. 1064\,nm, 1550\,nm and $\approx$2000\,nm), usually combined with long (hundreds  of meters to km-scale) filter cavities, implemented to give a effective  quantum noise reduction of  at least  10\,dB  across the full observation band .
\item Efficient implementation of QND schemes and squeezed-light technology relies on  low optical losses. For instance in order to obtain 6\,dB of observed squeezing the maximum loss on the full optical train from the creation of the squeezing up to the detection on the main photodiodes has to be below about 22\,\%. In order to increase the observed squeezing to the envisaged 3G levels of 10+\,dB of squeezing the total losses have to be reduced to below 8\,\% \cite{LSC_IS_WP}. Hence there are very stringent requirements to develop  low-loss optical components such as Faraday isolators (with less than 1\,\% of optical loss), reduction of optical scattering due to improved mirror surface figure errors, and adaptive optics for 3G detectors  which allow to reduce modematching losses to below 0.3\,\% . 
\item High-quantum-efficiency photodiodes (>99\%) at the operating light source wavelengths (e.g. 1064\,nm, 1550\,nm and $\approx$2000\,nm).
\item Heavy test masses of about 200\,kg,  possibly up to 500\,kg in the case of CE,  to suppress the effect of radiation pressure noise technical control noise and also to accommodate the large laser beams originating from the long baseline of the planned 3G detectors.
\item Adequate low-noise control systems for the main interferometer as well as auxiliary systems like filter cavities and readout system, compatible with astrophysical sensitivity in the sub-10Hz band.
For instance phase noise in the filter cavity control reduces the observable squeezing. For 3G detectors a phase noise level of smaller than 10\,mrads is required \cite{LSC_IS_WP}.
\item There are theoretical approaches to circumventing the SQL to any desired degree, but the limit to performance will always be set by the achievable optical losses in the system. Future R\&D will be able to quantify how the loss distribution and different optical layouts set this limit.
\end{itemize}


\section{Outlook and Recommendations}
A significant R\&D effort is required to make an informed selection of the optimal quantum noise reduction strategies for 3G detectors and to enable their risk free application to maximise the science return of 3G 
 observatories.

The R\&D roadmap in terms of quantum noise improvements for 3G detectors can be divided into 2 broad strands: 1) R\&D to realise the observation of frequency dependent squeezing at the 10+\,dB level. 2) R\&D to evaluate and ultimately qualify more innovative, but so far less developed, QND schemes, which have the potential to further improve the sensitivity of 3G detectors at the low frequency end.  
We recommend the following actions:
\begin{itemize}
    \item Parallel development of low loss optical components (e.g. Faraday isolators with loss of less than 1\,\%) and loss reduction techniques (e.g. for all wavelength currently under consideration for 3G detectors (e.g. 1064\,nm, 1550\,nm and $\approx$2000\,nm) 
    \item Parallel development of high-quantum efficiency photodiodes for 1550\,nm and $\approx$2000\,nm. A global coordination of the communication with photodiode manufacturers will increase the chances of success. 
    \item Development of unified classification scheme of QND techniques and a common approach to analysing and comparing their performances
    \item R\&D programme of table top proof of principle experiments, followed by demonstration in at least  10\,m class prototype facilities in order to qualify any promising QND schemes going beyond the Dual-Recycled Fabry-Perot Michelson with frequency dependent squeezing.  The quantum noise reduction techniques used for 3G detectors must reach maturity and be demonstrated by prototypes several years before 3G instrument installation.
\end{itemize}

%% file: Aux_Optics.tex
\chapterimage{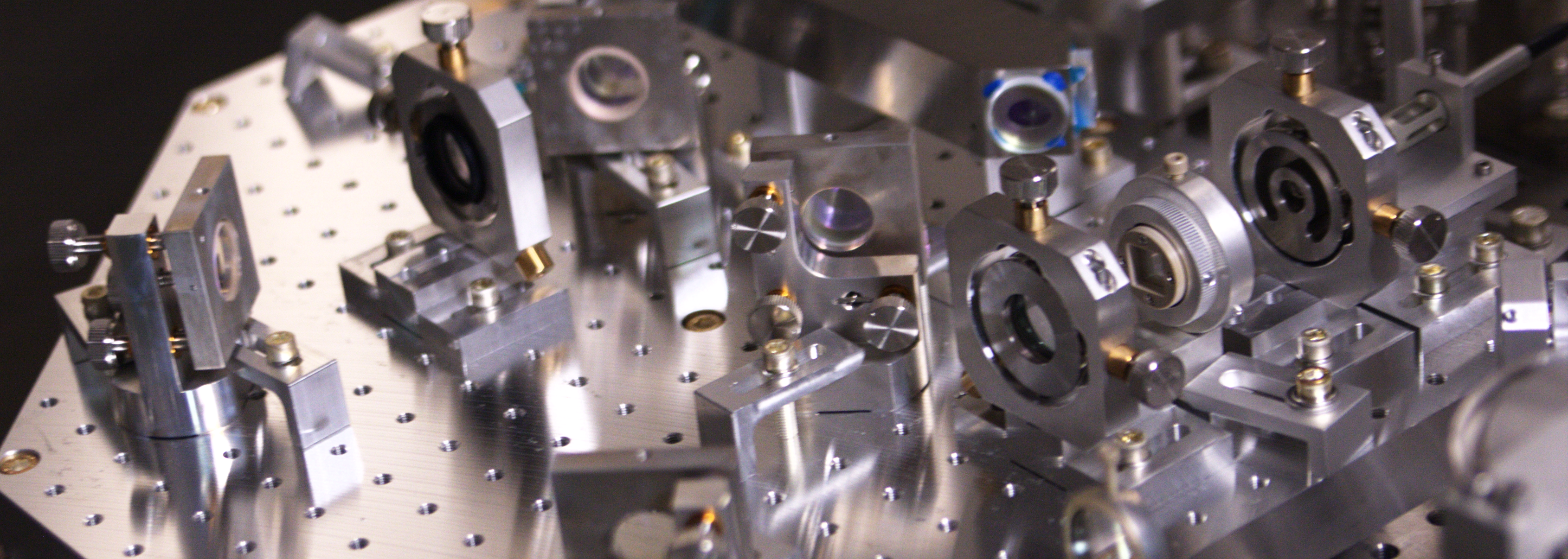} 
\chapter{Auxiliary Optics}
\label{sec:Aux-optics}

\begin{samepage} 

We use the term \emph{auxiliary optics} here for any optical subsystem not otherwise covered by a specific category (e.g. core optics or light sources). We further sub-categorize into these subsystems:
The {\bf Input Optics} subsystem includes all optics between the main light source and the input of the core interferometer, typically defined by the power recycling mirror. The {\bf Output Optics} subsystem includes all optics between the core interferometer output (typically defined by the signal recycling mirror) and the photo-detectors which detect the gravitational wave signal, with the exception of specific optical components used for the generation of squeezed light. The {\bf Active Wavefront Control} subsystem includes means for sensing and actively controlling the spatial properties of interferometer beams, with the typical goal of minimizing mode matching losses and contrast defects. The {\bf Stray Light Control} subsystem encompasses all design features which are intended to reduce the impact of stray light on detector operations. Finally, the subcategory {\bf Other Auxiliary Optics} catches optics that fall neither into the other larger categories or the above 
subcategories.

\section{State of the Art}
The {\bf Input optics} subsystem is responsible for delivering the laser light from the pre-stabilized laser (PSL) system to the core interferometer in the correct spatial mode, with the necessary phase modulation sidebands, and with the required frequency, intensity and alignment stability. This subsystem therefore typically includes an electro-optic modulator (EOM) for producing the phase modulation, one or more suspended input mode cleaner (IMC) cavities, a power control system, and an input Faraday isolator (IFI) to protect the PSL from light reflected from the main interferometer. Detailed descriptions of the IO subsystems of \ac{aLIGO}  and AdVirgo can be found in Refs.~\cite{aLIGO_IO,IOchapter}. 

\end{samepage} 

The {\bf Output optics} subsystem is responsible for filtering out higher-order spatial modes and unwanted control sidebands from the interferometer output light, and delivering the beam to the photodetectors where the gravitational-wave signal measurement is made. The recent and planned squeezed light upgrades expand the role of this subsystem to include delivery of the squeezed beam to the main interferometer. This subsystem typically includes an output Faraday isolator, an output mode cleaner cavity (OMC), and high quantum efficiency photodiodes.

\noindent {\bf \ac{AWC}} is used in gravitational-wave detectors to compensate for thermal effects caused by the absorption of light in the interferometer core optics. As the circulating light power goals increased from 1st to 2nd generation detectors, so too did the requirements for compensation of these thermal effects. The \ac{2G}  detectors have all implemented a range of active wavefront control methods including ring heaters, CO$_2$ laser heating, thermal radiation projection, and thermally deformable mirrors~\cite{aLIGO_AWC, AdVirgo_IO}. This subsystem is often conflated with the \ac{TCS}, although \ac{TCS} constitutes only a part of the \ac{AWC} system. 

\noindent {\bf Stray Light Control}
mitigates the effects on the sensitivity of the detector of light that is scattered out of the main beam and then scattered by another surface back into it. 
Many precautions are taken in gravitational-wave detectors to minimize the amount of light scattered out of the beam (by using high-quality smooth optical surfaces everywhere), and also to minimize the available paths for re-entry of scattered light into the main beam (by using baffles wherever possible). 

\noindent 
Examples of {\bf "Other Auxiliary Optics"} include the auxiliary length sensing (ALS) systems for pre-stabilizing the arm cavities, and optical levers for local sensing of the alignment of optics.

\section{Requirements, Challenges and Current/Planned R\&D }
\noindent{\bf{Input optics}}
The ways in which the \ac{3G}   detector input optics will differ from the current ones depends heavily on some of the design decisions for \ac{3G}   detectors, many of which have not yet been made. Most critical for the input optics will be the choice of laser wavelength; the \ac{EOM} and \ac{IFI} rely on unusual electro- and magneto-optic materials that may have significantly different characteristics at longer wavelengths. Choosing suitable materials is not expected to pose a great challenge if the wavelength is changed to 1550\,nm, but 2+$\mu$m will probably require considerable effort. The high power transmitted through these optics makes low absorption critical (due to the resulting thermal lensing and depolarization effects), thus further narrowing the choice of suitable material. At 2$\mu$m Faraday isolators already exist~\cite{EOTFI}, but a high power vacuum compatible one must be developed as it is not commercially available. Iron garnets should be considered as potential candidates for the magneto-optic material for 2+$\mu$m light, also having the advantage that the Faraday rotation effect is an order of magnitude larger than that of \ac{TGG}  in the near infrared range, allowing a reduction in magnet size and the overall footprint and weight of the device.
The \ac{IMC} design is not expected to fundamentally change beyond the 2nd generation.  However, AdVirgo and \ac{aLIGO}  both have significant noise contributions from beam jitter~\cite{aLIGOjitter,adVirgojitter}. Increasing the \ac{IMC} finesse or adding a second \ac{IMC} in series with the first could be a way to further suppress the beam jitter. Another possibility is to reduce the beam jitter at the source, either passively or with 
control loops.  

In addition to upgrades and modifications of existing IO components, several new components are also being researched. High bandwidth electro-optic beam deflectors could be used as actuators for beam jitter suppression loops and for providing alignment sidebands for an alternative method of alignment sensing in both the \ac{IMC} and in the core interferometer~\cite{RFJASC}. The use of higher-order modes for coating thermal noise reduction is still under consideration for \ac{3G}   interferometers~\cite{LGmodes}. If this technology is pursued further, the higher-order mode preparation path is likely to be incorporated into the IO remit. The use of complex modulation~\cite{complexmod} and parallel modulation~\cite{kagraMZI} are also under investigation for reducing sidebands-of-sidebands effects that may limit interferometer sensing and control performance.

\noindent{\bf{Output optics}}
With all \ac{3G}   concepts assuming significant enhancement from the use of squeezed light, the critical feature for the \ac{3G}   output optics will be ultra-low optical losses~\cite{squeeze_lossbudget}. \acp{OFI} with reduced optical losses are being developed~\cite{EGOLLFI,UFLLFI}.
\Acp{OMC} must also be designed for high throughput, and photodetectors must have high quantum efficiency (QE). High-QE photodetectors at longer wavelengths have been identified as a crucial \ac{RaD}  task; one which may be especially onerous for 2+$\mu$m light. 
Frequency dependent squeezing will also require the inclusion of a \ac{FC} in the path between the squeezed light source and the OFI. 300\,m long \acp{FC} are planned for the near-term upgrades to \ac{aLIGO} and \ac{AdVirgo}, following on from \ac{RaD} ~\cite{MITFC,TAMA_FDS2016}.
Alternative readout schemes such as \ac{BHD} (another project planned for inclusion in \ac{A+}) will require a redesign of the output optics chain~\cite{BHD}. There is also a need to develop robust length and angular control schemes for detuned filter cavities.

\noindent{\bf{Active wavefront control}}
In \ac{1G} and \ac{2G} detectors, \ac{AWC} has often been implemented in something of an \emph{ad hoc} way. In no cases has a feedback loop been used to maintain mode matching in a detector during normal operations. This is partly due to the low bandwidth of the typical actuators, but also due to the limitations in the \ac{AWC}  sensors available. As a result, \ac{AWC}  is typically employed in a set-and-forget manner, which requires frequent manual retuning as the circulating power in the interferometer changes. \ac{3G}   detectors will likely require higher performance from the \ac{AWC}  subsystem for several reasons: higher circulating powers, larger thermal gradients under cryogenic operation, larger beams with flatter wavefronts, and stricter requirements on optical losses. An \ac{LSC} white paper on \ac{AWC}  has been written to track the LSC \ac{RaD}  in this direction~\cite{aLIGO_AWC}, and similar work is planned for AdVirgo+. \ac{AWC}  \ac{RaD}  for \ac{3G}   detectors is expected to focus on new sensors (\ac{RF} bullseye wavefront sensors~\cite{bullseye}, improved Hartmann wavefront sensors~\cite{HWS}, phase cameras~\cite{phasecam}), as well as actuators (converting existing actuators for new optic materials, and developing new actuators). 

\noindent{\bf{Stray light control}}
Although it is often difficult to diagnose directly, stray light was likely limiting the sensitivity of both \ac{aLIGO}  and AdVirgo during O3. As such, improvements must clearly be made in order to reach \ac{3G}   sensitivities. 
Continued \ac{RaD}  into mirror polishing and coating techniques to give improved surface roughness will go some way to reducing stray light contributions, as will the development of better baffling materials with lower reflectivities. One concern is that if the laser wavelength is changed to 2+$\mu$m, it may be difficult to find good absorbers for baffle materials. The quieter \ac{3G}   infrastructure environments may allow relaxing some scattering performance requirements.

\noindent{\bf{Other auxiliary optics}}
AdVirgo, \ac{aLIGO}  and \ac{KAGRA}  all either use (or provision for) an auxiliary length sensing system in order to make the interferometer locking process deterministic. Since the optical layout of \ac{3G}   detectors is not expected to reduce in complexity, it seems natural that similar systems will be required in the future. Optical levers are omnipresent in \ac{2G}  detectors, and provide important information about angular motion of the optics. Increased sensitivity in these local sensors may be beneficial for \ac{3G}   detectors
and \ac{RaD}  is ongoing towards that end. \Ac{SPI}~\cite{SPI}, which is currently being used at the \ac{AEI} 10\,m prototype, may be a useful technique for reducing low frequency control noise in \ac{3G}   detectors, and will require an additional auxiliary optical subsystem.

\section{Pathways, Required Facilities, Collaborations and Mechanisms}

Much of the auxiliary optics subsystems \ac{RaD}, such as materials investigations for \acp{EOM} and \acp{FI}, can be performed in small laboratories. 
Some of the larger subsystems such as filter cavities, on the other hand, require larger integrated facilities for testing. In many cases, the development of \ac{3G}   auxiliary optics technologies is also beneficially incorporated into the modernisation of second-generation detectors. By virtue of being \emph{auxiliary}, modular, incremental upgrades in many of these technologies are possible. Modular upgrades are not feasible in the case of a change in the main laser wavelength, however. Such a change for \ac{3G}   detectors will necessitate a broader range of auxiliary optics \ac{RaD}, which would benefit from a more complete demonstration on prototype detectors or finally at full scale in Voyager.

\section{Outlook and Recommendations}

Activities on Input Optics and Output optics are relatively well covered and organized within the existing collaborations. Active wavefront control and stray light control would benefit greatly from increased global collaboration. Due to the ad hoc nature of the development of these subsystems to date many groups have taken individual paths to different but often similar solutions. Many of the auxiliary optics subsystems will be strongly impacted by the choice of laser wavelength. Beyond this, the input optics has a direct interface with the light source, and so information exchange between groups performing \ac{RaD}  in these two areas will be critical. The output optics subsystem will be closely linked to the quantum noise working groups as squeezing is foreseen in all future detectors. 

We recommend:
\begin{itemize}
\item the formation of global working groups across the different auxiliary subsystems to coordinate and focus development
\item  that mechanisms be put in place to facilitate close collaboration between groups working on auxiliary optics, light sources and quantum noise.  This would be facilitated by the formation of a global \ac{3G}   \ac{RaD}  planning committee.
\end{itemize}
While auxiliary optics are largely modular, enough time must be given between the selection of a laser wavelength and construction for the required \ac{RaD}  to conclude for auxiliary optics. Given the current \ac{RaD}  progress, this lead time could be five years. 

%% file: Sim_and_Control.tex
\chapterimage{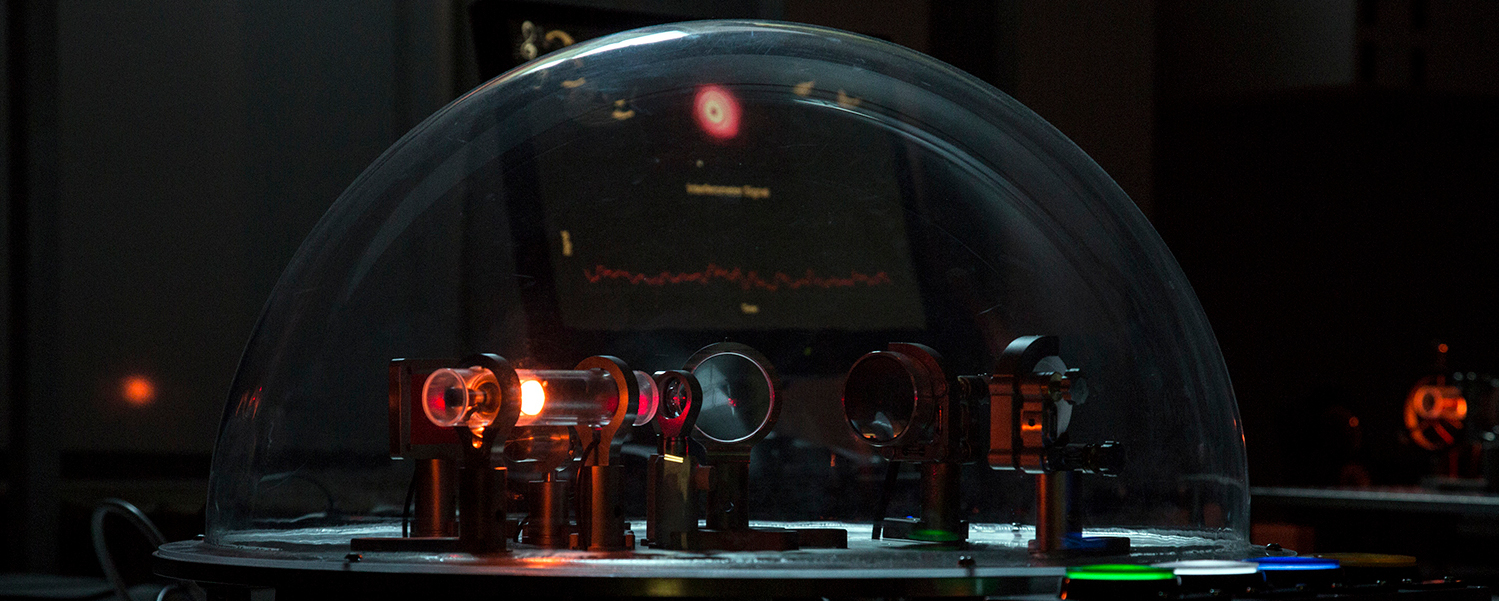} 
\chapter{Simulation and Controls}
\label{sec:Sim_Controls}

\vspace{1.5cm}

Interferometric gravitational-wave detectors are complex optical and mechanical systems. Their core instruments, Fabry-Perot and Michelson interferometers, can be used for high-precision measurements only if their parameters, especially their lengths and angular degrees of freedom, are careful controlled. 
The complete detectors have a large number of control loops for the various optics, their suspension systems and many other components, such as main and auxiliary lasers and active vibration-isolation systems. 
The \textbf{Simulations} section lists the requirements for modelling tools for design and commissioning of the complex opto-mechanical interferometers and the \textbf{Controls} section outlines the research required for developing adequate control schemes for future detectors.

\begin{samepage} 

\section{Simulations}
\subsection{Current Approach to Interferometer Modelling}
Gravitational-wave detectors require detailed modelling for design and performance studies; this also applies 
to prototype experiments (table-top and 10\,m scale). 
The detector behaviour cannot be modelled with commercially available optical simulations as our requirements differ significantly from those of conventional systems. 
The proposed third-generation detectors include either new technologies or envisage pushing detector parameters closer to their limits. 
A specific set of tools has emerged that can be used to achieve the required modeling for \ac{3G}  detectors, though work remains to expand and enhance these tools. Here we classify the most commonly used software:

\noindent{\textbf{Time-domain models}} describe the optical field and the optical components in the time domain, i.e. direct outputs will be signals over a discrete time step. Typically the time-domain models are the most powerful   simulations because they can include non-linear and dynamic behaviour.
The disadvantage, however, is that such simulations require significant computing power. Example software: SIESTA~\cite{SIESTA}, E2E~\cite{e2e_2000}.

\noindent{\textbf{Frequency-domain models}} are based on the approximation that the entire system is in a steady-state with only small disturbances or slow (quasi-static) changes. All the modelled features are linearised, thus the system can be described as a set of linear equations. 
This approach provides very fast and flexible models but obviously lacks the ability to model non-linear physical features. Frequency domain software can be further split into two categories: \emph{\ac{FFT} propagation models} and \emph{modal models}.

\end{samepage} 

\begin{itemize}
\item{\textbf{\ac{FFT}   propagation models}} have been developed to analyse the behaviour of optical fields in the presence of wavefront distortions by discretizing the transverse complex amplitude of the field on a 2D grid.
Example software: SIS~\cite{SIS}, \acs{OSCAR}~\cite{OSCAR}, \acs{DarkF}~\cite{DarkF, Vinet92}.
 
\item{\textbf{Modal models}} use Gaussian modes to describe the spatial properties of the beam, its propagation and scattering. The modal model expands the beam shape into orthogonal modes ordered by spatial frequency forming a complete basis. Thus for near perfect Gaussian beams only a few numerical values describing the mode amplitudes need to be computed resulting in faster simulations. The commissioning of advanced detectors, as well as the design of further upgrades have shown that such tools have become crucial to understand interferometer limitations. 
Example software: \acs{Optickle}~\cite{Optickle}, \textsc{\acs{Finesse}}~\cite{Finesse, Freise04}, \acs{MIST}~\cite{MIST}.
\end{itemize}
\noindent{\textbf{Interferometer noise calculators}} are configuration level simulations that operate at a higher level so that a given optical configuration is symbolically computed and parameterized. 
This approach is very effective for an initial exploration of a parameter space for a variety of optical configurations and for providing simple reference noise budgets. Example software: \ac{GWINC}  ~\cite{GWINC}.

\noindent{\textbf{\acs{SimPlant}}}~\cite{simplant}, is a virtual interferometer for commissioning in which the user can click a button to turn the machine over to Sim Mode, and then use the usual tools to measure, for example, the noise and transfer functions. This provides the fastest way to compare our theoretical knowledge with the real instrument.
The simulated interferometer runs in real time within the real time control system so that the usual controls system can be run. 

\noindent{\textbf{Ray tracing}} tools are used to find the exact position of beam axes and Gaussian beam parameters in real systems and can help with stray light investigations, often using \ac{CAD} data from the design of the instrument's mechanical components as a basis. Example software: \acs{IfoCAD}~\cite{IfoCAD, kochkina}, \acs{OptoCAD}~\cite{OptoCAD}.

\subsection{Requirements}
\label{sec:Sim:Req}


Significant support from interferometer modelling will be needed for tasks including:
\begin{itemize}
\item High-power operation at low optical loss.
Modelling of squeezed light in higher-order modes, improved thermal compensation systems, improved arm and mode matching techniques.
\item Scattered light control, modelling of backscatter of detection optics and interferometer scatter. Include injection of noise with specific coherence into interferometer modelling tools. Non-sequential ray-tracing. Monte-Carlo methods.
\item Control design: better models of control schemes can be achieved by developing more effective tools for the analysis of in-loop cross coupling of a mixed mechanical, optical and electronic system, and for the analysis of modern
control strategies.
\item Advanced quantum noise schemes, development of a robust `fundamental' quantum limit. Modelling of quantum correlations through complex \ac{MIMO}  systems.
\item Study of optical configurations which rely strongly on polarisation schemes, requires the addition of light polarisation to interferometer models.
\item Newtonian gravity noise reduction, advanced modelling of local sensing and global control strategies, require an advanced implementation of mechanical systems and seismic and gravitational noise coupling in interferometer models as well as the simulation of gravitational noise based on ground noise measurements.
\item A missing piece in detector simulation is a comprehensive mechanical simulation tool for the vibration isolation and suspension design which includes the capability to handle a variety of mechanical systems and the ability to compute thermal noise for any given configuration.
\end{itemize}

\subsection{Impact on Detector Upgrades}
Interferometer simulation tasks for upgrades in current facilities and for \ac{3G}  detectors are closely related and strongly benefit from each other. Simulation tools and interferometer modelling have to be advanced ahead of time in order to be able to provide the essential
support during the design and instrument development. The design of the advanced detectors triggered the development of new tools which then had a significant impact on the commissioning of the first generation. At the same time the interaction with commissioners (and scientists developing advanced detector
technology at prototypes) provided essential community interaction and feedback that resulted in tools with better capabilities, validated test results and expert users. The same synergy is expected now between advanced detectors and \ac{3G}  observatories; it should be encouraged and utilized as much as possible.

\section{Outlook and Recommendations}
Simulation tools are strongly defined through the context in which they are used. Many of the priorities for R\&D directly translate into a priority task for modelling work or a required effort in developing new capabilities for simulation tools. Also, modeling progress is not primarily limited by the development of technologies or methods, but by the available person power.

To address the above challenges, the current portfolio of software tools must be updated, either by extending the existing software or by providing new dedicated tools. The detailed list of code changes or required features goes beyond the scope of this document. The following are recommendations for the higher-level actions to support an effective and open environment for this effort.

\noindent{\textbf{Additional software}}
Most of the required modelling tasks can be performed by extending and updating the available tools, some of which is already well underway. However some missing functionality might be better achieved by developing new software. Needed packages include an easy to use and flexible time-domain simulation, a \acs{3D} beam tracing software dedicated to ground based detectors and a comprehensive modelling software for various suspension systems capable of computing the thermal noise of all elements. In addition, commercial tools should be reviewed to understand when these are superior to custom made tools, for example, for more common optics tasks such as modelling stray light.

\noindent{\textbf{Resources}}
We encourage collaborating institutions to increase efforts in the development and use of simulation tools over the next 5 years, a crucial time for the design of \ac{3G}  instruments and for upgrades to current detectors.
Experience has shown that individual post-docs and PhD students can provide effective tools that are quickly adopted and used by a wide community. However, those tools often come only with rudimentary or outdated documentation and code reviews or formal testing 
are not common practice. We recommend additional, dedicated post-doc support in this area to mitigate this.

\noindent{\textbf{Collaboration and coordination}}
Coordination between research groups and projects is important in three ways:
code development is often done by a few individuals in each project who will benefit from having a forum to discuss priorities and technical challenges. 
Similarly, collaboration between people doing modeling to understand a new or not-understood behavior with other scientists investigating related issues has been shown to greatly reduce the time needed to reach a conclusion. 
We recommend to continue (or establish) working groups within projects dedicated to interferometer modelling and to organize workshops adjacent to international meetings.


\noindent{\textbf{Software and code distribution}}
Accessibility of the software, and ideally the source code, should be improved. Each software tool should have: a) an active maintainer who is responsible for the current code base and who can be identified and reached by any users of the software, b) a single, discoverable web page hosting the master version of the tool or code under a clear and permissible software license, c) documentation about the implemented models and their limitations, including descriptions of mathematical algorithms, or parameter sets used and d) training material, such as a set of examples and tutorials for new users, especially graduate students. Where possible (without breaking existing compatibility) the adoption of common input and output formats, for example, for files storing interferometer parameters should be encouraged.

The effectiveness of specific tools is often not defined by the a single feature but by a network effect based on many factors. Any tool benefits greatly from a large user base, for example, through receiving bug reports and the availability and diversity of examples as well as experts on how to use the specific tool. And the impact of a modelling tool is improved significantly by a strong track record and trust by the wider community. We recommend that software maintainers adopt the aim of making the software as accessible as possible without compromising its core functionality. 

Those software tools that aim at providing standard results for the wider community must also provide the official data sets or model files, such as, e.g., the default \ac{LIGO}  models for \textsc{Finesse}. The \ac{GWINC}  software package should provide standard noise budgets for all envisaged detectors and the
international community should establish a mechanism to review these models.

\noindent{\textbf{Source code maintenance}}
Some important simulation tools date back to their original development in the 80s and 90s. Given the rapid development of computing and computer languages, it could be beneficial to re-implement these in modern frameworks. At the same time the experience and knowledge acquired with the originals should not be lost in the process. A careful development process and code design that finds a balance between modern technology and backwards compatibility
is recommended. Some codes, such as \textsc{Finesse} and \ac{GWINC}  are currently undergoing such a process. Other software should be reviewed for similar updates.

In summary we recommend
\begin{itemize}
\item that the current portfolio of software tools must be updated, either by extending the existing software or by providing new dedicated tools.  
\item  that collaborating institutions increase efforts in the development and use of simulation tools over the next 5 years, a crucial time for the design of \ac{3G}  instruments and for upgrades to current detectors.
\item continuation (or establishment) of working groups within projects dedicated to interferometer modelling and the organization of workshops adjacent to international meetings. Code developers in projects will benefit from discussion fora. Detector modelers can better share experiences, e.g. noise behavior
\item  improved accessibility of the software, and ideally the source code;  software maintainers adopt the aim of making the software as accessible as possible without compromising its core functionality. 

\end{itemize}

\section{Controls}
\label{sec:Controls}
Control systems are a fundamental part of all interferometric gravitational wave detectors. The optical and interferometric methods used to achieve the strain and displacement sensitivity required for gravitational wave detection are all non-linear; actively nulling the sensing signal with a feedback control system linearizes the output. 
The control system thus enables the low noise, stable, and linear operation of the detector in the presence of seismic, acoustic, and radiation pressure disturbances.
The detection of gravitational waves poses stringent requirements on the operations of the feedback and feed-forward control systems.
In the past, meeting these requirements has proved to be extremely challenging.
This is partially because the detectors were designed without sufficient consideration of the controls challenges, and partially because the controls challenges are so extreme and gravitational wave detectors so unique in the field of controls.
Some plans for future detectors aim to strongly improve the sensitivity at frequencies below 10\,Hz. This frequency region is currently dominated by excess noise associated with control systems, and the envisaged noise reduction represents a significant challenge for the control system design and implementation.

\subsection{Current Approach}
Gravitational wave detectors have many degrees of freedom,  including the  \ac{DARM}, which is sensitive to gravitational waves, and many auxiliary degrees of freedom, which are not. \ac{DARM} must be stabilized to linearize the signal, and the many auxiliary degrees must be well stabilized to avoid coupling noise into the gravitational wave readout. Achieving this second requirement is the primary challenge, due to the large number of coupled degrees of freedom, where the couplings may be both non-linear and non-stationary, and the variety of timescales involved (which range from seconds to hours).
In most cases, classical control methods are used to minimize some quantity, typically pole-zero based linear filters in feedback or feedforward systems.
The filter design is based on modelled or measured response functions of the respective degree of freedom (composed by optical, mechanical and electronic parts). There is often a trade-off between robustness of the control loop and the overall sensitivity of the detector.
There are also a few implementations of more modern techniques including:
global feed-forward of seismic noise to platforms and suspensions; $\mu$-synthesis approach for limited angular control; feed-forward sensor noise subtraction (removal of seismic noise from wave-front sensors); and parametric instability damping by phase-locked loops and damping through aliasing of high frequency signals.

\subsection{Requirements}
{\bf Low noise operation:} all control systems must be able to operate on the main and auxiliary degrees of freedom without introducing additional noise. 
This is a point often missed in design studies, where only ``fundamental'' noises are considered (thermal noise, quantum noise, etc.). However, different design choices of the interferometer translate to different requirements for the control system. As an example, second generation detectors are still limited at low frequencies by (mostly angular) control noise. 
This is directly related to the interplay between performance of the suspension and seismic isolation design and the trade-off between stability and low-noise in angular controls. 
If all suspension and seismic platform resonances could be moved to lower frequencies, and if the residual motion due to microseism could be reduced by the seismic isolation, then all angular control systems could be relaxed, improving the low frequency noise of the detectors. 
With the \ac{3G}  detectors aiming at considerably lower frequencies, this becomes a vital design aspect. Research is required to understand how best to integrate realistic control limitations into interferometer designs.\par
\noindent{\bf Non-linear lock acquisition:} the term `lock' refers to a stable state of the control systems in which all relevant degrees of freedom are stable at their nominal operating position. In the uncontrolled state the interferometer is a highly non-linear system, in the sense that all the optical signals that can be used to estimate the resonance conditions depend in a complex non-linear way on the relative mirror position. The lock acquisition scheme is an algorithm designed to bring the system in a deterministic way from the uncontrolled state to the final low noise state. There is significant room for improvement in this area, since all lock acquisition strategies developed so far are somehow sub-optimal. An example of a design choice that resulted from prototype research and vastly improved the lock acquisition process is the installation of the arm-length stabilization in Advanced LIGO~\cite{Mullavey:12}. The use of similar techniques and modern control theory could improve upon the current lock acquisition schemes
and reduce the down time of the detectors.\par
\noindent{\bf Robustness:} control systems must be able to withstand external perturbations due to environmental disturbances, such as earthquakes, without losing lock and with minimal reduction in sensitivity. Any loss of control translates directly into downtime and a reduction of the observed time-volume. Understanding the mechanisms by which earthquakes cause detectors to lose lock, developing controls system states which might be able to ``ride out'' earthquakes, and developing methods to quickly transition from a low-noise state to a robust state all require further research.\par
\noindent{\bf Noise cancellation:} feed-forward and linear stationary noise cancellation are techniques already implemented routinely in second generation detectors. However, as those techniques reach their limit, the residual noise couplings are bound to be either non-linear or non-stationary. Development of control strategies, e.g., using machine learning, to cope with such systems is needed. 
\par
\noindent{\bf Optimisation:} In the currently operating detectors, the controls systems are tuned manually, which yields sub-optimal results.  Considering the large number of coupled degrees-of-freedom and the long timescales involved, brute force explorations of the parameters is also impossible. The following all need to be developed or modified to suit the unique needs of gravitational wave detectors: systematic methods to explore the parameter spaces of control systems; robust automated filter design; optimal multiple-input multiple-output sensing, control, and system identification methods. These will likely require combination of offline analysis of data collected while the 
detector is controlled in a non-optimal way, coupled with sophisticated simulations, and carefully chosen sets of test parameters that sample the global parameter space.
\par
\noindent{\bf High optical power:}
High circulating light power introduces a number of challenges related to control systems, such as instabilities of the optical cavities due to increased opto-mechanical coupling, and thermal distortions of the laser beam. Auxiliary control systems have been and are being developed to mitigate these effects post-hoc in the first and second generation detectors. These should be fully integrated into the global control strategy.
\par
\noindent{\bf Low frequency control:}
At low frequencies (below $\approx$30Hz), the gravitational-wave output is dominated by `technical' noise sources. While the coupling mechanisms are non-linear and complex, the underlying origin is clear: excess motion between 0.1 and 5Hz. This motion must be reduced through active control to an \ac{RMS} level that is different for each subsystem, but which is determined by the interferometer's peak operating point and the linearity around that point.
Future interferometers will need to consider the design of low frequency sensors and control systems fundamentally. It is no longer sufficient to say that low-frequency noise is a `controls problem', it must instead be solved through clear and traceable requirements for all relevant systems. 
It is necessary to work on a wide range of issues and solutions, including: the development of reliable models for low-frequency noise couplings, construction of new (local) sensors 
to reduce environmental couplings and sensor-noise, new control strategies to reduce the motion most critical to interferometer noise and operation, and improved seismic isolation and interferometer control to move the detection band below 10 Hz.


\section{Outlook and Recommendations}

Similarly to simulations, progress in the development of the required controls is not defined by future technical breakthroughs in the topic itself. Instead we can adapt techniques that have been developed in other areas, such as modern control techniques. Simulation tools can play a crucial role for designing control techniques and they should be further developed or extended for this purpose. More advanced techniques, such as adaptive control or machine learning, might need different and more powerful hardware to run, for example dedicated \acp{FPGA}s or \acp{GPU}s. It is important to advance the control design compared to other subsystems, so that controls can inform the overall system design, to avoid designs that make the control harder or impossible. 
In addition to recommendations listed under Simulation, we recommend that
\begin{itemize}
\item control designs be undertaken in advance of other subsystems so that controls can inform the overall design .  Control schemes should be developed and tested early as an integral part of detector designs.
\item Prototype interferometers be used to validate control schemes when these incorporate a more fundamental change
\end{itemize}

Ongoing commissioning of current detectors includes the improvement and optimization of the control systems. We expect that many if not most new control techniques or ideas will be tested and implemented in \ac{2G}  interferometers first. However, this is not true for schemes that require a very different sensing or actuation approach.


%% file: calibration.tex
\chapterimage{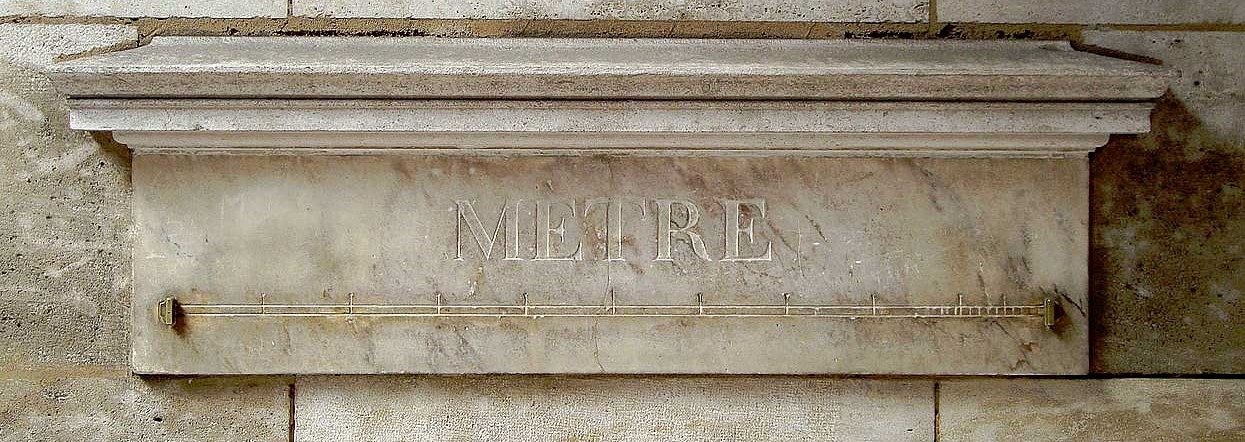} 
\chapter{Calibration}
\label{sec:Calibration}


Extracting new science from the observed gravitational waves requires accurate knowledge of the amplitude and timing of the signals. With the very high \ac{SNR}, up to 1000, expected in the \ac{3G}   era, extremely low calibration uncertainty will be necessary for observations to be noise limited.

\section{Science-Driven Calibration Requirements}
The exact calibration requirements for the scientific objectives of \ac{3G}   gravitational-wave observatories are not yet known, but can be estimated. There are two aspects of calibration uncertainty, the absolute uncertainty and the relative uncertainty. The first tells us how well we understand the total calibration in absolute numbers, while the latter is a frequency-dependent calibration uncertainty with respect to some fixed reference frequency. 
Looking for deviations from a modeled waveform template, we are mainly concerned with relative calibration uncertainty. In order to estimate distances to the sources, we are most interested in absolute calibration.
In addition to the calibration error of the detector instrument, the waveform models have an uncertainty.
Studies are underway to determine the exact calibration requirements integrating both detector and waveform uncertainties.

The third generation science that will set our calibration requirements include \ac{BNS} tidal deformation, deviations from \ac{GR}, and measurements of the Hubble constant. The first two look for deviations from a modeled waveform template, while the latter is based on absolute distance measurements. With $\mathcal{O}(1000)$ low-redshift binary neutron-star events, the Hubble constant may be determined to a ${\sim}1\%$ level with Einstein Telescope~\cite{Cai:2016sby}.
This would require a sub-percent systematic absolute amplitude calibration.
The signals with the highest \acp{SNR} give the narrowest limits for \ac{GR}. With a network of three \ac{3G}   detectors we expect about one event per year with $\text{\ac{SNR}} \sim 1000$. To avoid being dominated by calibration errors, we need a calibration to an amplitude of <\,0.5\% and
about 0.1 Radian in phase integrated over 10\,Hz to 400\,Hz \cite{SathyaPers2019}.


\section{State of the Art}
Overall calibration relies on both technologies and methods of measuring the detector’s control systems. The development of technologies important for calibration are discussed in the following subsections. The state of the art in calibration methods involve the use of \ac{MCMC} simulation to determine unknown systematic uncertainties.
\pagebreak
The detector's control systems (sensing and actuation) have a considerable influence on the total calibration accuracy and therefore must be precisely characterized for the calibration of each gravitational wave detector.
Calibration lines are applied using the detector's length actuators in order to compute time dependent correction factors.
For Advanced LIGO, this process results in an absolute calibration uncertainty of a few percent in amplitude and a few degrees in phase across the majority of the frequency band~\cite{PhysRevD.96.102001}.

\subsection{Photon Calibrators}
Photon calibrators provide a calibration reference starting from a traceable reference power meter. They use an amplitude modulated laser to apply a photon-recoil force to the test mass. For all interferometric gravitational-wave detectors with arm cavities, this is the current method of choice for absolute calibration.
Ultimately, the current photon calibrator implementation in Advanced LIGO has an absolute systematic uncertainty of 0.5\%, set mainly by a combination of uncertainty in the calibration of the \ac{NIST}-traceable power standard and uncertainty in where the photon-calibrator beams and the interferometer beams impinge on the test masses~\cite{ALIGOPhotCalib2016, NISTWorkshop2019}. The national metrology institutes are improving the primary laser power calibration standards. On request of \ac{LIGO}, \ac{NIST} has improved the laser power standard from 0.44\% error to 0.31\% and is envisioning a level of 0.05\% in the next few years.

\subsection{Newtonian Calibrators}
Newtonian calibrators rely on the Newtonian gravitational interaction between an interferometer test mass and a known arrangement of rapidly rotating calibration masses, with the arrangement often approximating a dipole, hexapole, or other multipole distribution~\cite{Matone:2007vk,Inoue:2018okt}.
These devices have absolute systematic uncertainties related to how well the geometry of the calibrator, and its distance and orientation with respect to the test mass, can be characterized.
A Virgo prototype~\cite{0264-9381-35-23-235009} has already been deployed, and an improved prototype may achieve a 1\% systematic uncertainty.
\ac{KAGRA} is designing a dual-mass-distribution calibrator that, when combined with a photon calibrator, may achieve an absolute uncertainty of 0.17\%~\cite{PhysRevD.98.022005}.
A Newtonian calibrator prototype is also being developed for Advanced LIGO.
These Newtonian calibrator technologies are still in the very early stages of development, but there is significant effort in this direction.


\subsection{Other Calibration Methods}
Laser frequency can also be used as a reference against which to calibrate an interferometer \cite{Leong2012, PhysRevD.95.062003}. Further \ac{RaD}   is required to cut down uncertainties from current levels of about 10\% (in Advanced LIGO) to the requirements of \ac{3G}   observatories.
It is also possible to fix certain aspects of the calibration using an astrophysical source whose properties are sufficiently well known~\cite{CalibrationGW170817,Pitkin:2015kgm}; more work is needed to understand how this technique will inform the calibration for \ac{3G}   detectors.


\section{Outlook and Recommendations}

Requirements on the absolute and relative calibration for third-generation detectors are not yet fully known, and dedicated modeling efforts are needed to understand the constraints set by various \ac{3G}   scientific objectives. \ac{3G}   detectors will likely require much better than sub-percent amplitude calibration, potentially necessitating \ac{RaD}   to improve the capabilities of the current calibration apparatuses; such \ac{RaD}   is already underway and cooperation activities between the different gravitational-wave collaborations are considered to be at an adequate and sufficient level.
The exact nature and combination of calibration methods to be used in \ac{3G}   can be designed soon, but left flexible enough such that the best performing technologies can be installed in the instruments with only a few years lead time before operations.

%% file: Outlook.tex
\chapterimage{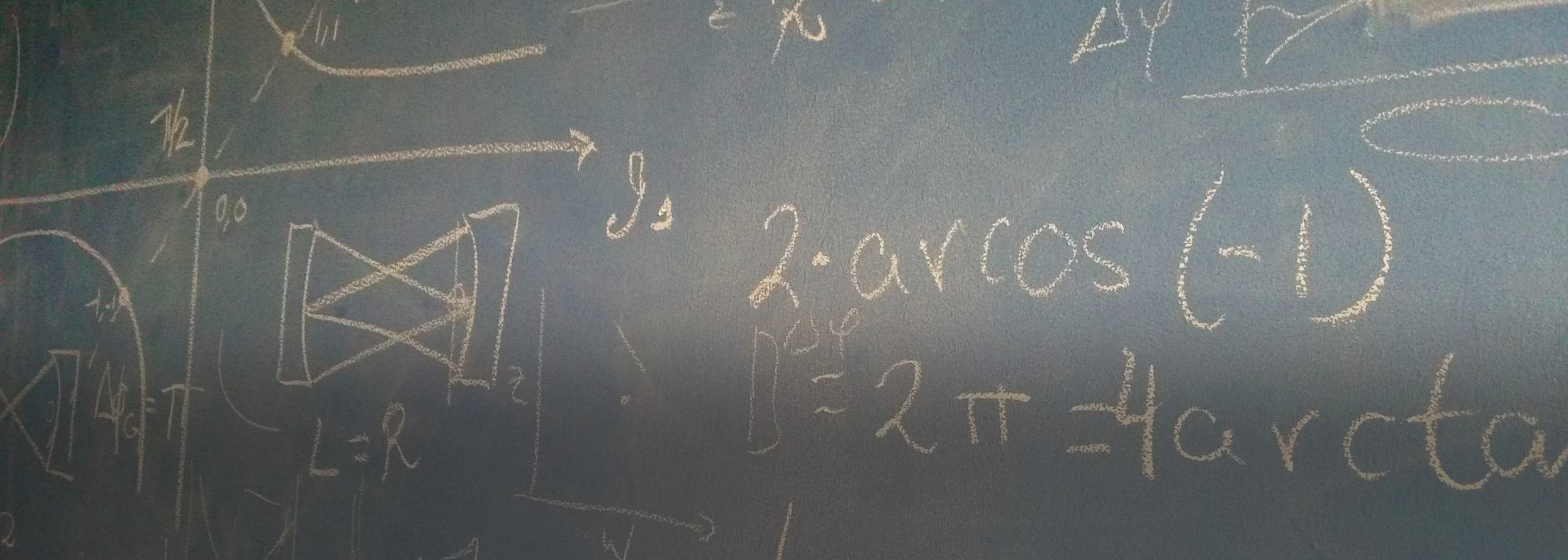} 
\chapter{Outlook and Key recommendations}
\label{sec:Outlook}

\begin{figure}[h]
\centering
\includegraphics*[width= \textwidth]{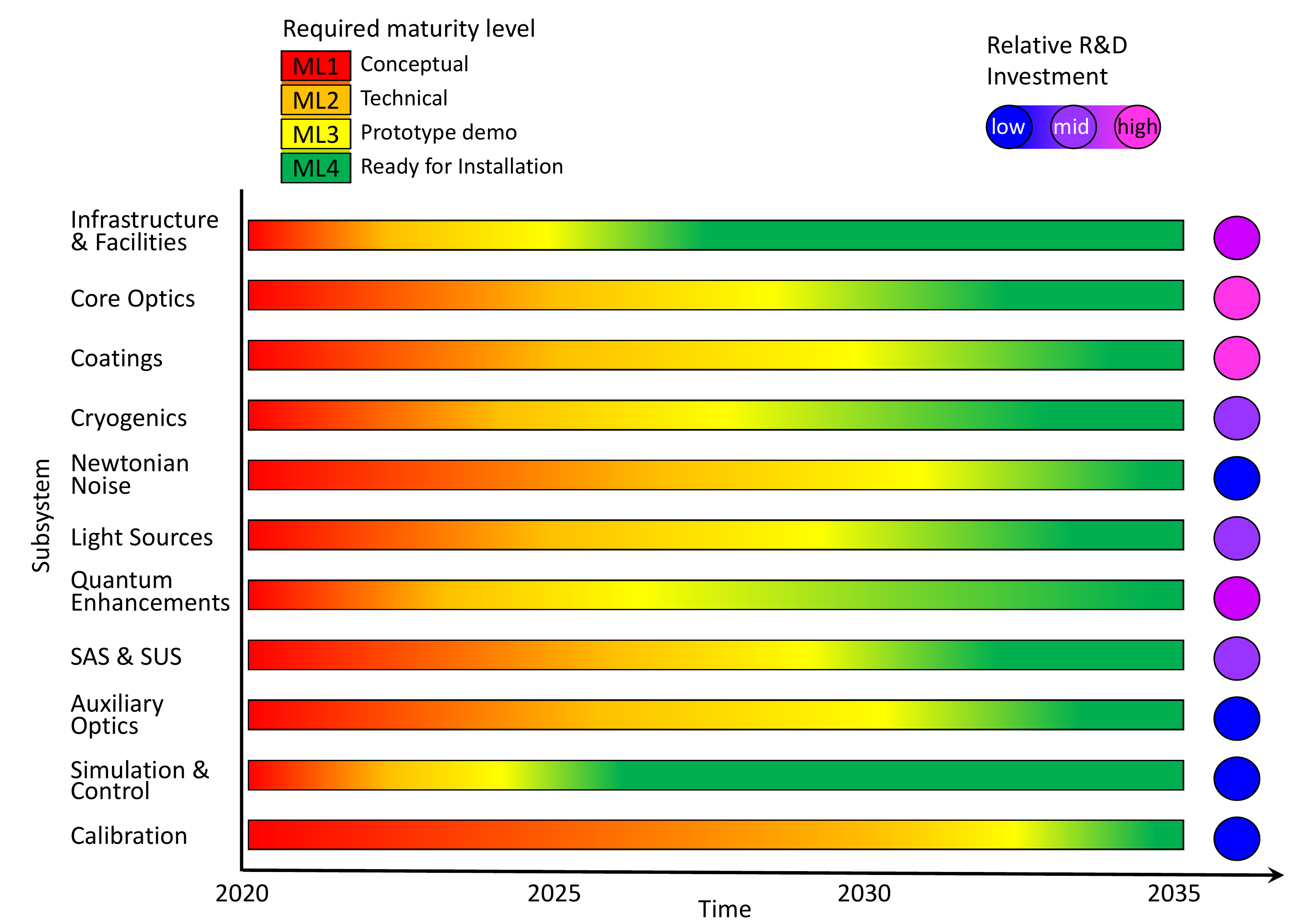}
\caption{Approximate timelines for the required maturity levels for \ac{3G} instruments and resource levels needed from now to installation of the first phase. The estimates for the required \ac{RaD} resource level shown on the right hand side only include investment (not \ac{RaD} person costs, assumed to be supplied form the labs) and are roughly categorized into \textit{low, mid} and \textit{high}.\\
}
\label{fig:maturity}
\end{figure}

This chapter presented the goal sensitivities of the first detectors to be installed in planned \ac{3G} facilities and reviewed the \ac{RaD} required  to deliver  subsystems capable of facilitating  the required performance. Figure~\ref{fig:maturity} depicts the required maturity levels for the various subsystems, depending on the foreseen time of installation and anticipated lead times.  



 Infrastructure and facilities have the longest development lead times and thus will have to reach maturity earliest, followed by the other subsystems in the sequence of installation. The timings for the lower maturity level (ML) ML1 - ML3 (relative to the highest level ML4) depend on the duration required between the individual steps; e.g. once technical readiness is achieved for core optics it still takes a few years to demonstrate full scale prototypes and manufacture the final optics substrates through a pathfinder process. Despite considerable differences in designs for the proposed \ac{3G} observatories (\acs{ET} and \acs{CE}), large variations within the subsystems and an inevitable uncertainty in timelines, we nonetheless attempt to summarize the timelines for each subsystem in a single bar in figure~\ref{fig:maturity}. The resources required to reach operational readiness are roughly divided into low, mid and high, with indicative financial investments for \ac{RaD} for the various subsystems over the entire period from now to the start of installation. 
The highest costs for the detector elements (as distinct from the civil and vacuum construction) are expected for developing the capabilities to manufacture the main optics (presumably fused silica and silicon) and for the development of coatings and coating facilities. Producing ultra-pure optics substrates of approx. 200-470\,kg weight requires an international effort and tight collaboration with industry. Developing coatings of outstanding optical quality and uniformity over the whole mirror surface, combined with the required low mechanical losses at room temperature and cryogenic temperatures is currently regarded as the biggest hurdle to overcome for building \ac{3G} gravitational wave observatories. International collaboration and building redundancy in coating capabilities is essential for success

For underground infrastructures and facilities, \ac{RaD} will incur significant costs for exploration and prototyping. Exploratory efforts have already been started at the \ac{ET} candidate sites. In the construction phase, building the infrastructure and facilities will be the biggest cost items and consequently have the largest cost saving potential. \ac{RaD} efforts to minimise costs while satisfying the strict technical demands is mandatory.

Achieving timely progress in the development of enabling technologies for \ac{3G} detectors will require global collaboration and coordination. A broad, global coherent detector \ac{RaD} program is needed now addressing key technological challenges in the next 5-7~years. In order to facilitate timely development, we propose four broad recommendations on top of the earlier presented subsystem specific recommendations. 
\begin{itemize}
\item \textbf{Recommendation 1}:  An international \ac{3G} \ac{RaD} coordination committee should be formed, with broad and inclusive membership representing \ac{GW} groups across the world.

 A series of workshops on enabling technologies shall be held in order to stimulate exchange of ideas and allowing (if deemed useful) for coordination of the person-power intensive experimental activities.  Each of the major \ac{RaD} tasks should generate a list of  goals with quantitative metrics,  timelines and required resources.   Activities requiring global collaboration and coordination should be laid down and pathways identified.

\item \textbf{Recommendation 2}:  International consortia should be formed to work on key \ac{RaD} challenges and opportunities with industrial partners, establish a governance and organisational structure with authority and seek funding through joint proposals submitted across funding agencies.

\item \textbf{Recommendation 3}: 
National funding agencies should maintain a proactive role in ensuring that the \ac{RaD} activities are well-focused and effective.  Coordination of funding plans through\ac{GWAC} would be a first step, followed by supporting the role of the international \ac{3G} \ac{RaD} coordination committee (recommendation 1) in organizing the global effort.  This is likely to require increased funding for staff and instrumentation to enable the required long-term research programs at relevant laboratories and prototype interferometers.

\item \textbf{Recommendation 4}: Existing \ac{GW} collaborations embrace \ac{3G} \ac{RaD} tasks and integrate them into their programs and deliverables, in order to ensure sufficient support for the long-term future of the field (as it has been done so successfully for \ac{2G} \ac{RaD} during the operation of the initial detectors). Over the next 5 years, mature \ac{2G} enabling technologies (e.g. 1064\,nm laser, fused silica optics, coatings) will need to be scaled up and shown ready for application in \ac{3G} facilities.
\end{itemize}

%% file: Appendix_Core_Optics.tex
\chapter{Core Optics}
\label{sec:Appendix_Core_optics}

\section{Candidate Materials for Low Noise Substrates}
At room temperature fused silica is the best material that can be used as substrate due to its very low thermoelastic effect and ultra low optical and mechanical losses.
At cryogenic temperatures crystalline materials have to replace fused silica. The main challenge is the production of homogeneous large volume substrates with low enough defects and optical absorption. In general, crystalline materials have much less Brownian noise at cryogenic temperatures due to their ordered lattice structure.
\subsection{Silica}
The homogeneity of the refractive index is the main issue here. For the substrates of the cavity mirrors, excellent homogeneity is only important in the two dimensions perpendicular to the beam axis. This can be achieved even for large volumes (corresponding to a total mass of several hundred kilograms). However, the beam splitter requires a very high homogeneity in all three dimensions and this can currently only be guaranteed by the manufacturers for masses up to 40 kg (diameter 55 cm, thickness 7 cm). The company Heraeus has planned some tests to push this limit to about 100 kg. Whether such large beam splitters are required depends on the optical layout of the detector and is under investigation.

\subsection{Silicon}
Silicon has a low mechanical loss at cryogenic temperatures (below 10 K is similar or lower than that of fused silica at room temperature), resulting in low substrate thermal noise. In addition, the thermal expansion coefficient of silicon is zero at $\sim$123\,K and 18\,K, leading to the ability to eliminate substrate thermoelastic noise and thermal expansion effects due to absorbed laser power.

Silicon is not transparent at the currently used wavelength of 1064\,nm. Initially, the telecommunications wavelength of 1550\,nm was proposed for use with silicon mirrors, due to wide availability of high-powered lasers and optical components. More recently, there has been growing interest in using a wavelength close to 2000\,nm. A major driver towards 2000\,nm is the development of amorphous silicon as a possible low thermal noise cryogenic coating material. Amorphous silicon exhibits significantly lower absorption (a factor of $\sim$7) at 2000\,nm than at 1550\,nm. It seems likely, therefore, that the choice of mirror coatings will be a major factor in the choice of wavelength for future detectors. 

Sufficiently low optical absorption can be obtained from silicon produced using the Float Zone technique. However, the maximum size of ingot which can be produced with this method is $\sim$200\,mm. This is too small for the requirements of future detectors (e.g. ET-LF requires 450\,mm diameter, 550\,mm thick optics). While larger diameter silicon pieces can be produced using the Czochralski method, the optical absorption of this type of silicon is too high, due to impurities related to the production method. A magnetic Czochralski process (MCz) exists, in which a magnetic field is used to reduce the impurity concentration in the centre of the ingot. This process can produce diameters of up to 450 mm, and a production line for manufacturing silicon of this diameter does exist at the company Shin Etsu, but is currently not operational. Initial studies of the optical absorption have shown low values at room temperature of $\sim$3\,ppm/cm at 1550\,nm and $\sim$5\,ppm/cm at 2000\,nm. The measurements showed and increase towards lower temperatures, reaching approximately 10\,ppm/cm at 50\,K. This is promising for meeting the requirement of 15\,ppm/cm for cryogenic silicon mirrors. However, initial studies indicate that the absorption of this material can vary significantly, both along the radius and along the length of an ingot, and more studies of the homogeneity of the absorption and its dependence on the thermal history of the sample are required. 

It will be important to test the optical scattering from MCz silicon, particularly as the MCz growth process is known to produce a high void content in the material. Work on this is underway at Glasgow and at Caltech. Initial scattering estimates at Glasgow \cite{SiliconScatter2017} suggest that the scattering is higher than in fused silica, but is likely to be within the required limits.

There is evidence that polishing silicon surfaces can increase their optical absorption. The presence of surface absorption was confirmed in a study in the IGR in Glasgow \cite{SiliconSurfaceAbsorpBell2017} and it was shown that a particular polishing process can be used which does not produce this effect. While this has been consistently demonstrated, further work is required to test whether a silicon surface can be polished to the specifications required for a GW detector without resulting in surface absorption.

Non-linear absorption in silicon is not expected to set a major limit to performance, contributing <\,0.5\,ppm/cm absorption for a wavelength of 2\,$\mu$m at the light intensity assumed for inside the Voyager ITM. At the significantly lower intensity within an ET-LF ITM, these effects are even less significant. Two-photon absorption generates free carriers in silicon: the absorption due to these free carriers depends crucially on the carrier life time. Experiments to measure this for magnetic Czochralski silicon are underway in a collaboration between Stanford and Glasgow.

Two sources of phase noise in silicon optics have not yet been studied experimentally: thermo-refractive noise and carried density noise. These noise sources should be experimentally verified to ensure that they do not set unexpected limits of silicon mirror performance. An experiment targeted at measuring thermo-refractive noise in silicon is currently being constructed in Glasgow.

\subsection{Sapphire}
Sapphire is transparent at 1064\,nm and hence does not require a change from the currently used detector wavelength. Mechanical quality factors of sapphire samples are as high as few $10^8$ at room temperature\cite{Rowan_2000a} and they increase at cryogenic temperatures\cite{uchiyama1999mechanical}. 
Sapphire's elastic constants are about 3 times higher than silicon's, helping to reduce thermal noise. The thermal conductivity of sapphire increases with decreasing temperature and reaches a peak of several $10^3$\,W/(m \,K) around 20-40K.  Thermoelastic noise is also quite low due to the high thermal conductivity at low temperatures.  This high conductivity (and the low temperature coefficient) make the thermal lens effect negligible~\cite{Tomaru:2002}. The high Young's modulus of elasticity has two additional advantages: fewer parametric instabilities~\cite{Yamamoto:2008} and a higher calibration frequency in a gravitational wave detector.
The optical absorption of sapphire has been found to vary strongly from crystal to crystal and for crystals from different suppliers. There is only a very small chance of finding sapphire crystals with low absorption by cherry-picking among a large number of products on the market. In KAGRA, it turned out that it was necessary to develop sapphire crystals with low absorption by working closely with the crystal manufacturers. Although it took some time, sapphire crystals were developed to meet 50ppm/cm with sufficient margin \cite{Hirose_2014a}. The production success rate is two out of twelve bulk.
Theoretical work on scattering in sapphire sets a lower limit of 0.21\,ppm/cm, with higher measured values of around 13 ppm/cm being attributed to impurities and vacancies.
Sapphire's high Mohs hardness of 9 and its crystalline structure, resulting in orientation dependent machinability, makes it harder to process sapphire substrates. Polishing the KAGRA test masses took much longer than polishing the aLIGO test masses, although the size is smaller. Despite this hardness, no degradation in micro roughness or surface figure of the polished surfaces has been observed in KAGRA sapphire test masses\cite{Hirose_2014a}, compared to aLIGO or advanced VIRGO test masses made of fused silica. 
Sapphire is known to be birefringent. Since sapphire is a uniaxial crystal, the alignment of the optical axis, called the c-axis, to the beam axis should theoretically make the refractive index uniform in the plane perpendicular to the beam axis. In reality, however, even with optimized alignment, it is common that the inhomogeneity of the refractive index of sapphire is one order of magnitude worse than that of fused silica. It is therefore essential to compensate for this by locally adjusting the thickness of the substrates. This can be achieved by aspherically polishing the back of the mirror, and the ion beam figuring technique (IBF) has been applied to the KAGRA ITMs.

Dielectric multilayer coating of SiO2 and Ta2O5 on sapphire substrates appear to be as good as such coatings deposited on silica substrates\cite{Yamamoto:2006,Hirose_2014b}.
The biggest c-axis sapphire window currently available in the market for gravitational wave detectors is about 220mm in diameter and 150mm in thickness (mass of 23 kg) although the ingot size is actually much larger. These are grown by either HEM or TSMG methods which assure the lowest dislocation density among all the production methods. If we allow crystals to have bubbles or apparent defects inside, the size could be much larger. The ingot size has been simply limited by size of furnace where sapphire crystal is grown.

The KAGRA monolithic suspensions feature sapphire components for the core optic, the ears, fibers, and blade springs. As attached techniques, hydroxide catalysis bonding, indium~\cite{Kumar:2016_KAGRA} and gallium bonding.
KAGRA has demonstrated that sapphire can be successfully bonded through hydroxide catalysis bonding. 
Sapphire technology has also been extensively developed in order to produce mono--crystalline fibres of length greater than 1\,m and large plates of a large variety of thickness.

A collaborative research project with a Japanese company has commenced  targeting 100kg sapphire windows whose diameter is 400mm and thickness is 200mm. Developing that large size windows compatible with lower absorption and higher homogeneity will be a key to success toward 3G detectors with sapphire test mass mirrors.

At the end of 2019 the Universit\'e de Lyon and the CNRS has funded a project called OSAG submitted by the groups g-MAG at iLM and LMA at iP2i. The aim of the project is to design and build an oven to grow along the c-axis 500 kg of crystalline sapphire. These type of ingots will be used to develop 450 mm diameter sapphire substrates for cryogenic GW detectors, reducing the absorption down to 10 ppm/cm. The g-MAG group is already equipped with $\mu$-pulling down setups to draw sapphire fibres of length up to 1.2 m.

%% file: Appendix_Coating.tex
\chapter{Coatings}
\label{sec:Appendix_Coatings}

Different techniques for minimizing thermal coating noise for the candidate wavelengths (1064\,nm, 1550\,nm, 2+\textmu m) and envisaged temperatures (293\,K, 124\,K, 18\,K) are listed in the following section. \\

\begin{figure}[ht]
\centering
\includegraphics*[width=\textwidth]{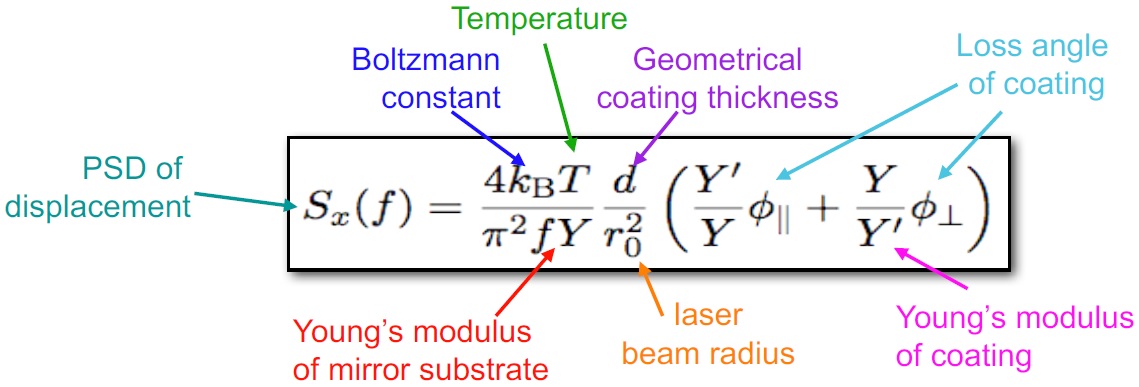}
\caption{Parameters contributing to Thermal noise in GWDs. Equation from \cite{Harry:CQG2002}}
\label{fig:Thermal_Noise}
\end{figure}

\section{Amorphous Oxides}
Empirical results for amorphous oxide mirrors deposited by IBS show that the mechanical loss at room temperature is reduced by post-deposition annealing, and that the reduction is greater at higher annealing temperatures \cite{vajente2018effect}. The maximum annealing temperature is limited by crystallization of the film. One approach under investigation is to chemically frustrate crystallization through introduction of an appropriate dopant into the coating to increase the temperature at which crystallization occurs. Another approach is to use the oxides mixing available on the market that do not crystallize and that have high refractive index.

There is also some evidence that low deposition rates correlate with reduced losses. A promising result combining these two approaches, the low rate deposition of Zr:tantala and subsequent high temperature annealing, yielded a film with the 4-fold reduction in mechanical loss at room temperature required for aLIGO+ and AdVirgo+ mirrors, though this result is not fully reproducible. It should also be noted that no measurements of cryogenic losses are available and changes in room temperature losses and cryogenic losses are often anticorrelated, so the effects of this material on cryogenic 2.5G and 3G detectors are not yet clear \cite{martin2010effect}. Some measurements have begun to support a different approach to suppress crystallization and thus allow higher annealing temperatures, geometrically frustrating crystallization using nano-layers, where each quarter-wave layer consists of alternating layers of two materials with a thickness of several nanometers. Suppression of crystallization has been observed in such titania/silica nanolayers \cite{pan2014thickness}. Concerns such as homogeneity, scattering and interface effects in the nanolayers have not yet been characterized.

Another line of research is the one that is trying to use the crystallization as a way to organize the material to medium range order. The aim is to reduce the TLS density and at the same time to keep the nano-crystals dimensions at a level that are not relevant for the scattering.

Until recently, progress in reducing mechanical losses has been based on empirical studies alone. Recently, there has been progress in theoretical tools to guide the empirical work. Molecular dynamics methods enable calculation of the atomic structures that correspond to the energy spectrum of TLS in amorphous materials. Theoretical results correctly predict empirical data for the temperature dependence (at cryogenic temperatures) of losses in model systems like silica, and the trend in loss vs Ti-doping in tantala \cite{trinastic2016molecular}. Extension to room-temperature regimes is currently under development. These methods also have suggested dopants whose promise has been borne out experimentally, e.g. Zr in tantala. Closely related are experimental methods to determine atomic structure in amorphous films via electron or X-ray diffraction techniques \cite{bassiri2013correlations,hart2016medium,shyam2016measurement}. A “virtuous circle” of theoretical modeling, atomic structure characterization, and macroscopic property measurements (in particular mechanical loss) has reached the stage where the methods mutually reinforce and speed developments in the respective methods.

A theoretical concept that has emerged is that of an ultra-stable glass, i.e. one that has an atomic structure of low internal energy with greatly reduced density of TLS, which is inaccessible to conventional annealing processes on realistic time scales, but can be reached through vapor deposition \cite{singh2013ultrastable}. According to this picture, deposition at elevated temperatures (to increase surface mobility) and at low rates (to give longer times for surface atoms to explore the energy landscape) should lower the density of TLS. The first empirical example of such an ultra-stable inorganic material, elevated temperature deposition of amorphous silicon with two-order-of-magnitude reduction in cryogenic mechanical losses compared to room-temperature deposition, indicates the physical reality of the concept \cite{liu2014hydrogen}. This result both shows that a-Si is a promising material for long-wavelength mirrors (though unacceptable levels of optical absorption, discussed below, remain an issue), and motivates the search for ultra-stable amorphous oxides suitable for 1\,$\mu$m wavelength operation. Recent results for amorphous alumina show lower cryogenic losses even after annealing than in room-temperature deposited films, which appears, subject to further verification, the first example of an ultra-stable amorphous oxide. It is also a promising material for a low-index layer in a cryogenic mirror, since its mechanical loss is almost an order of magnitude lower than silica at cryogenic temperatures. These studies are in an early stage; further work (theoretical and empirical) is necessary to establish whether the ultra-stable glass concept is a route to a low-noise cryogenic mirror.  

\section{Semiconductor Materials}

\noindent The results for the mechanical loss of amorphous silicon deposited at elevated temperatures show that a mirror consisting of a-Si as a high-index layer together with silica as a low index layer would meet the thermal noise requirements for cryogenic detectors \cite{steinlechner2018silicon}. The issue that remains in this case is the optical loss, which is more than an order of magnitude higher than typical requirements. This absorption is associated with ``dangling bonds'' in silicon atoms that are three-fold rather than four-fold coordinated. As such, this mechanism is not intrinsic, and depends strongly on deposition conditions and post-deposition processing. For example, low deposition rates, post-deposition annealing and annealing in hydrogen can all reduce the absorption significantly, and are topics of active current research \cite{birney2018amorphous}. The absorption cross-section decreases with increasing wavelength, resulting in a strong preference for 2\,$\mu$m vs 1.5\,$\mu$m wavelength operation, and probably eliminating this material for use at 1\,$\mu$m. It is also worth noting that the best results for cryogenic losses in a-Si have been obtained with films fabricated by evaporation rather than sputtering \cite{liu2014hydrogen}. Verification of these results in sputtered films is important, since meeting optical quality requirements is not practical via evaporative deposition methods.

Silicon nitride is another material widely used in the semiconductor industry with potential for low-noise mirrors. Low-pressure chemical vapor deposition (LPCVD) produces films with adequate mechanical loss that, together with silica low-index layers, could form a mirror that meets thermal noise specifications for ET-LF or Voyager \cite{pan2018silicon}. IBS deposited silicon nitride has been also produced with a 1.8-fold reduction of mechanical losses at room temperature with respect to titania doped tantala \cite{pan2018silicon}. Both processes result in excess optical absorption, which is currently more than order of magnitude too large to meet specifications, but, with further development, silicon nitride, unlike amorphous silicon, has the potential to be used with 1\,$\mu$m wavelength operation. Another potential problem is the high Young's modulus, which makes the material more suitable for silicon or sapphire than for silica substrates.

\section{Multi-Material Coatings}

In this sense, the problem for semiconductor materials is the converse of that for oxide materials: the mechanical properties are adequate, but the optical absorption is too high. A general approach that can address this problem is the use of multi-material coatings, in which the top layers, where the optical intensity is highest, consist of materials with low optical absorption but too large mechanical loss, while the lower layers consist of materials with low mechanical loss but too large optical absorption \cite{yam2015multimaterial,steinlechner2015thermal}. In this way, the limitations of the types of material can be traded off against each other. Examples exist of multi-material coatings based on demonstrated material properties that could meet the requirements of ET-LF; the method is reviewed in \cite{Craig2018ETmultimaterialDCC}. While the additional complexity of depositing such multi-material coatings could pose a fabrication challenge, especially if the optimal deposition method differs for the different materials involved, it remains an important option if simpler material combinations fail to meet requirements.

\section{Crystalline Coatings}

Epitaxially grown coatings consisting of alternating layers of high and low index layers of crystalline materials present an alternative method that avoids the mechanical loss issues associated with TLS in amorphous materials. The best developed of these materials are AlGaAs/GaAs mirrors grown by molecular beam epitaxy (MBE) on GaAs substrates and then transferred by wafer-bonding methods to mirror substrates such as silica, silicon or sapphire \cite{cole2013tenfold}. The mechanical and optical losses measured in small cavities with mirrors made in this fashion appear adequate for 2.5G and 3G detectors. As compressional mechanical losses are considerably higher than shear losses, it needs to be clarified whether thermal noise is as low for larger beam diameters, where compressional losses become more important.

The issues with respect to applying these as general solutions for gravitational wave interferometry are related to scaling. One scaling issue is associated with developing the infrastructure necessary for fabrication of $\mathrm{\sim}$\,40\,cm (or larger) diameter optics: growth of the necessary GaAs crystal substrates, and development of MBE and wafer-bonding tools meeting the optical homogeneity requirements. An order of magnitude estimate of the cost of that effort is $\mathrm{\sim}$\,\$\,25\,M \cite{Cole2018AlGaAsCost}. The other issue is related to MBE growth, in which an areal density of discrete defects generally appears. While these defects can be avoided in experiments with small beams by alignment in small cavities, this would not be the case with beam sizes characteristic of full-scale detector configurations. The optical scatter, absorption, and elastic loss associated with these defects are not yet well characterized, nor is their anticipated density known. Samples with several inch diameters are currently being characterized to better understand these issues, which presumably should be clarified before the investment in scaling the tooling is seriously considered. The effects of the electro-optic and piezoelectric sensitivities of these crystalline materials must also be investigated. 

A less well-developed approach is the use of GaP/AlGaP crystalline coatings \cite{lin2015epitaxial}. These have the advantage of being able to be grown directly onto silicon, removing the need for transferring the coating between substrates. Initial measurements of the cryogenic mechanical loss of these coatings appear very promising for use in cryogenic gravitational wave detectors. However, significant work on refining deposition parameters and reducing optical absorption is likely to be required.

\section{Current Research Programs}

There are several current and planned programs devoted to developing low-thermal-noise mirror coatings suitable for enhanced 2G, 2.5G, and 3G detectors. Participants are involved in all aspects of coating research, including various deposition methods, characterization of macroscopic properties at room and cryogenic temperatures, and atomic structure modeling and characterization. The recent incorporation into the project of several groups involved in coating deposition is particularly important, as the costs and time delays associated with commercial deposition of research coatings have been significant impediments to rapid progress. The efforts in the LSC and Virgo are presently somewhat independent although with good information exchange. 

\subsection{LSC}

There are approximately ten U.S. university research groups participating in various aspects of coating research. In late 2017, a more coordinated effort and additional funding for these groups were initiated under the Center for Coatings Research (CCR), jointly funded by the Gordon and Betty Moore Foundation and the NSF. Work in the U.S. also importantly includes that in the LIGO Laboratory. These efforts are complemented by groups not formally affiliated with the CCR, notably that by the large optical coating group at U. Montreal. Other major LSC coatings research programs are those in GEO (U. Glasgow, U. Strathclyde, U. West of Scotland, U. Hamburg, U. Hannover). The efforts of these groups are coordinated through biweekly telecons of the LSC Optics Working Group. In 2020, the Australian National University commenced re-commissioning of the IBS coating chamber used successfully by the CSIRO for aLIGO optics.  This will hopefully provide a large optic coater for the collaboration.  

\subsection{VIRGO}

In Europe about ten universities are involved in various aspects of the VIRGO Coatings R\&D Project, including new material research, metrology and more recently simulation. The ViSIONs project, supporting six French laboratories, including the LMA and ILM, is focused on studying the relation between the physical properties of sputtered or evaporated materials and the structural and macroscopic properties of the deposited films. LMA is the only facility currently operational that is capable of producing coatings of the size and optical quality required for gravitational-wave interferometers. LMA has already started improving the uniformity of coating deposition in order to meet the challenges of the Advanced+ detectors. The AdVirgo+ project is considering the use of end cavity mirrors of 55 cm diameter. Therefore LMA has developed plans to upgrade their coaters and tools to deal with such increase of diameter and weight.

%% file: acronyms.tex
\chapterimage{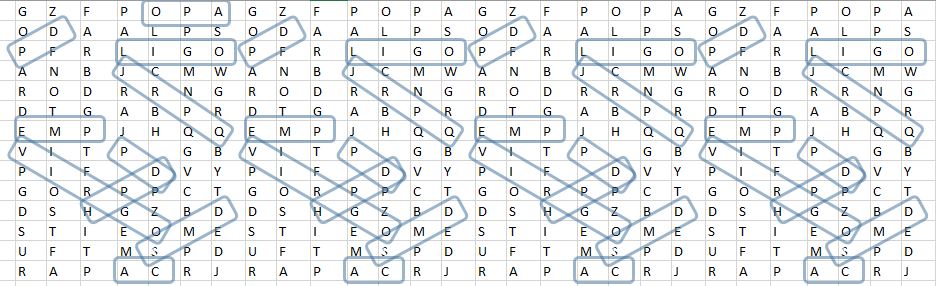} 
\chapter{List of Acronyms}
\label{sec:AcronymList}
\pagenumbering{roman}
\renewcommand*{\thepage}{AL\roman{page}}
\textbf{1G} first generation\\
\textbf{2G} second generation\\
\textbf{2.5G} an improved version of the second detector generation with incrementally new technology\\
\textbf{3D} three dimensional\\
\textbf{3G} third generation\\
\textbf{A+} the upgraded advanced interferometer generation\\
\textbf{LIGO A+}  upgraded advanced Laser Interferometer Gravitational wave Observatory\\
\textbf{AdVirgo} advanced Virgo detector\\
\textbf{AdVirgo+} upgraded advanced Virgo detector\\
\textbf{AEI} Albert Einstein Institute\\
\textbf{AlGaAs/GaAs} Aluminium-Gallium-Arsenide / Gallium Arsenide\\
\textbf{aLIGO} advanced LIGO\\
\textbf{Alphanov} Centre Technologique Optique et Lasers, France\\
\textbf{ANR} Agence nationale de la recherche\\
\textbf{ANU} Australian National University\\
\textbf{AWC} Active Wavefront Control\\
\textbf{BHD} balanced homodyne detector\\
\textbf{BNS} binary neutron star\\
\textbf{CAD} computer aided design\\
\textbf{CCR} Center for Coatings Research, Stanford\\
\textbf{CE} Cosmic Explorer\\
\textbf{CE1} Cosmic Explorer phase 1, using 2G technology\\
\textbf{CE2} Cosmic Explorer phase 2\\
\textbf{CERN} European Centre for Nuclear Research\\
\textbf{CLIO} cryogenic 100\,m prototype interferometer in the Japanese Kamioka mine\\
\textbf{CNRS} Centre National de la Recherche Scientifique (the French national research centre for scientific research) \\
\textbf{CSIRO} Commonwealth Scientific and Industrial Research Organisation (an Australian federal government agency responsible for scientific research)\\
$d$ coating thickness\\
\textbf{DarkF} an optical simulation code in FORTRAN 90\\
\textbf{DARM} differential arm motion\\
$\sqrt{S_x(\omega)}$ Fourier components of the displacement of the system\\
\textbf{EOM} electro optic modulator\\
\textbf{ET} Einstein Telescope\\
\textbf{ET-HF} Einstein Telescope high frequency interferometer\\
\textbf{ET-LF} Einstein Telescope low frequency interferometer\\
\textbf{ETM} End Test Mass\\
\textbf{FC} filter cavity\\
\textbf{FEA} finite element analysis\\
\textbf{FFT} fast Fourier transform\\
\textbf{FI} Faraday isolator\\
\textbf{Finesse} Frequency-domain interferometer simulation with higher-order spatial modes\\
$F(\omega)$  Fourier components of the force leading to the displacement of the system \\
\textbf{FP} Fabry-Perot cavity, a linear optical resonator\\
\textbf{FPGA} field programmable gate array\\
\textbf{GaP/AlGaP} Gallium Phosphide / Alluminium Gallium Phosphide\\
\textbf{GeNS} Gentle Nodal Support System\\
\textbf{GEO} the GEO collaboration\\
\textbf{GEO 600} the German/British 600m gravitational wave detector in Germany\\
\textbf{GPU} graphics processing unit\\
\textbf{GR} general relativity \\
\textbf{GW} gravitational waves\\
\textbf{GWAC} gravitational wave agency correspondents\\
\textbf{GWD} gravitational wave detector\\
\textbf{GWIC} gravitational wave international committee\\
\textbf{GWINC} Gravitational wave interferometer noise calculator\\
\textbf{GWADW} Gravitational Wave Advanced Detector Workshop\\
\textbf{HEPI} hydraulic external pre-isolator\\
\textbf{Heraeus} a German company producing ultra-pure fused silica glass\\
\textbf{HPL} long = high power laser\\
\textbf{IBS} ion-beam sputtering\\
\textbf{ICRR} institute for cosmic research, University Tokyo\\
\textbf{IFI} input Faraday isolator\\
\textbf{IfoCAD} 3D ray tracing tool using Gaussian beams\\
\textbf{IIT} Indian Institute of Technology\\
\textbf{ILM} nsitut Lumi\`ere Materi\`e\\
\textbf{IMC} input Mode cleaner\\
\textbf{ITM} long = Inner Test mass\\
\textbf{JLab} Jefferson Lab\\
$k_b$ the Boltzmann constant\\
\textbf{KAGRA} Japanese underground gravitational wave detector\\
$L$ interferometer arm length\\
\textbf{LASTI} LIGO advanced system test interferometer at MIT\\
\textbf{LIDAR} light detection and ranging\\
\textbf{LIGO} Laser Interferometer Gravitational wave Observatory\\
\textbf{LIGO Lab} laboratory running the LIGO Project\\
\textbf{LMA} Laboratoire des Matériaux Avanc\'es \\
\textbf{LSC} LIGO Scientific Collaboration\\
\textbf{LVC} LIGO Virgo Collaboration\\
\textbf{LZH} Laser Zentrum Hannover\\
\textbf{MBE} Molecular Beam Epitaxy\\
\textbf{MCMC} Markov-Chain-Monte-Carlo method\\
\textbf{MCz} magnetically assisted Czochralski\\
\textbf{MIMO} multiple in, multiple out\\
\textbf{MIST} Matlab-based fast modal simulation of paraxial optical systems \\
\textbf{MIT} Massachusetts Institute of Technology\\
\textbf{Mitsubishi} Japanese company\\
\textbf{ML} maturity level\\
\textbf{MOPA} master oscillator power amplifier\\
\textbf{NEMO} Neutron Star Extreme Matter Observatory, a plan for a high frequency  GW detector in a 2G scale facility\\
\textbf{neoVAN4S} a solid state laser of the German company NEOLASE\\
\textbf{NIST} National Institute of Standards and Technology\\
\textbf{NN} Newtonian noise, a gravitational coupling of surrounding mass (especially of moving mass due to seismic waves or infra-sound) to the mirrors\\
\textbf{NSF} National Science Foundation\\
\textbf{Nufern} Fibre laser company \\
\textbf{OFI} Output Faraday isolator\\
\textbf{OMC} output mode cleaner\\
$\omega$ the angular frequency\\
\textbf{Optickle} a general model for the electro-opto-mechanical part of a GW detector\\
\textbf{OptoCAD} 2D ray tracing tool using Gaussian beams\\
\textbf{OSCAR} optical FFT code to simulate Fabry Perot cavities with arbitrary mirror profiles\\
$\varphi$ mechanical loss angle\\
\textbf{PR} Power Recycling\\
\textbf{PSL} pre-stabilized laser system\\
\textbf{QND} Quantum Non-Demolition\\
\textbf{QRPN} quantum radiation pressure noise\\
\textbf{R\&D} research and development\\
\textbf{RF} radio frequency \\
\textbf{RMS} root mean square \\
\textbf{SA} Super Attenuator\\
\textbf{SAS} Seismic attenuation system\\
\textbf{SimPlant} virtual interferometer for commissioning \\
\textbf{SIS} seismic isolation system\\
\textbf{SN} shot noise\\
\textbf{SNR} signal to noise ratio\\
\textbf{SPI} suspension point interferometer\\
\textbf{SQL} standard quantum limit\\
\textbf{SQZ} squeezing, a technology to modify quantum noise distribution over the two quadratures of the light field\\
\textbf{SR} Signal Recycling\\
\textbf{SUS} suspension system\\
$T$ Temperature\\
\textbf{TCS} thermal compensation system\\
\textbf{TGG} Terbium gallium garnet\\
\textbf{TLS} two level system\\
\textbf{UHV} Ultra High Vacuum\\
\textbf{USA} United States of America\\
\textbf{UWS} University of the West of Scotland\\
\textbf{VCR\&D} long = Virgo Coating R\&D collaboration\\
\textbf{Virgo} Virgo gravitational wave detector\\
\textbf{ViSIONs} a coating project funded by \acs*{ANR}\\
\textbf{Voyager} Voyager, a detector with 3G technology in the 2G infrastructure\\
 $w$ laser beam radius\\
 $Y$ mechanical admittance\\